\documentclass[apj,iop]{emulateapj-rtx4}
\usepackage{apjfonts}
\usepackage{lscape}

\usepackage{amsmath}
\usepackage{amssymb}

\usepackage{xspace}

\newcommand{\bzk}{$BzK$\xspace}

\newcommand{\sbzk}{$sBzK$\xspace}
\newcommand{\sbzks}{$sBzK$s\xspace}

\newcommand{\pbzks}{$pBzK$s\xspace}
\newcommand{\ha}{H$\alpha$}
\newcommand{\hasp}{H$\alpha$\xspace}
\newcommand{\hb}{H$\beta$}
\newcommand{\hbsp}{H$\beta$\xspace}
\newcommand{\fha}{\ensuremath{f(\text{H}\alpha})}
\newcommand{\fnii}{\ensuremath{f(\text{[\ion{N}{2}]$\lambda6583$})}}

\newcommand{\ebv}{\ensuremath{E(B-V)}\xspace}
\newcommand{\pegase}{P\'EGASE.2\xspace}
\newcommand{\tausf}{\ensuremath{\tau_{\text{sf}}}}
\newcommand{\oh}{\ensuremath{12+\log(\text{O}/\text{H})}\xspace}
\newcommand{\kms}{\ensuremath{\,\text{km}\ \text{s}^{-1}}}

\newcommand{\lz}{{$L$-$Z$}\xspace}
\newcommand{\mz}{{$M_\star$-$Z$}\xspace}
\newcommand{\mssfr}{{$M_\star$-SSFR}\xspace}

\defcitealias{kobulnicky:2004}{KK04}
\defcitealias{kong:2006}{K06}
\defcitealias{tremonti:2004}{T04}

\slugcomment{Accepted for publication in Astrophysical Journal}

\shorttitle{NIR Spectroscopy of \sbzks at $z\sim2$}

\shortauthors{Onodera~et~al.}

\begin{document}

\title{
  A Wide Area Survey for High-redshift Massive Galaxies. II. 
  Near-infrared Spectroscopy of B\lowercase{z}K-selected 
  Massive Star-forming Galaxies\,\altaffilmark{1}
}

\author{
  Masato Onodera\altaffilmark{2,3}, 
  Nobuo Arimoto\altaffilmark{4,5}, 
  Emanuele Daddi\altaffilmark{2}, 
  Alvio Renzini\altaffilmark{6}, 
  Xu Kong\altaffilmark{7}, 
  Andrea Cimatti\altaffilmark{8}, 
  Tom Broadhurst\altaffilmark{9}, 
  and 
  Dave M. Alexander\altaffilmark{10}
}


\altaffiltext{1}{Based on data collected at Subaru Telescope, 
  which is operated by the National Astronomical Observatory of Japan 
  (S04A-081, S05A-098), and on observations collected at 
  the European Southern Observatory, Paranal, Chile (075.A-0439).}

\altaffiltext{2}{CEA, Laboratoire AIM-CNRS-Universit\'e Paris Diderot, Irfu/SAp, 
  Orme des Merisiers, F-91191 Gif-sur-Yvette, France; \texttt{masato.onodera@cea.fr}}

\altaffiltext{3}{Department of Astronomy, Yonsei University, 
  Sinchon-dong 134, Seodaemun-gu, Seoul, South Korea}

\altaffiltext{4}{National Astronomical Observatory of Japan, 
  Osawa 2-21-1, Mitaka, Tokyo, Japan}

\altaffiltext{5}{Graduate University for Advanced Studies, Osawa 2-21-1, Mitaka, Tokyo, Japan}

\altaffiltext{6}{INAF-Padova, Vicolo dell'Osservatorio 5, I-35122 Padova, Italy}

\altaffiltext{7}{CfA, University of Science and Technology of China, Hefei 230026, China}

\altaffiltext{8}{Dipartimento di Astronomia, Universit\`a di Bologna, Via Ranzani 1, 40127 Bologna, Italy}

\altaffiltext{9}{School of Physics and Astronomy, Tel Aviv University, Tel Aviv 69978, Israel}

\altaffiltext{10}{Institute for Computational Cosmology, Department of Physics, Durham University, Durham DH1 3LE, UK}

\begin{abstract}
Results are presented from near-infrared spectroscopic observations of a sample
of $BzK$-selected, massive star-forming galaxies (\sbzks) at
$1.5<z<2.3$ that were obtained with OHS/CISCO  at the
Subaru telescope and with SINFONI at the VLT.  Among the 28 \sbzks
observed, \hasp emission was detected in 14 objects, and 
for 11 of them the [\ion{N}{2}]$\lambda6583$ flux was also measured. Multiwavelength
photometry was also used to derive stellar masses and extinction
parameters, whereas \hasp and [\ion{N}{2}] emissions have allowed us
to estimate star-formation rates (SFR), metallicities, ionization
mechanisms, and dynamical masses. In order to enforce agreement between
SFRs from \hasp with those derived from rest-frame UV and mid-infrared,
additional obscuration for the emission lines (that originate in
\ion{H}{2} regions) was required compared to the extinction derived
from the slope of the UV continuum.  We have also derived  the stellar
mass-metallicity relation, as well as the relation between stellar
mass and specific SFR, and compared them to the results in other
studies. At a given stellar mass, the \sbzks appear to have 
been already enriched to metallicities close to those of local star-forming
galaxies of similar mass. The \sbzks presented here tend to have higher metallicities
compared to those of UV-selected galaxies, indicating that
near-infrared selected galaxies tend to be a chemically more evolved
population. The \sbzks show specific SFRs that are systematically higher, by up
to $\sim2$ orders of magnitude, compared to those of local galaxies of
the same mass. The empirical correlations between stellar mass and
metallicity, and stellar mass and specific SFR are then compared with those of
evolutionary population synthesis models constructed either with the simple
closed-box assumption, or within an infall scenario. Within the
assumptions that are built-in such models, it appears that a short
timescale for the star-formation ($\simeq100$~Myr) and large initial
gas mass appear to be required if one wants to reproduce both
relations simultaneously.
\end{abstract}


\section{Introduction}
Much action takes place during the $\sim 2$ billion years of cosmic
time between $z=3$ and $z=1.4$: both the global star-formation rate
(SFR) and nuclear (AGN) activity peak at $z\sim 2$, while much of the
galaxy assembly and morphological differentiation are taking place and
a population of massive, passively evolving galaxies is gradually
emerging. Thus a full exploration of this redshift interval is
critical for our understanding of galaxy evolution. Unfortunately,
this is also a relatively difficult redshift range to penetrate
observationally, as strong spectral features such as \ion{Ca}{2} H\&K
and the 4000 \AA{} break (for passive galaxies) and
[\ion{O}{2}]$\lambda3727$, [\ion{O}{3}]$\lambda5007$, \hasp and \hb\
(for star forming ones) have all moved to the near-infrared (NIR).

Still, most spectroscopic surveys of galaxies at $1.4\lesssim z\lesssim
3$ have used optical spectrographs, relying on much weaker
spectroscopic features for redshift determination and galaxy
characterization, with these limitations being partly compensated by
the possibility to achieve a fairly high multiplex
\citep[e.g.,][]{steidel:2004,mignoli:2005,lefevre:2005,vanzella:2005,
lilly:2007,cimatti:2008,popesso:2009}.
For passive galaxies at $z>1.4$ such features practically restrict to
a set of absorption features at $\lambda\lambda \sim 2600$--$2850$ \AA{} due
to neutral and singly ionized Mg and Fe, and for star-forming galaxies
to several weak absorption lines over the rest-frame UV continuum,
most of which due to the interstellar medium (ISM)  of these galaxies.

The intrinsic weakness of the absorptions and/or of the continuum made
such spectroscopic observations very demanding in terms of telescope
time.  Therefore, several studies of large samples of galaxies at
$1.4\lesssim z\lesssim 2.5$ have relied on color selections and
photometric redshifts.  Particularly effective has proven the $BzK$
criterion introduced by \citet{daddi:2004bzk}, which is able to select
both star-forming (called \sbzks) as well as passively evolving
galaxies (\pbzks) over this redshift range. This has enabled 
estimates of SFRs, stellar masses, and clustering properties of such
galaxies, with samples from $\sim 100$'s to over $\sim 30,000$ objects
\citep[e.g.,][]{kong:2006,daddi:2007sfr,dunne:2009,mccracken:2010}.
The $BzK$ technique ensures a nearly unbiased selection of $z\sim 2$ galaxies, 
including UV-selected galaxies and single color, NIR-selected galaxies 
\citep[e.g.,][]{reddy:2005,mccracken:2010}.

Whereas many aspects concerning the evolution of galaxies can be
investigated using only photometric redshifts, spectroscopy remains
indispensable for a variety of investigations. These include full
mapping of the local environment (locating clusters, groups, filaments
and voids), refining SFR and mass estimates, measure stellar and ISM
metallicities, and finally map the internal dynamical workings of
galaxies via 3D spectroscopy. In this respect, the required telescope
time is not the only drawback of the optical spectroscopy of galaxies
at $1.4\lesssim z\lesssim 3$.  
Indeed, optical spectroscopy down to a
limit as faint as $B\sim 25$ does not recover 
but a minor fraction of the global SFR and stellar mass at $z\sim 2$, in
particular missing galaxies that are among the most massive and most
intensively star-forming ones \citep{renzini:2009b}.  
Moreover, high spatial resolution, 3D spectroscopy of high redshift galaxies requires
the knowledge of the spectroscopic redshift, to make sure
that interesting emission lines (e.g., H$\alpha$) are free from OH and
other atmospheric contaminations \citep[see e.g.,][]{genzel:2006,forsterschreiber:2009}.

In the case of star-forming (\sbzk) galaxies, the poor correlation of
mass and SFR with $B$ magnitude is a result of high
extinction. Therefore, the situation should appear more favorable in
the NIR, and not only because moving to longer wavelengths should
reduce the impact of extinction, but also because the most active
star-forming and most massive galaxies are also among the brightest
objects at these longer wavelengths, and one can access strong emission lines such as [\ion{O}{2}]
and H$\alpha$. Yet, NIR spectroscopy of $z\gtrsim 1.4$ galaxies is
still in its infancy, especially for NIR selected
samples. \citet{erb:2006mz,erb:2006mass,erb:2006sfr} have presented
results for over 100 UV-selected galaxies at $z\sim 2$, deriving SFRs
from the strength of H$\alpha$. NIR spectroscopic observations of
samples of $z\sim 2$ galaxies selected in the NIR have been also presented
by \citet{kriek:2006a,kriek:2006b,kriek:2007,kriek:2008a,kriek:2008b},
focusing mainly (but not exclusively) on passive galaxies by
detecting the 4000 \AA{} break, and deriving spectro-photometric
redshifts from it.  NIR spectroscopy of \sbzk{} galaxies has been carried out by
\citet{hayashi:2009} for a sample of 40 \sbzks, and detected \hasp
emission from 15 of them.  Their detections, however, are limited to 
$z<2$.  Finally, integral field NIR spectroscopy for  some 60 star forming galaxies 
at $z\sim 2$ has been obtained by \cite{forsterschreiber:2009},  for partly UV-selected, partly $BzK$-selected targets.  
Therefore, there is still just scanty spectroscopic information
in the rest-frame optical wavelength for actively star-forming and
heavily obscured galaxies at $z\simeq2$, many of which would be missed
by the UV-selection, and or are virtually unreachable  by current optical spectroscopy. 

In the perspective of improving upon this situation, in 2004 we started
NIR spectroscopic observations of $z\sim 2$ galaxies primarily selected
on \bzk{} technique, and using a variety of NIR instruments, 
namely OHS, CISCO and MOIRCS, at the Subaru telescope and SINFONI 
at the VLT.  Our intent was to explore the effectiveness of
NIR spectroscopy to improve our characterization of $z\sim 2$ galaxies
using a relatively small pilot sample of them, while assessing the
feasibility of wider surveys with future instruments with higher
multiplex. In this paper we present the results of observations with
the OHS/CISCO instruments on Subaru and SINFONI instrument on VLT 
of \sbzk galaxies, leaving the results obtained with the MOIRCS instrument 
for a future paper. With the observations presented here we have attempted to measure for each
galaxy the dust extinction, SFR, ISM metallicity, and dynamical mass,
while checking for a possible AGN contribution. Physical quantities are
derived assuming the concordance cosmology, i.e.,
$\Omega_{\text{M}}=0.3$, $\Omega_{\Lambda}=0.7$, and $H_0 = 70$
\kms~Mpc$^{-1}$ and photometric magnitudes are expressed in the AB system
\citep{oke:1983} if it is not explicitly noted otherwise. For the solar oxygen
abundance, we use $12+\log(\text{O/H})_\odot=8.69$ \citep{allendeprieto:2001}.
Emission line width are measured  assuming a Gaussian profile 
and the FWHM and the velocity dispersion ($\sigma$) is given by $\text{FWHM}=2.355\sigma$.

\section{Sample Selection, Observations, and Data Reduction}

\subsection{Sample Selection\label{section:sample}}
The \sbzk galaxies which are the object of the present study have been
culled from the $K$-band selected catalog of objects in the EIS Deep3a
field and Daddi field, which are described along with the photometric
data in \citet[hereafter \citetalias{kong:2006}]{kong:2006}.  We have
primarily selected \sbzks with spectroscopically confirmed redshifts
from the rest-frame UV spectroscopy obtained with the VIMOS instrument at
the VLT (Daddi et al., in preparation).  Then we have selected those
for which the \hasp emission line falls at wavelengths where
atmospheric and instrumental transmission are high and is not
contaminated by OH-airglow lines.  Additional criteria have then been
applied to select objects which (1) are bright in the $K$-band
($K_{\text{AB}}\lesssim22$) to select massive galaxies, 
i.e., $M_\star\gtrsim5\times10^{10}\,M_\odot$ 
\citep[][see also \S\ref{section:massdust}]{daddi:2004bzk}, (2) have red $z-K$
colors, $(z-K)_{\text{AB}}\gtrsim2$ to select possibly more reddened
ones, (3) are not detected in X-rays down to
$(2$--$7)\times10^{-16}$~ergs s$^{-1}$ cm$^{-2}$ (if in the Daddi
field) and do not show strong AGN features in rest-frame UV spectra,
so to exclude galaxies with an obvious AGN contribution, and (4) are
bright in \textit{Spitzer}/MIPS 24 \micron{} images, with flux
$\gtrsim70\,\mu$Jy (for the Daddi field) and
$\gtrsim100\,\mu$Jy (for the Deep3a field), to make sure that they
are actively forming stars.

Our selected objects in the Daddi field are not detected in
\textit{XMM-Newton}/\textit{Chandra} X-ray data with 0.5--2 keV
luminosities $L_X \lesssim10^{43}$ ergs\,s$^{-1}$, which ensures the
exclusion of luminous unobscured AGNs (M. Brusa, private
communications).  However, one \sbzk in our sample shows mid-IR (MIR)
excess which suggests it may host an obscured AGN 
\citep[see \S\ref{section:compsfr};][]{daddi:2007sfr,daddi:2007mirx}.  We note
that only 10 of our 28 objects comply to all the above constraints,
and pairs of \sbzks that can be put in the same slit have been
preferentially included to have a larger sample.  Among the 28 objects
selected for NIR spectroscopy, 20 objects have a redshift from
rest-frame UV spectra, and 13 objects satisfy all 4 criteria above and
have a spectroscopic redshift.  Object IDs, coordinates, redshifts
from rest-frame UV spectra, and magnitudes are listed in Table
\ref{table:objsample}.

\subsection{OHS/CISCO Spectra\label{section:ohsspec}}
Among the selected objects, 7 \sbzks were observed on May 1, 2004 and 
April 24--26, 2005,  by using the OH-airglow suppressor
\citep[OHS;][]{iwamuro:2001} with the Cooled Infrared Camera and
Spectrograph for OHS \citep[CISCO;][]{motohara:2002} mounted on the
Nasmyth focus on the Subaru telescope.  The $JH$ grism covering the $1.1$--$1.8\,\mu\text{m}$ 
spectral range and a $1''$ slit  were used, which give a spectral resolution $R \simeq 200$.
On May 6, 2004 and April 30, 2005 $K$-band spectroscopy with the $wK$
grism covering the $1.85$--$2.5\mu\text{m}$ spectral range was carried out for 5 \sbzks with CISCO 
(i.e., without OHS), which with a slit width of $1''$ gives a resolution $R \simeq 300$.
Position angles were selected to put bright nearby objects 
into the slit, so  to facilitate target acquisition since 
our targets are usually too faint to be seen on acquisition images.  
In case no nearby object was available, target acquisition was made with 
blind offset from bright objects as close as possible to the target object.
Each spectrum was obtained by taking $4\times(\text{900s or 1000s})$
exposure sequences with slightly modified ABBA standard nodding
pattern to avoid bad pixels.  As spectroscopic standards to correct for
atmospheric and instrumental transmission, the A-type stars SAO 120721,
SAO 121153, SAO 122123, SAO 180911, and SAO 180521 were observed for
the $JH$ grism, while the white dwarf GD 153 and the A-type star SAO
121856 were observed for the $wK$ grism.

All the data were reduced with standard procedures.  Two dimensional
spectra were produced with flat fielding, distortion correction, bad
pixel rejection, and sky subtraction by making a median sky from
adjacent detector areas.  Sky residuals were subtracted by fitting
polynomials in the spatial direction, and then co-addition was carried
out with appropriate offsets for the dithering width.  The resulting
2D spectra were then collapsed to 1D spectra.  The tilt of the spectra
on the array was corrected by adopting the tilt of standard star
spectra.  The wavelength calibrations were carried out by using the
standard pixel-wavelength relation of OHS/CISCO with systematic error
of $<0.5$ pixels \citep[3\AA{} and 5\AA{} for OHS and CISCO, respectively;][]{motohara:2001}.  
Atmospheric and instrumental transmissions were corrected by using the 1D spectra of the standard
stars reduced in the same way as object frames.
  
The absolute flux calibrations were carried out by comparing the
photometric $J$- and $K$-band fluxes with those derived from the
object spectra convolved with the filter transmission curves.  Here we
have adopted $2''$ aperture magnitudes, hence assuming that in the
observed spectra both continuum and lines come from same region of a
galaxy.  Noise spectra were derived by measuring rms of counts in the
blank sky region.

\subsection{SINFONI Spectra\label{section:sinfonispec}}
On April 14--16, 2005 we observed 16 \sbzks with the Spectrograph for
Integral Field Observations in the Near Infrared
\citep[SINFONI;][]{eisenhauer:2003,bonnet:2004} at the VLT/UT4.  The
$H+K$ grating covering the $1.45$--$2.45\mu\text{m}$ spectral range
with $R\simeq 1500$ and a pre-optics giving a $125\times250$ mas
pix$^{-1}$ spatial resolution were used.  Each object was observed by
1 or 2 sequence(s) of $4\times(450\text{s or }900\text{s})$ exposures
with 4 arcsec dithering both in $x$- and $y$-direction.  The B-type
stars, HIP 007873, HIP 026816, HIP 031768, HIP 068100, HIP 068372, HIP
072367, HIP 075711, HIP 083861, HIP 092957, HIP 094333, HIP 095806,
and HIP 099244, were observed as spectroscopic standards.

The reductions of SINFONI data were carried out with the SINFONI 
pipeline%
{\footnote{http://www.eso.org/sci/data-processing/software/pipelines/}}
and custom scripts, including flat-fielding, sky-subtraction by median
sky of a sequence, bad pixel rejection, distortion correction,
wavelength calibration by arc lamp frames, cube reconstruction,
residual sky subtraction by subtracting the median in $x$-direction from
each pixels, and finally co-addition with appropriate offsets.  Telluric and
instrumental absorptions were corrected with spectroscopic standard
star frames.  Since neither the standard stars nor the objects are affected by slit losses, the
absolute flux calibrations were obtained by dividing the object
spectrum by the standard star spectrum, scaled to the broad
band photometric magnitudes.

\section{Emission Line Measurements\label{section:linemes}}
\begin{figure}
  \begin{center}
    \plotone{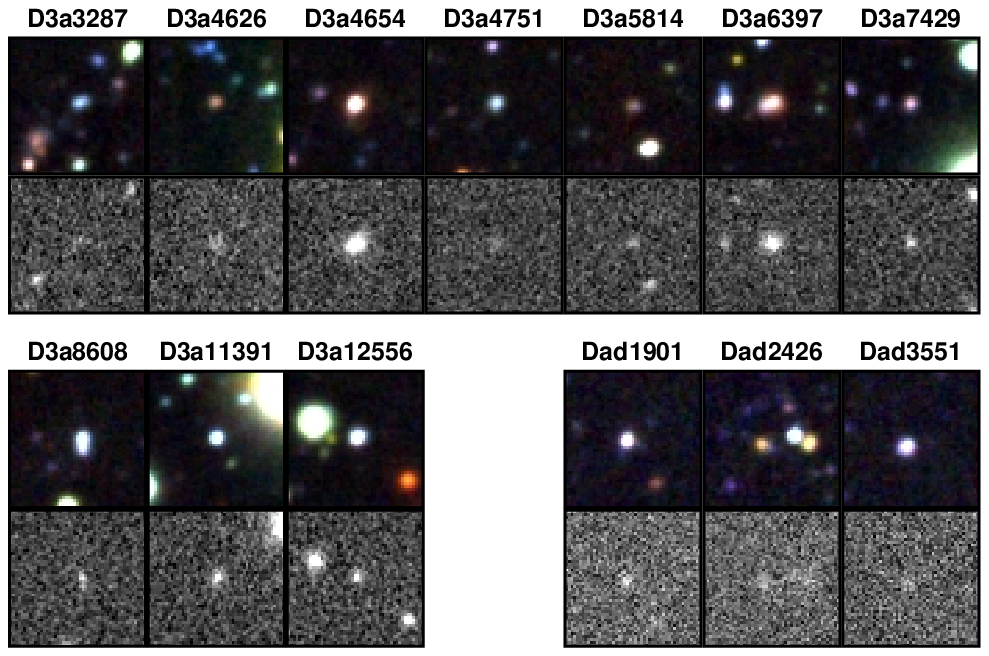}
  \end{center}
  \figcaption{
    Cutout images ($\simeq 15''\times15''$) of \sbzks with
    \hasp detection in our sample. North is up
    and east is to the left.  Each pair consists of composite color
    image (top) and $K$-band image (bottom).  Composite color images
    are created with Subaru/Suprime-Cam $BRz'$ and $BIz'$ for Deep3a
    and Daddi fields, respectively. Objects are located at the center
    except for Dad 2426-b which can be seen at 3 arcsec west from Dad 2426.
    \label{fig:bzkcutout}
  }
\end{figure}

\begin{figure}
  \begin{center}
    \includegraphics[width=0.32\linewidth]{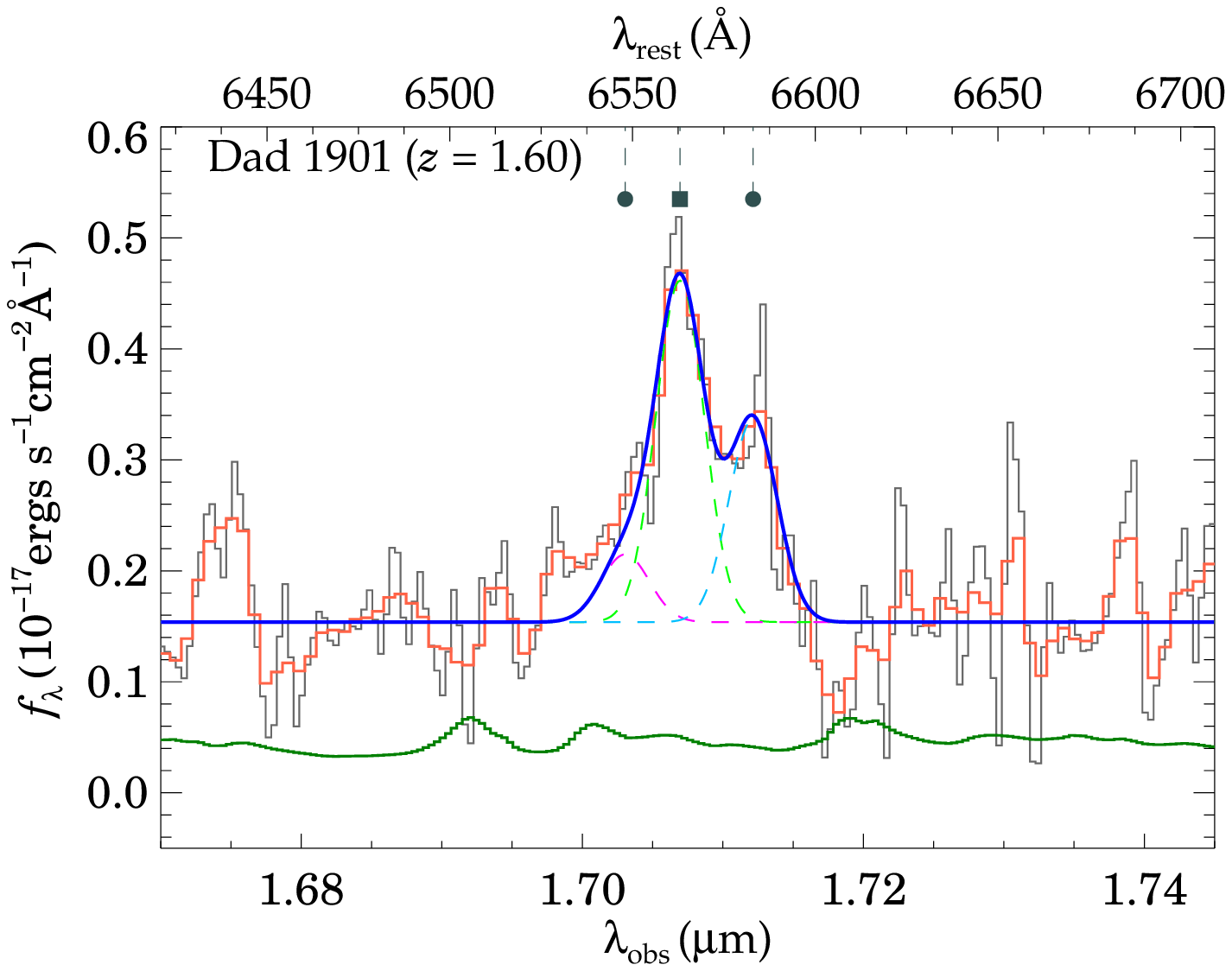}
    \includegraphics[width=0.32\linewidth]{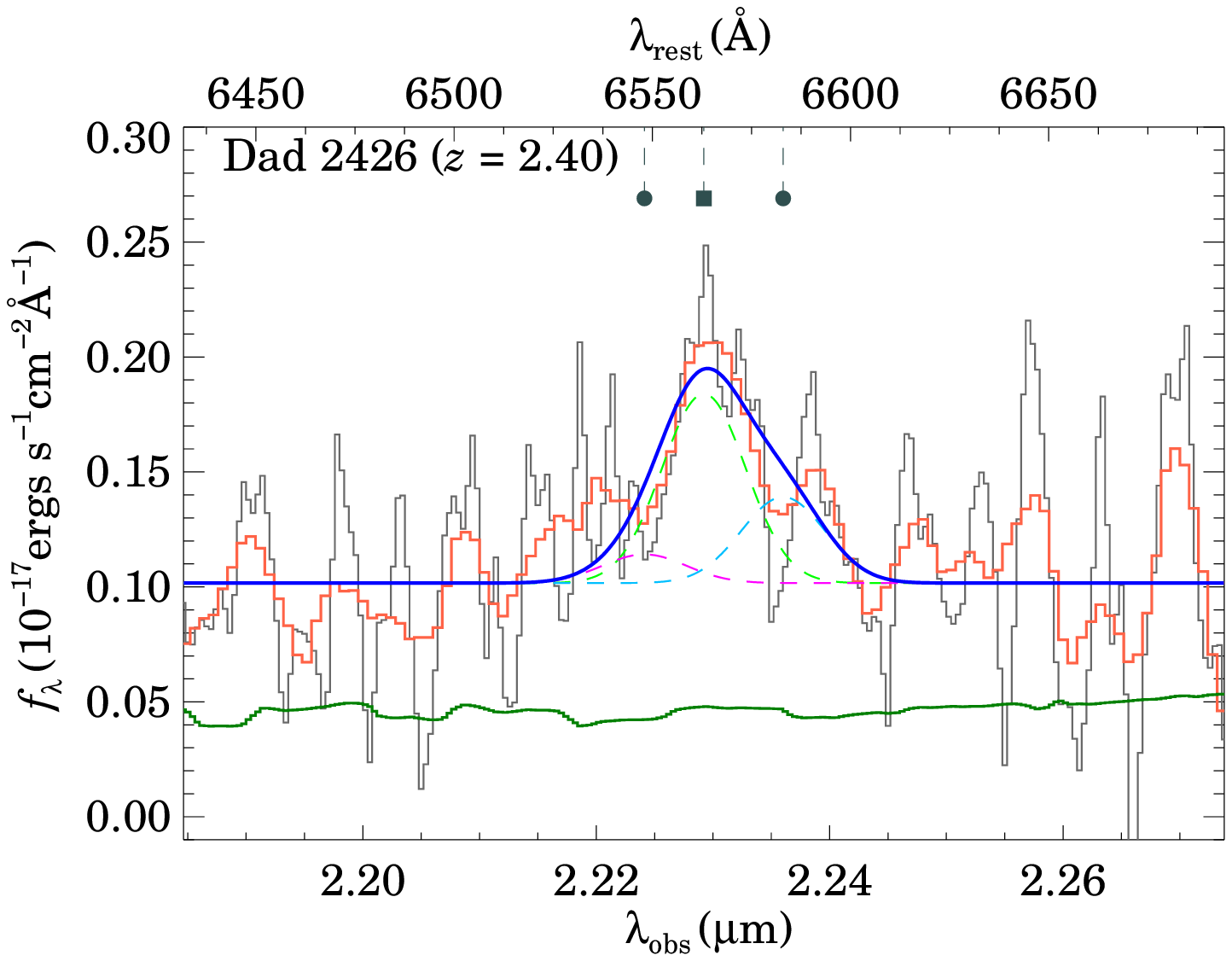}
    \includegraphics[width=0.32\linewidth]{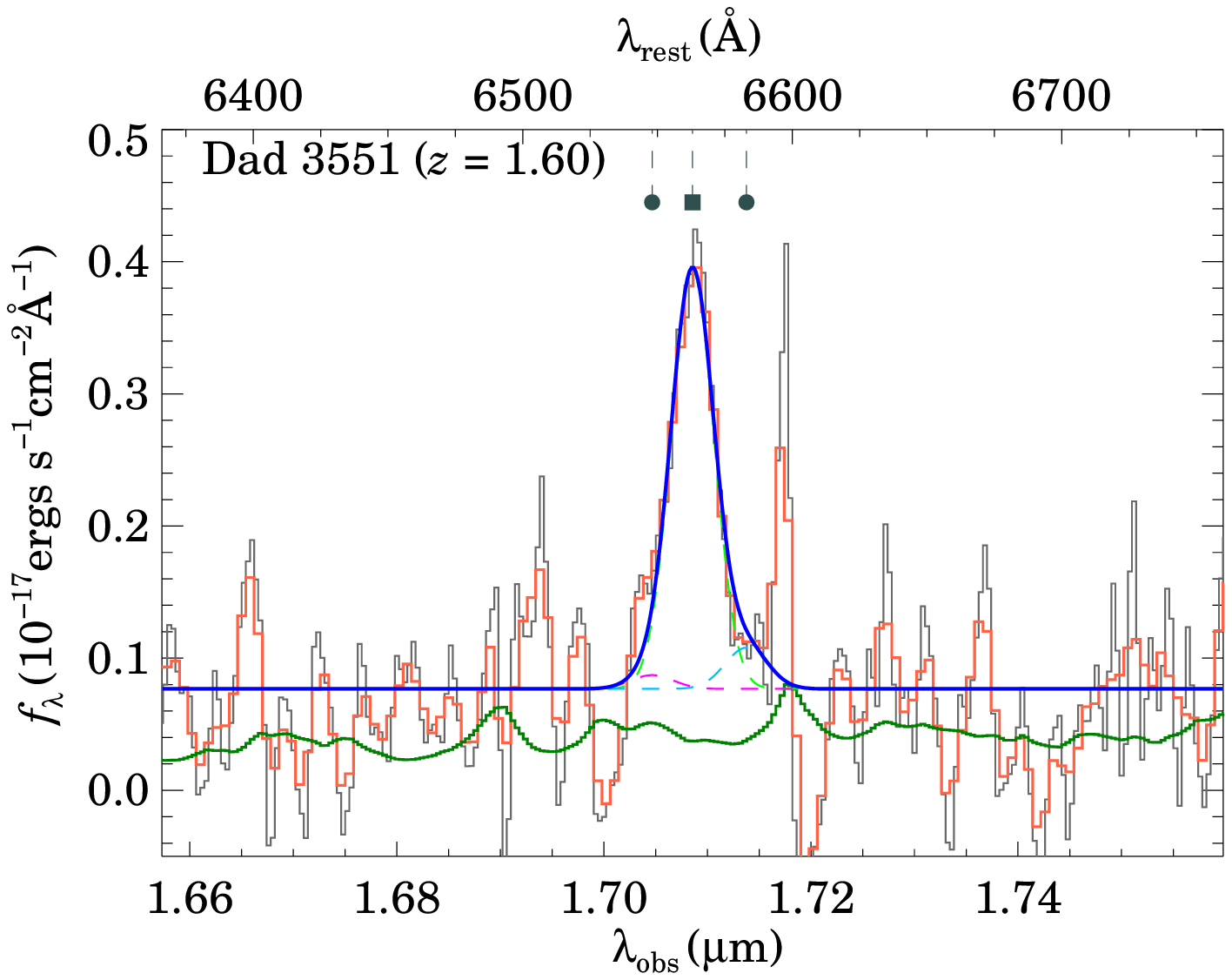}
    \includegraphics[width=0.32\linewidth]{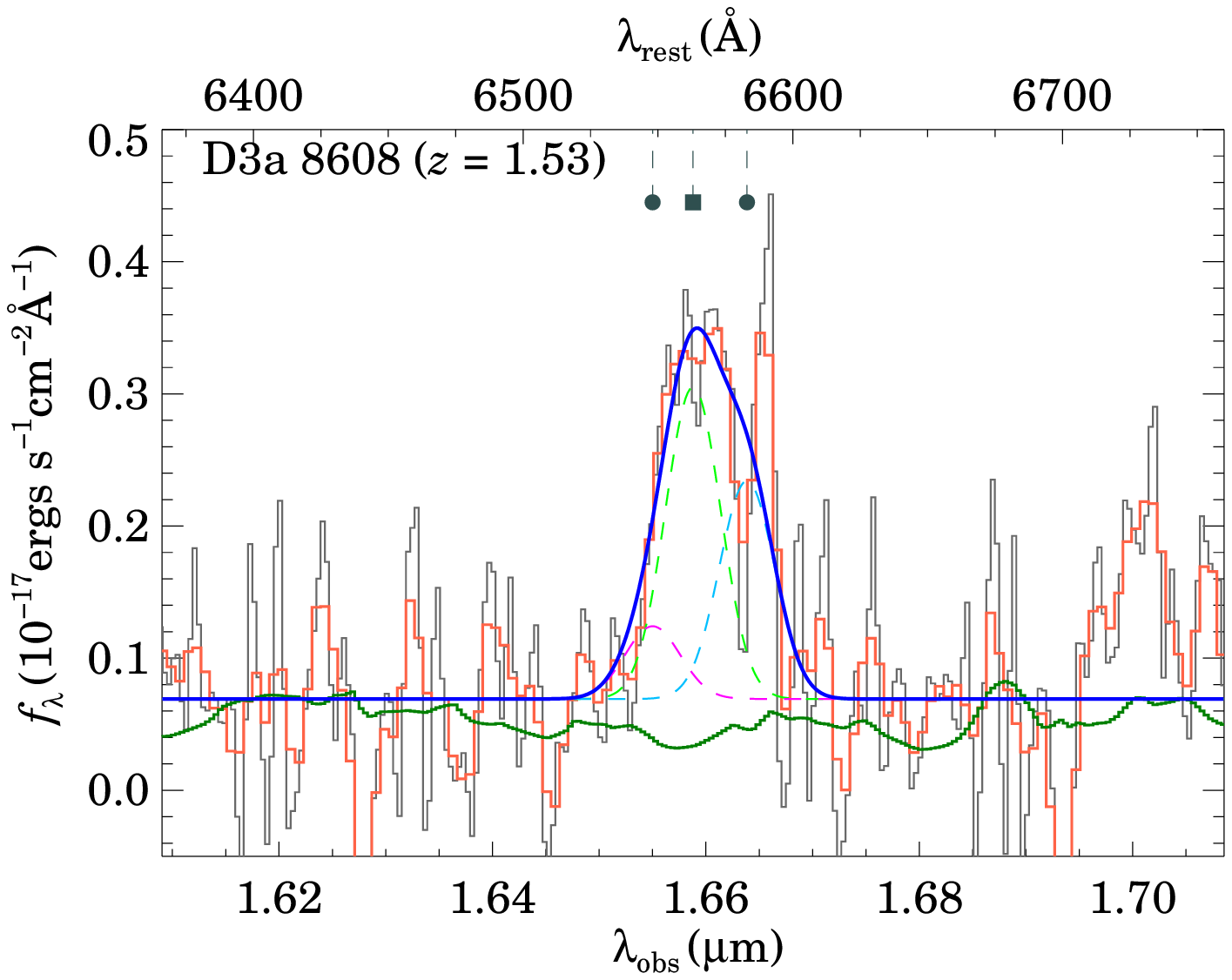}
    \includegraphics[width=0.32\linewidth]{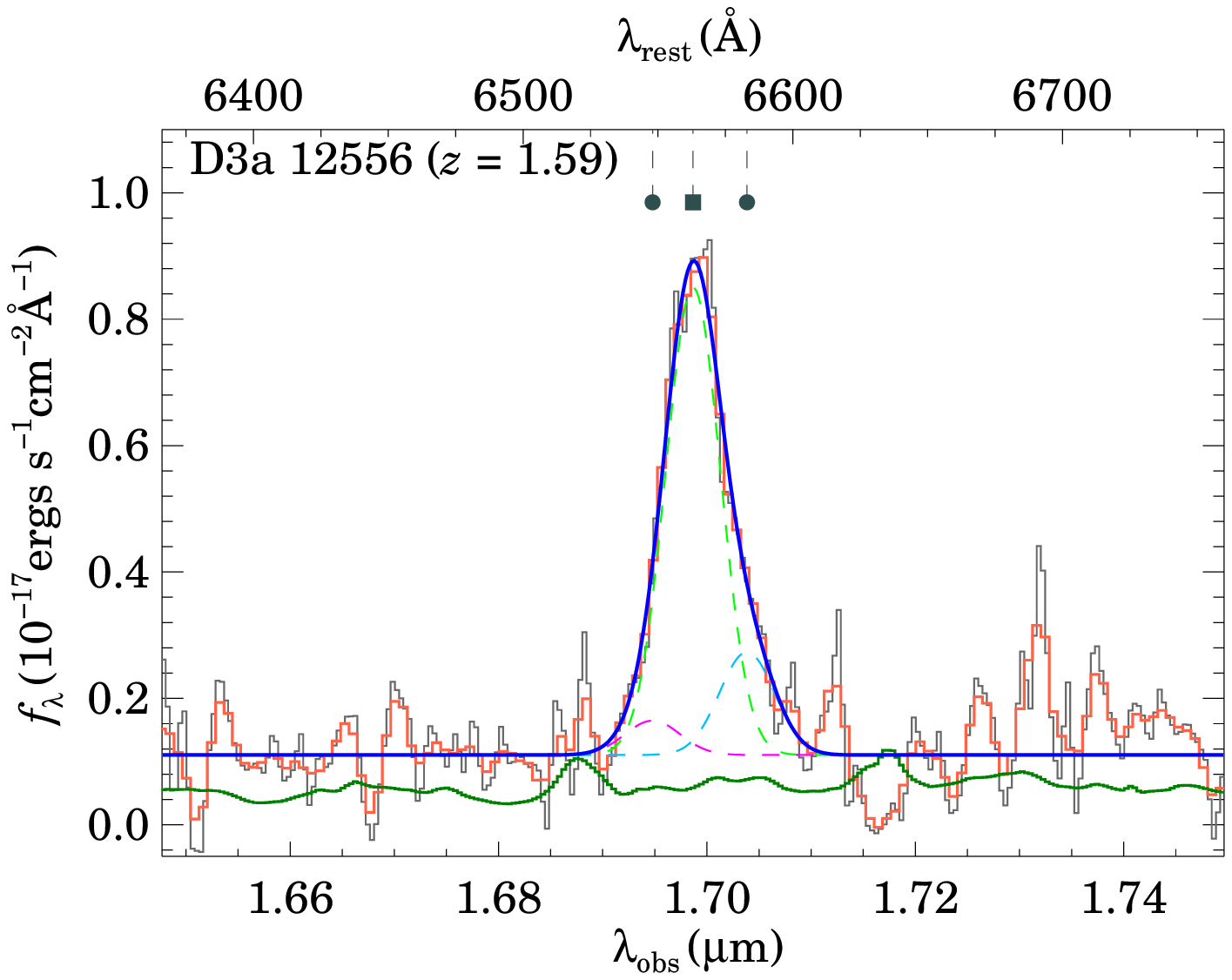}
    \figcaption{
      One-dimensional spectra around the \hasp emission line
      for the spectra taken with OHS/CISCO.  Black, orange, and green
      lines correspond to raw, 40 \AA{} boxcar smoothed, and $1\sigma$
      sky fluctuation spectra, respectively.  Gaussian functions fitted
      to the smoothed spectrum are also shown with dashed lines which
      respectively correspond to \hasp (light green),
      $\text{[\ion{N}{2}]}\lambda6548$ (magenta) and
      $\text{[\ion{N}{2}]}\lambda6583$ (light blue).  The sum of these three
      Gaussian is shown with blue solid line.  Positions of
      $\text{[\ion{N}{2}]}\lambda\lambda6548,6583$ and \hasp are
      indicated by gray dashed lines with circles and squares,
      respectively.
      \label{fig:nirspeczoomohs}
    }
  \end{center}
\end{figure}

\begin{figure}
  \begin{center}
    \includegraphics[width=0.32\linewidth]{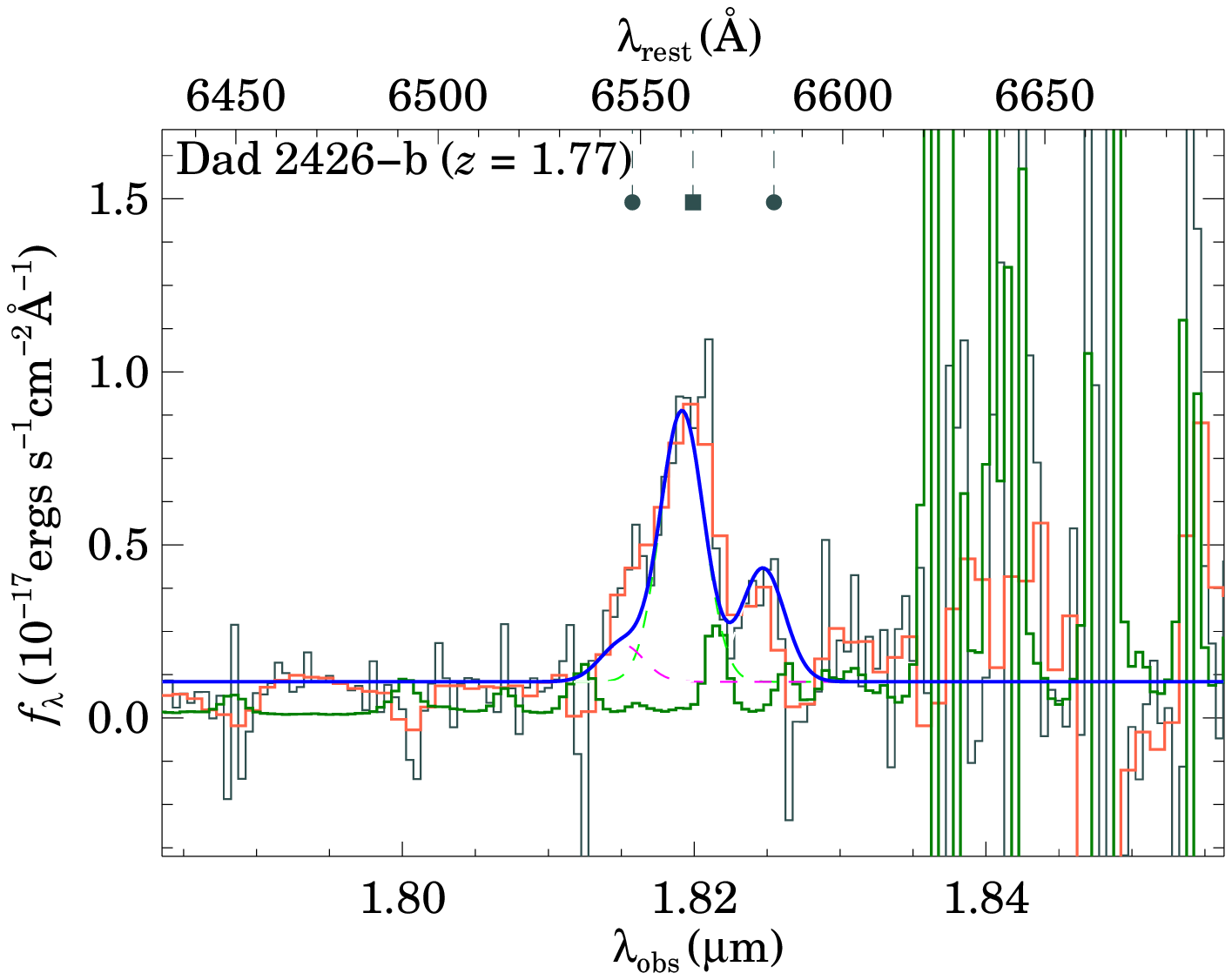}
    \includegraphics[width=0.32\linewidth]{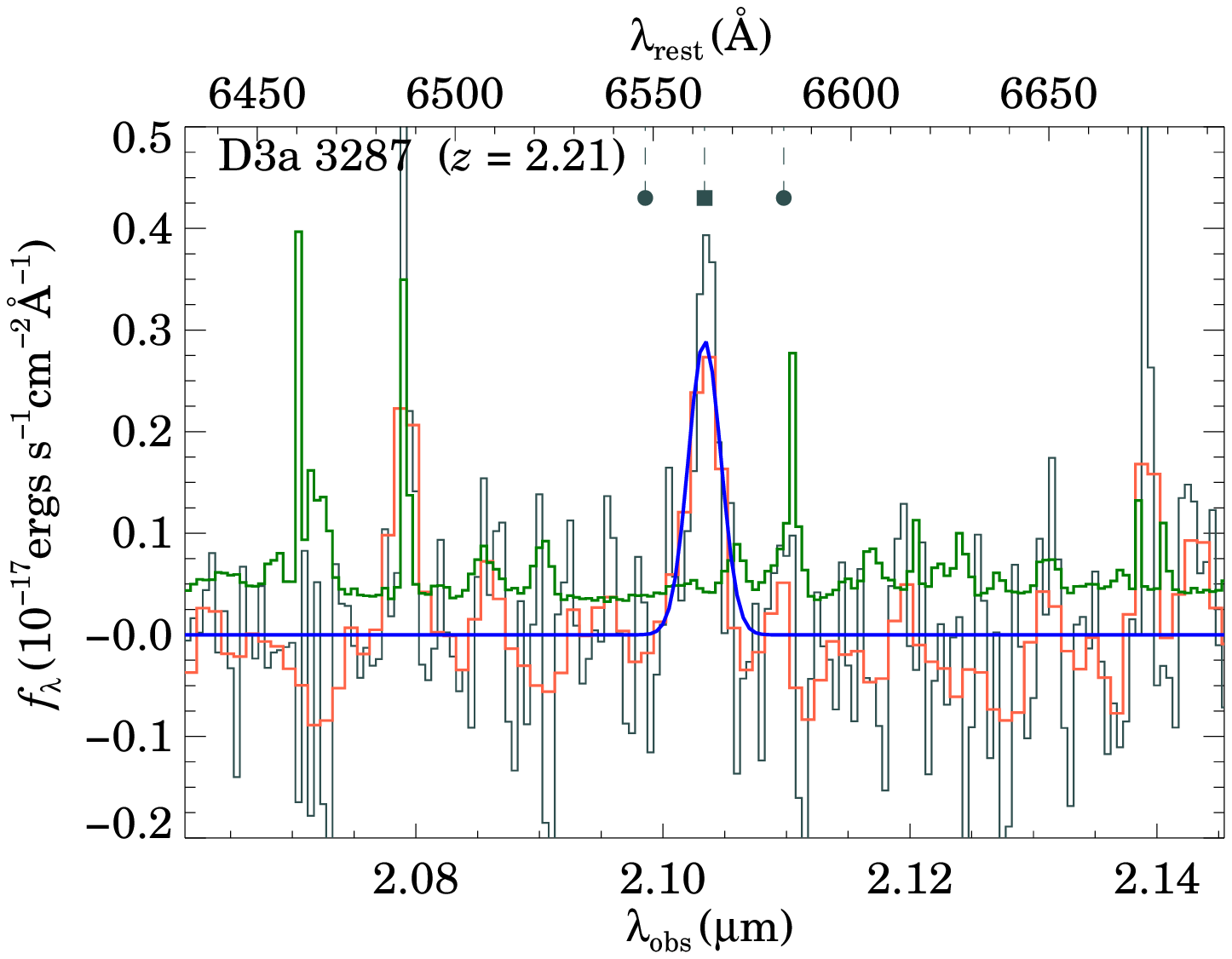}
    \includegraphics[width=0.32\linewidth]{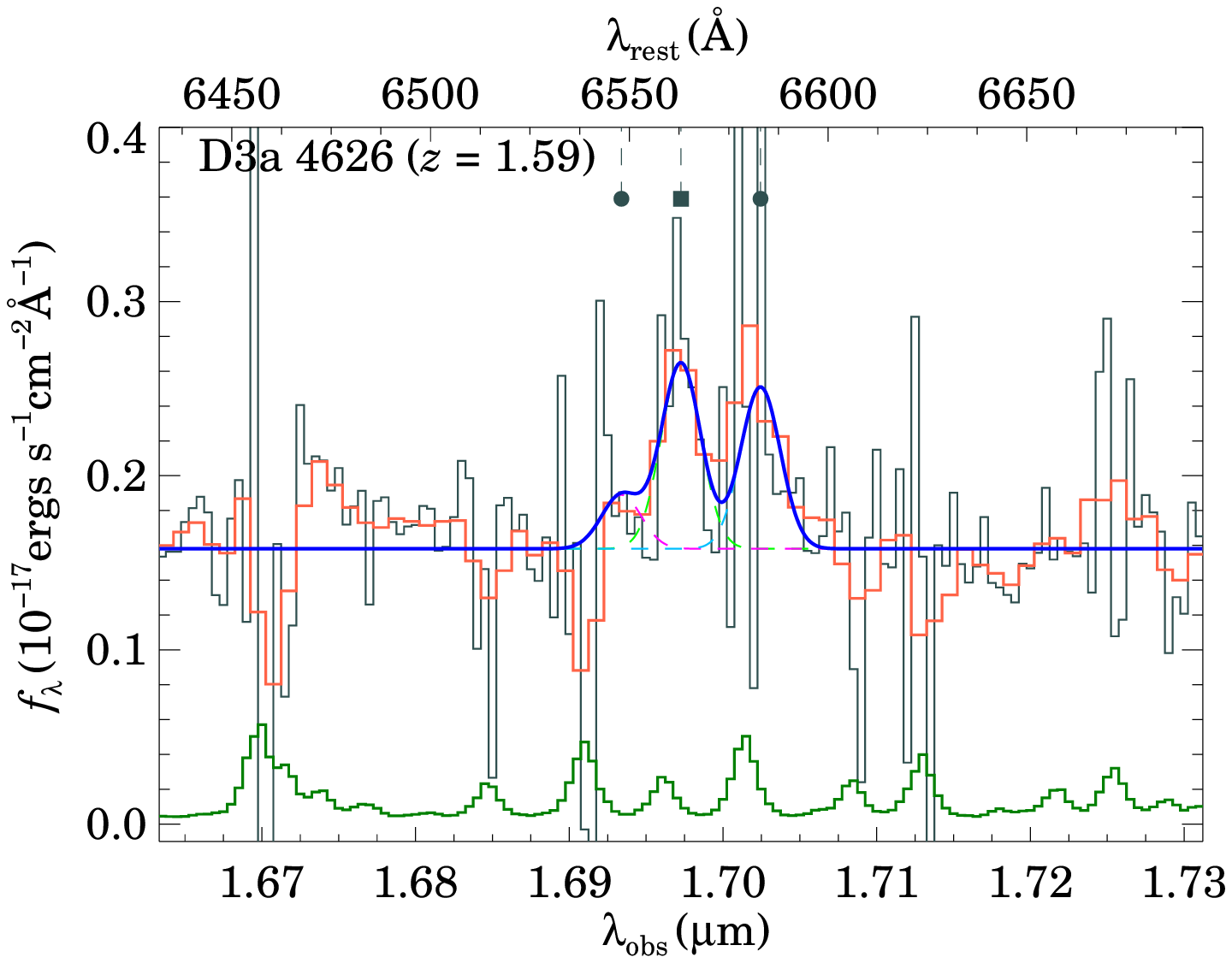}
    \includegraphics[width=0.32\linewidth]{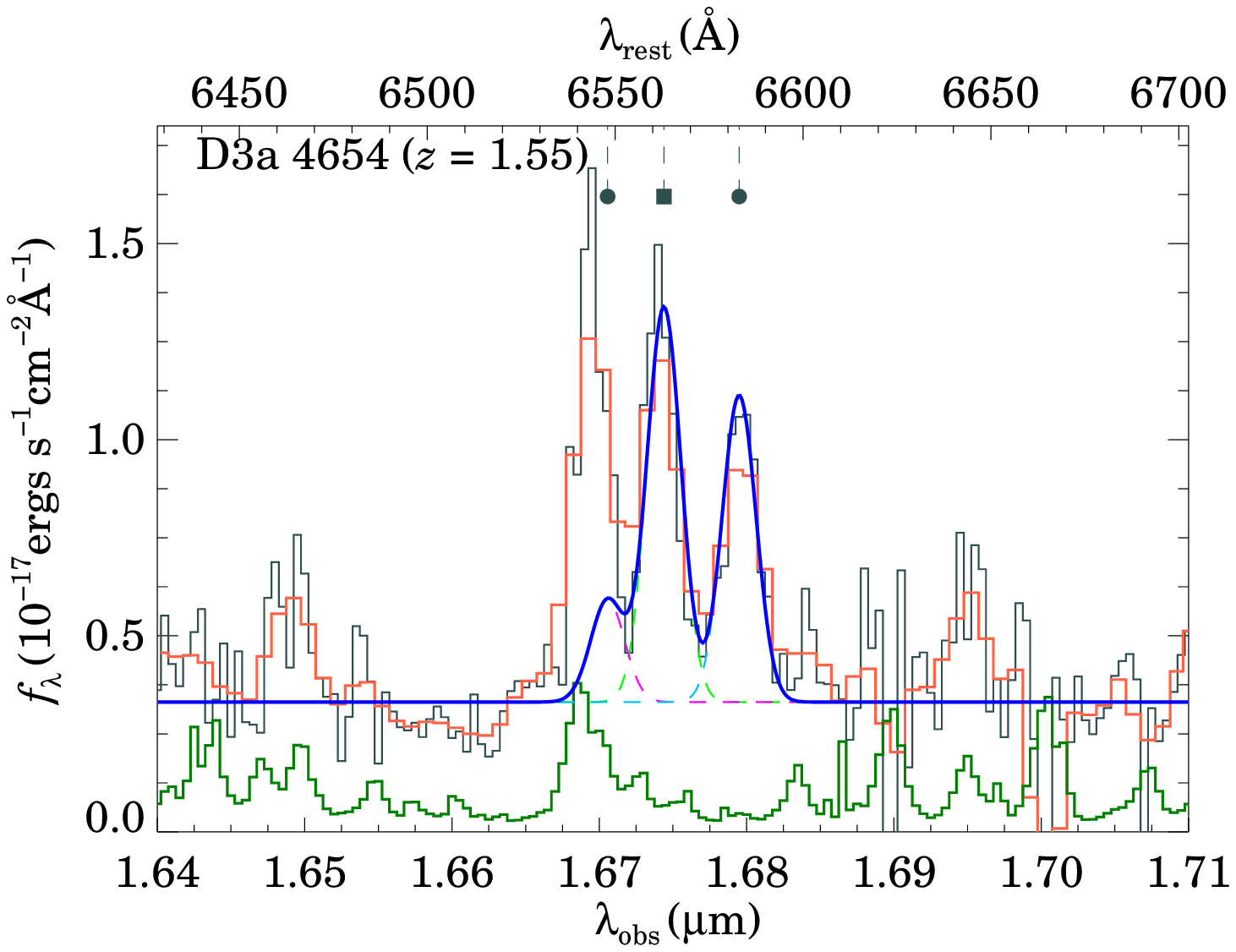}
    \includegraphics[width=0.32\linewidth]{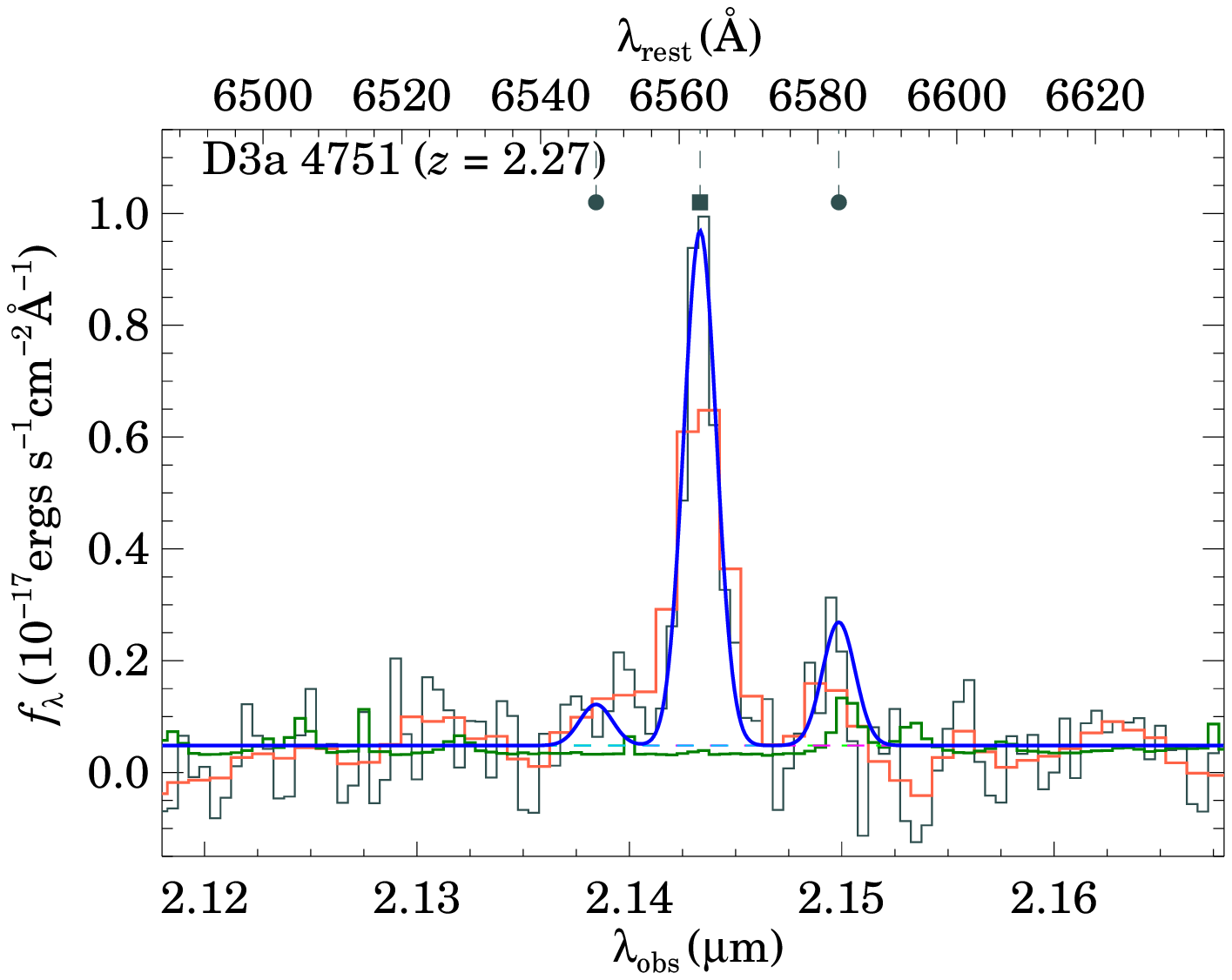}
    \includegraphics[width=0.32\linewidth]{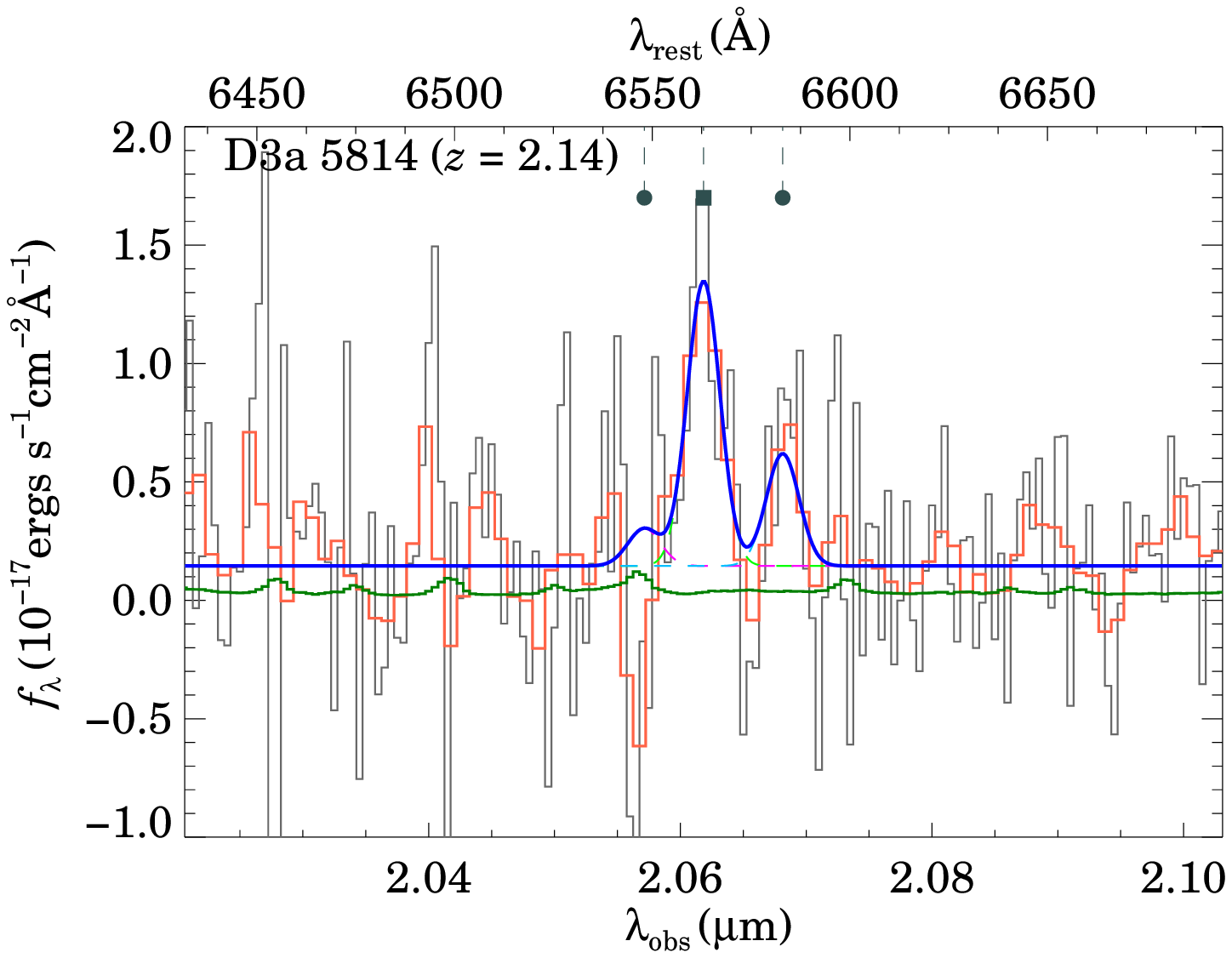}
    \includegraphics[width=0.32\linewidth]{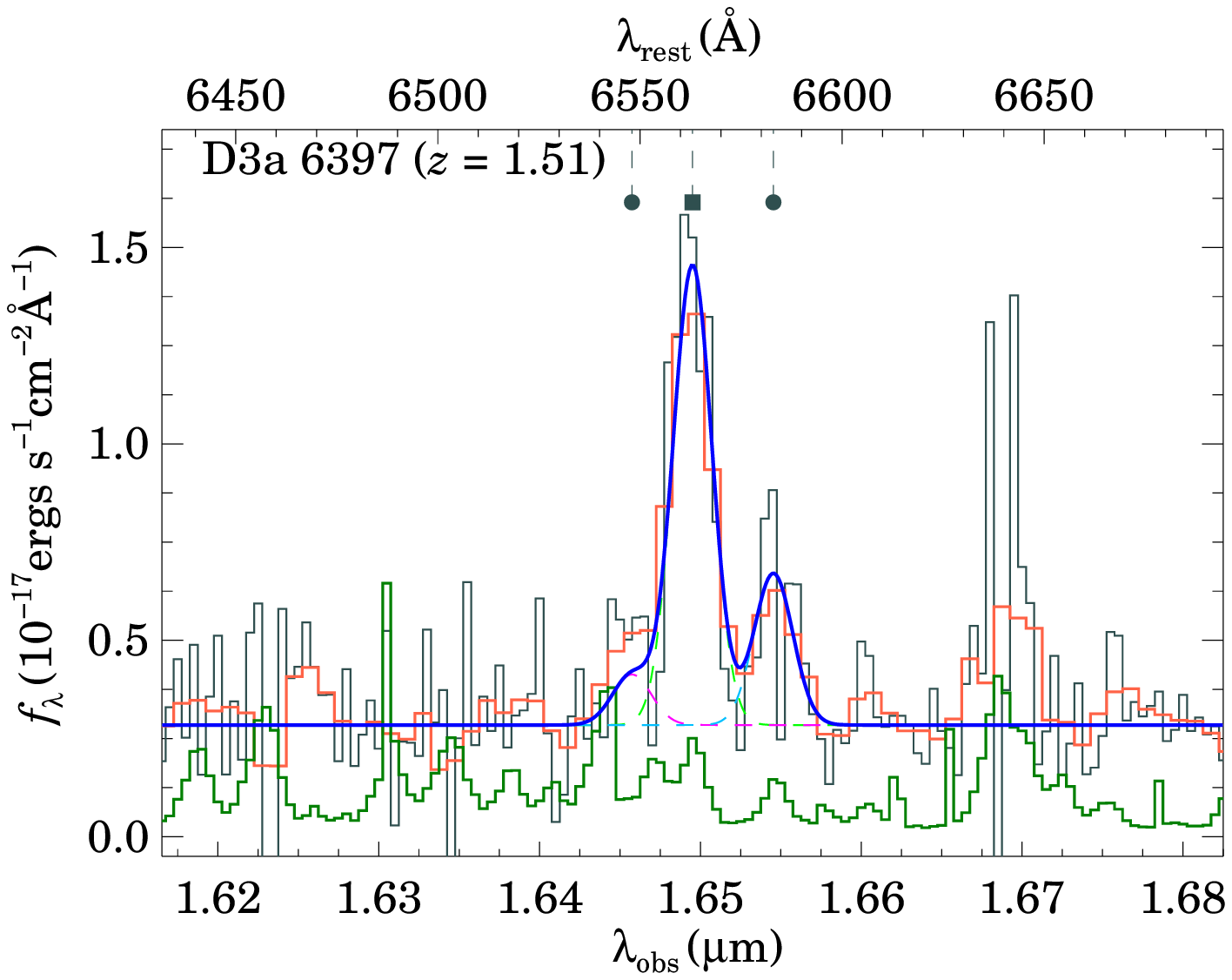}
    \includegraphics[width=0.32\linewidth]{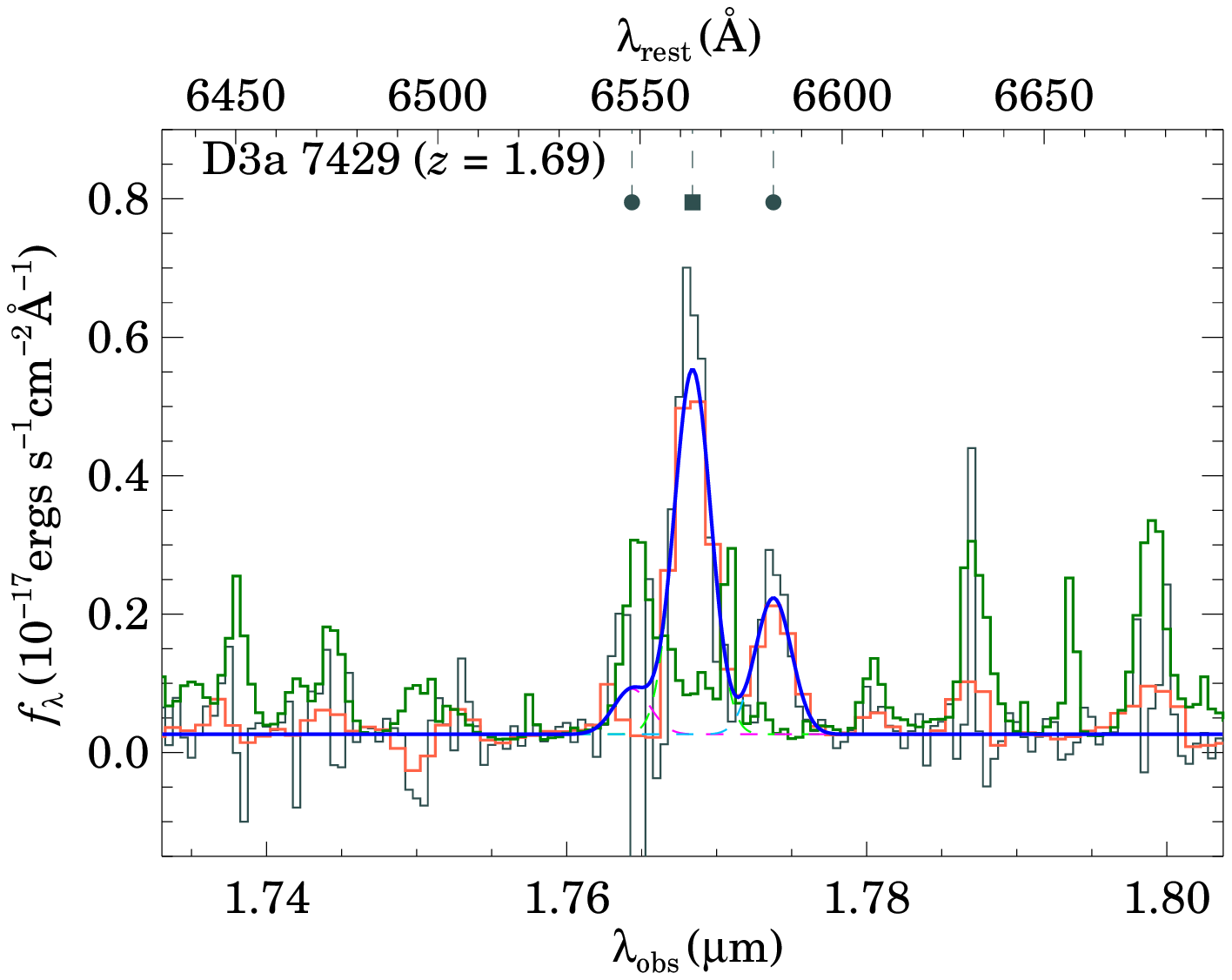}
    \includegraphics[width=0.32\linewidth]{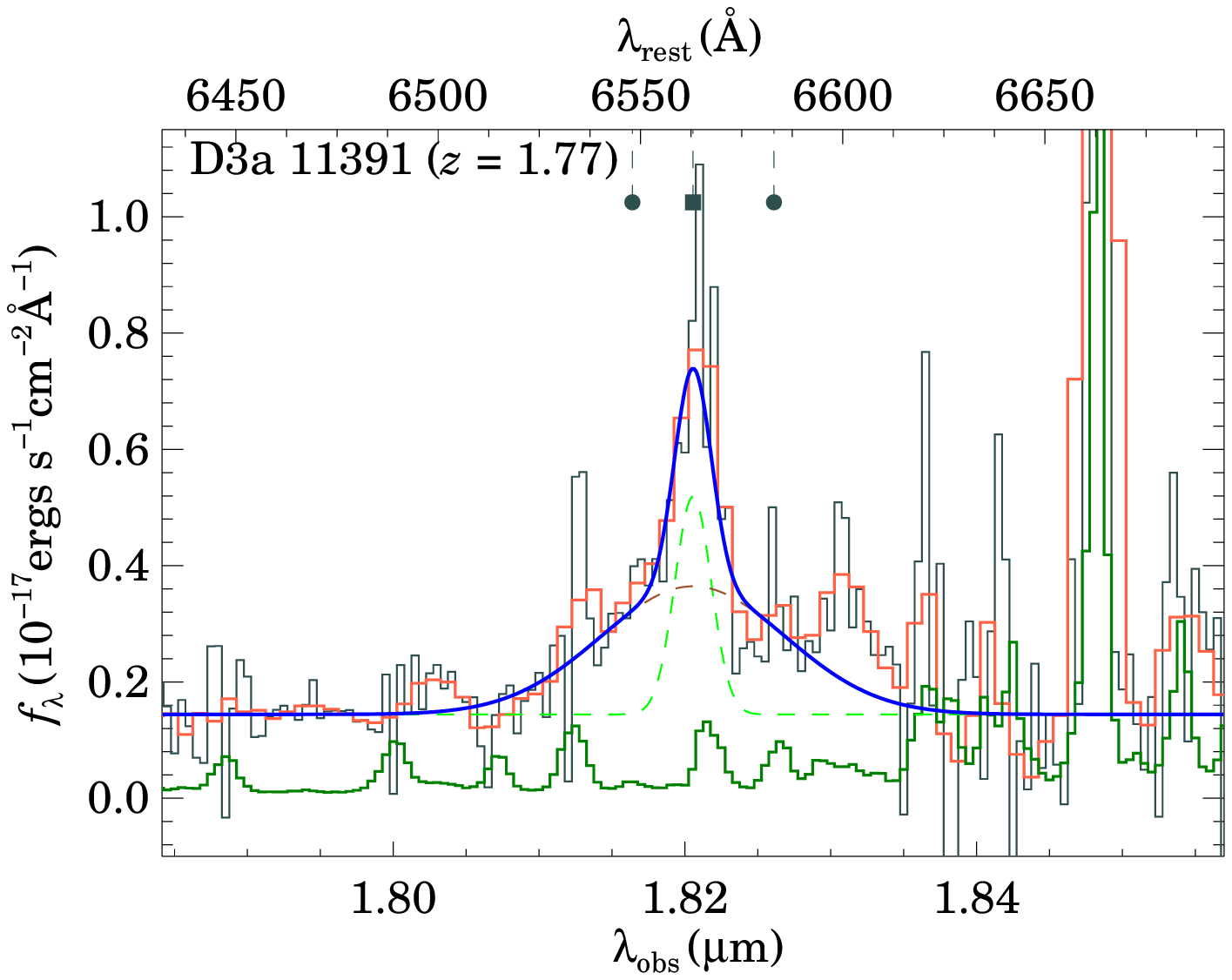}
    \figcaption{
      One-dimensional spectra around the \hasp emission line
      for the spectra taken with SINFONI.  The same line coding as in Figure
      {\ref{fig:nirspeczoomohs}} is used to indicate observed and
      fitted Gaussian functions.  For the D3a-11391 the narrow and broad \hasp
      components of the fit are indicated as light green and brown dashed lines,
      respectively.  The positions of
      $\text{[\ion{N}{2}]}\lambda\lambda6548,6583$ and \hasp are also
      indicated by dashed lines with circles and squares, respectively.
      For the smoothed spectra, a 20 \AA{} boxcar smoothing kernel was
      used.
      \label{fig:nirspeczoomsinfoni}
    }
  \end{center}
\end{figure}

Among the 28 \sbzks observed, emission lines were detected in 13 of them,
and identified as \hasp in all cases.  Emission lines 
in the SINFONI data cube have also been detected in an object in the vicinity of Dad-2426 (hence named Dad-2426b),  
and the lines are identified as \hasp and 
[\ion{N}{2}]$\lambda6583$ at $z=1.772$ according to the separation 
between them. 
Optical composite images and $K$-band images of the galaxies 
with \hasp detections are shown in Figure \ref{fig:bzkcutout},
and their spectra around \hasp are shown in Figures \ref{fig:nirspeczoomohs} 
and \ref{fig:nirspeczoomsinfoni} for OHS/CISCO and SINFONI, respectively. 

Due to their low resolution in wavelength, the \hasp emission is always
blended with [\ion{N}{2}]$\lambda\lambda6548,6583$ in the OHS/CISCO
spectra, while these lines are resolved in the SINFONI spectra.  We
have measured the \hasp line fluxes by fitting multiple-Gaussians for
\hasp and [\ion{N}{2}]$\lambda\lambda6548,6583$, assuming that all
three lines have same  width,
[\ion{N}{2}]$\lambda6583$/[\ion{N}{2}]$\lambda6548=3$, and a constant
continuum flux. This leaves redshift, \hasp emission line flux \fha,
flux ratio \fha/\fnii, line width $\sigma$, and constant continuum
flux as free parameters.  For most of the objects, the fitting converged
to a single solution against various initial guesses, and the derived
line widths agree with those expected from instrumental resolution in
the case of OHS/CISCO spectra.  However, there are several exceptions.
For D3a-8608 and Dad-2426 the procedure did not converge to stable
solutions, and we fixed the line width to that of the instrumental
profile (measured on the OH lines close to the position of emission
lines).  Small perturbations to the assumed line widths did not change
fluxes and equivalent widths (EWs) appreciably.  Since the [\ion{N}{2}]$\lambda
6583$ line is below the detection limit in D3a-3287 and D3a-4626, only a
single Gaussian to the \hasp line was fitted.  D3a-11391 appears to have both
broad and narrow \hasp  components. This object also shows MIR-excess
\citep{daddi:2007mirx} as discussed later (\S\ref{section:compsfr}),
hence the spectrum was fitted with both broad and narrow \hasp lines, which
resulted in a better $\chi^2$ compared to the fit with only narrow line
components.  We could not find a good fit when including the
[\ion{N}{2}] lines, possibly due to residual of OH sky lines and
rapidly variable atmospheric transmission at the edge of $H$-band.
The best fit Gaussian functions are overplotted in Figure \ref{fig:nirspeczoomohs} 
and \ref{fig:nirspeczoomsinfoni}. 

The resulting redshifts derived from \hasp, along with the fluxes and
EWs of \hasp and [\ion{N}{2}]$\lambda6583$, and velocity dispersions 
are listed in Table \ref{table:lineprop}. 
The EWs listed in Table \ref{table:lineprop} are 
derived with the measured line fluxes and continuum fluxes 
from the best-fit spectral energy distribution (SED) (see \S\ref{section:massdust}). 
\hasp EWs are also corrected for stellar \hasp absorption,
derived from the synthetic spectrum that best-fits the SED. 
The error in the continuum flux from the best-fit SED around 
the position of \hasp is estimated to be 20\%{}.
For Dad-2426b, which is not listed in our $K$-selected catalog, 
the continuum flux is calculated from its $5\sigma$ limit of $K_\text{AB}=21.5$,
and the stellar \hasp EW is assumed to be 4.4~\AA{}, which is the average 
of stellar \hasp EWs estimated for the other objects. 
These redshifts agree well with those from
rest-frame UV spectra, except for 2 objects, Dad-2426
($z_{\text{UV}}=2.36$ and $z_{\text{H}\alpha}=2.40$) and D3a-6397
($z_{\text{UV}}=2.00$ and $z_{\text{H}\alpha}=1.51$).  The spectrum of
Dad-2426 is located a little far from the detector center and since
detector distortion increases with distance from the center, we first
doubted that the adopted pixel-wavelength relation may not apply in
such case. However, a cross-check of wavelength with OH lines showed
that the wavelength was well calibrated. This object was also observed
with SINFONI and a feature is marginally seen at 2.23\micron{} which
if due to \hasp corresponds to $z=2.40$, consistent with the redshift
from the CISCO spectrum.  For D3a-6397, the discrepancy is very 
large, but its $z_{\text{UV}}$  is quite uncertain whereas
[\ion{N}{2}]$\lambda6583$ is well detected at the expected position
relative to \hasp, hence we consider $z_{\text{\ha}}$ as more
reliable.

From the \ha/[\ion{N}{2}]$\lambda6583$ emission line ratio it is
possible to estimate whether there is a significant contribution from
an AGN component, though more emission lines such as \hbsp and
[\ion{O}{3}] are required for a more robust diagnostic
\citep[e.g.,][]{veilleux:1987,baldwin:1981,kauffmann:2003agn}.  Here
we classify \sbzks with
$\text{[\ion{N}{2}]}\lambda6583/\text{\ha}<0.7$ as star-formation
dominated and those with
$\text{[\ion{N}{2}]}\lambda6583/\text{\ha}>0.7$ as AGN dominated
\citep{swinbank:2004}.  Adopting this criterion, one \sbzk,
D3a-8608, is classified as AGN-dominated. 
Besides the emission line ratio diagnostics, D3a-11391 could host an AGN because
of its broad-line component.  This would read an AGN fraction of
$\simeq15\%$ for our sample of 14 galaxies with detected emission lines.  
Note that true AGN fraction among \sbzks
might be even higher, since part of the objects have been pre-selected
for lacking AGN features.  In any case, this AGN fraction
is not far from that ($\simeq 25\%$) estimated for the full \sbzk
population in the Deep-3a and Daddi fields \citepalias{kong:2006},
while up to $\simeq 50\%$ of massive
($M_{\star}\simeq10^{11}M_{\odot}$) \sbzks show MIR-excess, possibly
related to heavily obscured AGN \citep{daddi:2007mirx}.

\section{Physical Properties of \sbzk{} Galaxies}

\subsection{Stellar Masses and Reddening\label{section:massdust}}
Stellar masses  ($M_{\star}$) and the reddening \ebv were
derived by SED fitting to the $BRIzJK$- and
$BRIK$-band data for the \sbzks in the Deep3a field and the Daddi field,
respectively.  Spectroscopic redshifts determined from \hasp emission
lines were used for the fit.

Template SEDs were generated by using
\pegase{} \citep{fioc:1997,fioc:1999}, assuming a Salpeter initial
mass function \citep[IMF;][]{salpeter:1955} for stars with $0.1
M_\odot$ to $120 M_\odot$.  These models assume exponentially
declining SFRs and incorporate chemical evolution using chemical
yields from \citet{woosley:1995}. Real galaxies at $z\sim 2$ are
unlikely to have evolved with exponentially declining SFRs
\citep[cf.][]{cimatti:2008, renzini:2009a, maraston:2010}, and their gas accretion
histories may be radically different compared to those assumed in the
\pegase{} models. Although the use of these models introduces some
rigidity in the SED fits, we believe that the derived stellar masses
should be correct within a factor of $\sim 2$ \citep{drory:2004}.  A
systematic overestimate by up a factor $\sim 3$ may be present for
those galaxies in which the bulk of stars have ages around $\sim 1$
Gyr, i.e., at the peak of the contribution by the TP-AGB phase of
stellar evolution, that was not adequately included in the \pegase{}
\citep[e.g.,][]{maraston:1998,maraston:2005,kajisawa:2009,magdis:2009}. Furthermore, a systematic
reduction by a factor $\sim 1.6-2$ of both SFRs and masses would be
produced adopting a {\it bottom light} IMF, such as those of
\citet{kroupa:2002} or \citet{chabrier:2003}.

The adopted models assume SFR $e$-folding times from $\tausf=100$ Myr
to 500 Gyr with various age grids ranging from 10 Myr to 15 Gyr.
The synthetic SEDs are then dust attenuated according to the Calzetti law
\citep{calzetti:2001} with $0<\ebv<1.5$.  Absorption by neutral
hydrogen intervening along the line of sight is also applied
\citep{madau:1995,madau:1996}.  Fitting is finally obtained by a
$\chi^2$-minimization, with fixed redshift derived from \hasp , and with
the constraint that age cannot exceed the age of the universe at the observed
redshift.  Stellar masses are then derived  by multiplying the luminosity by
the $M/L$ ratio of the model which best-fits the observed SED. 

\begin{figure}
  \begin{center}
    \includegraphics[width=0.32\linewidth]{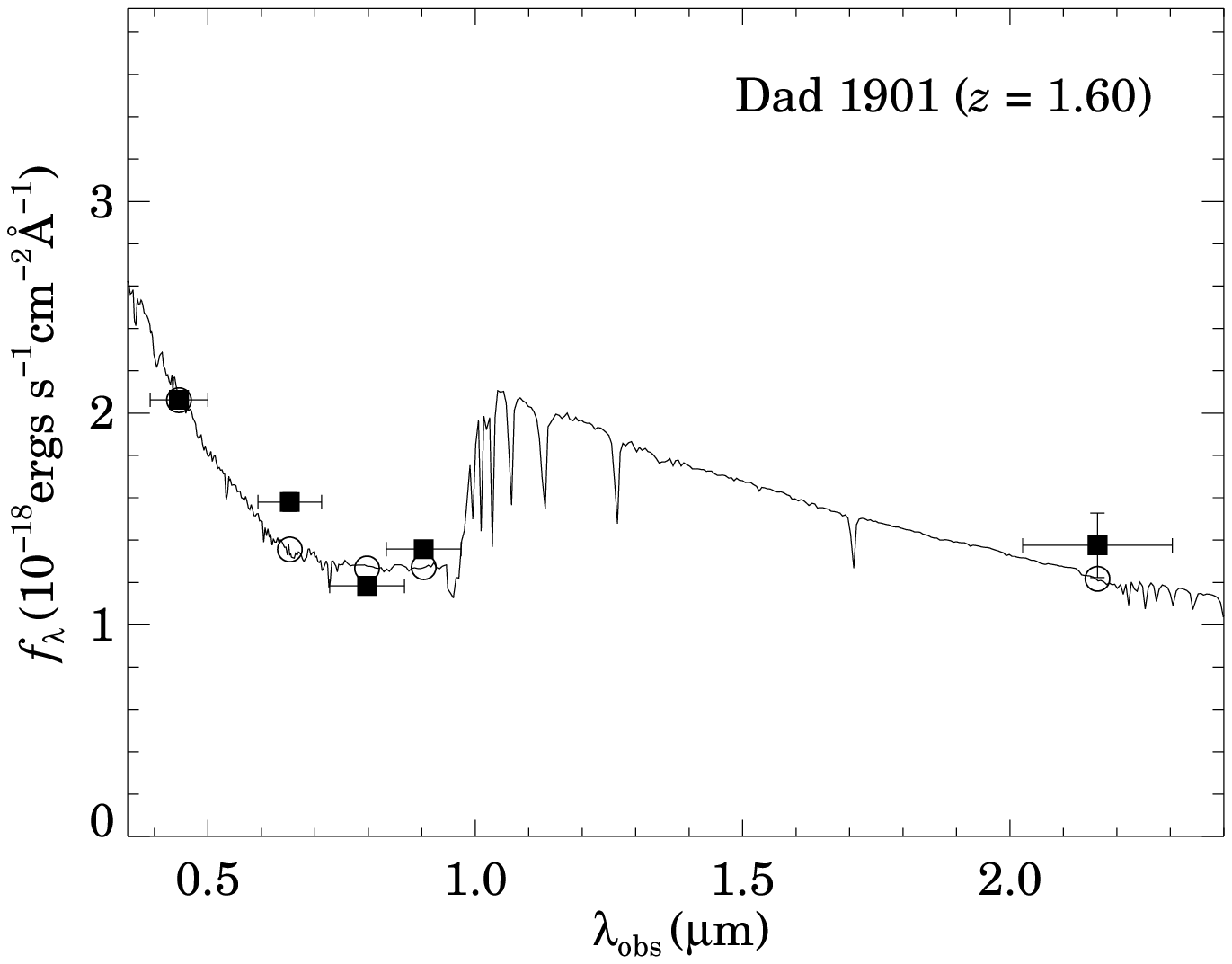}
    \includegraphics[width=0.32\linewidth]{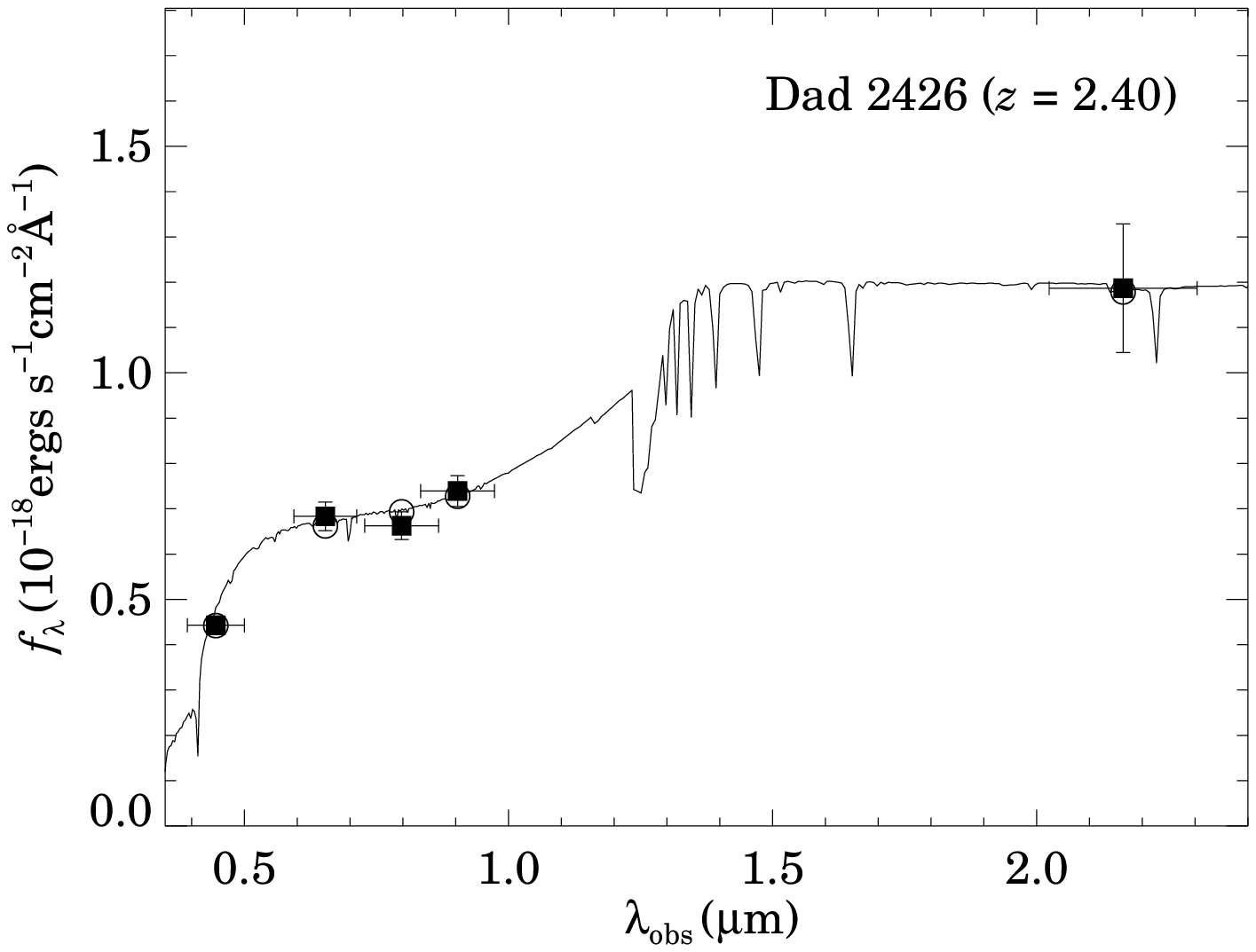}
    \includegraphics[width=0.32\linewidth]{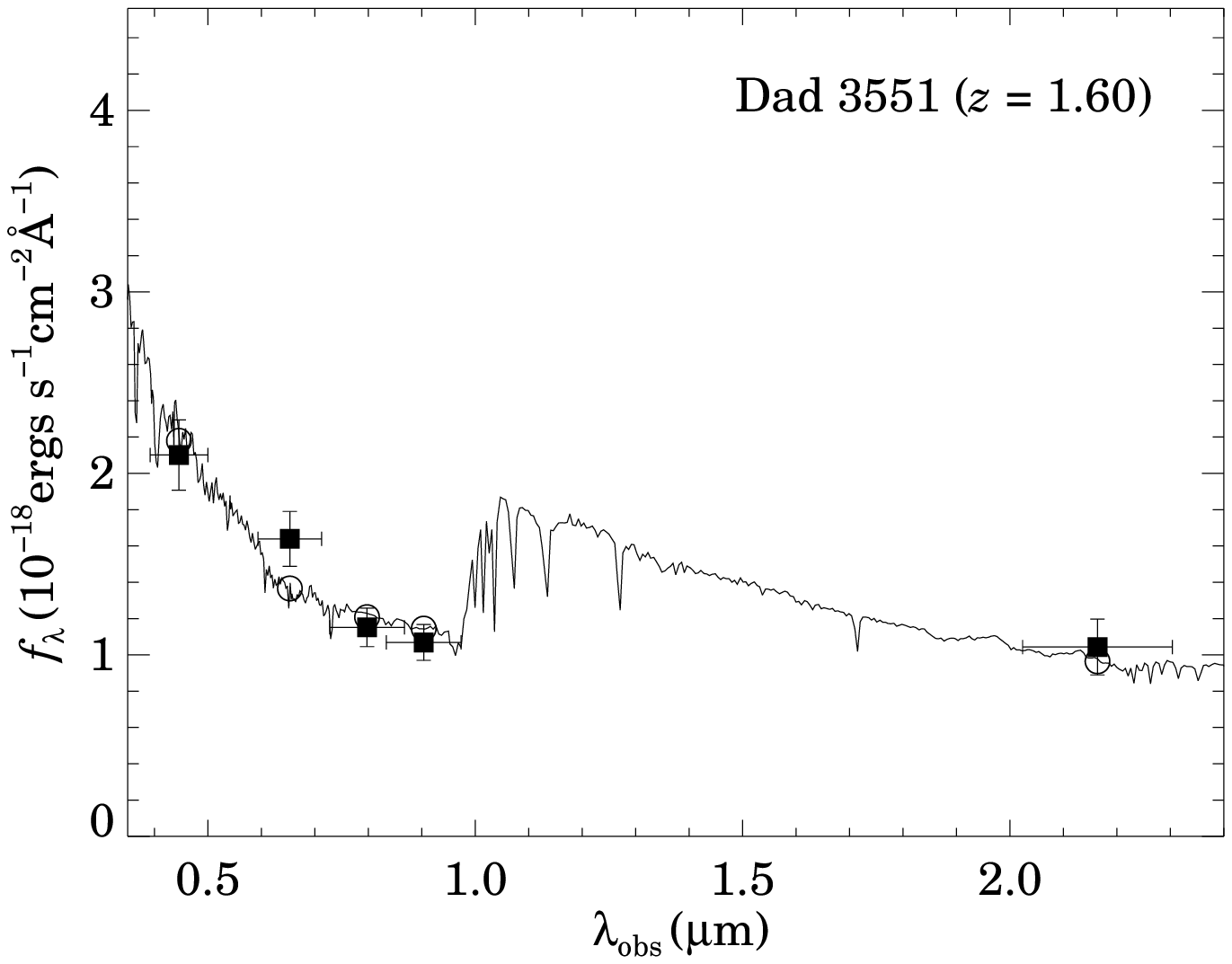}
    \includegraphics[width=0.32\linewidth]{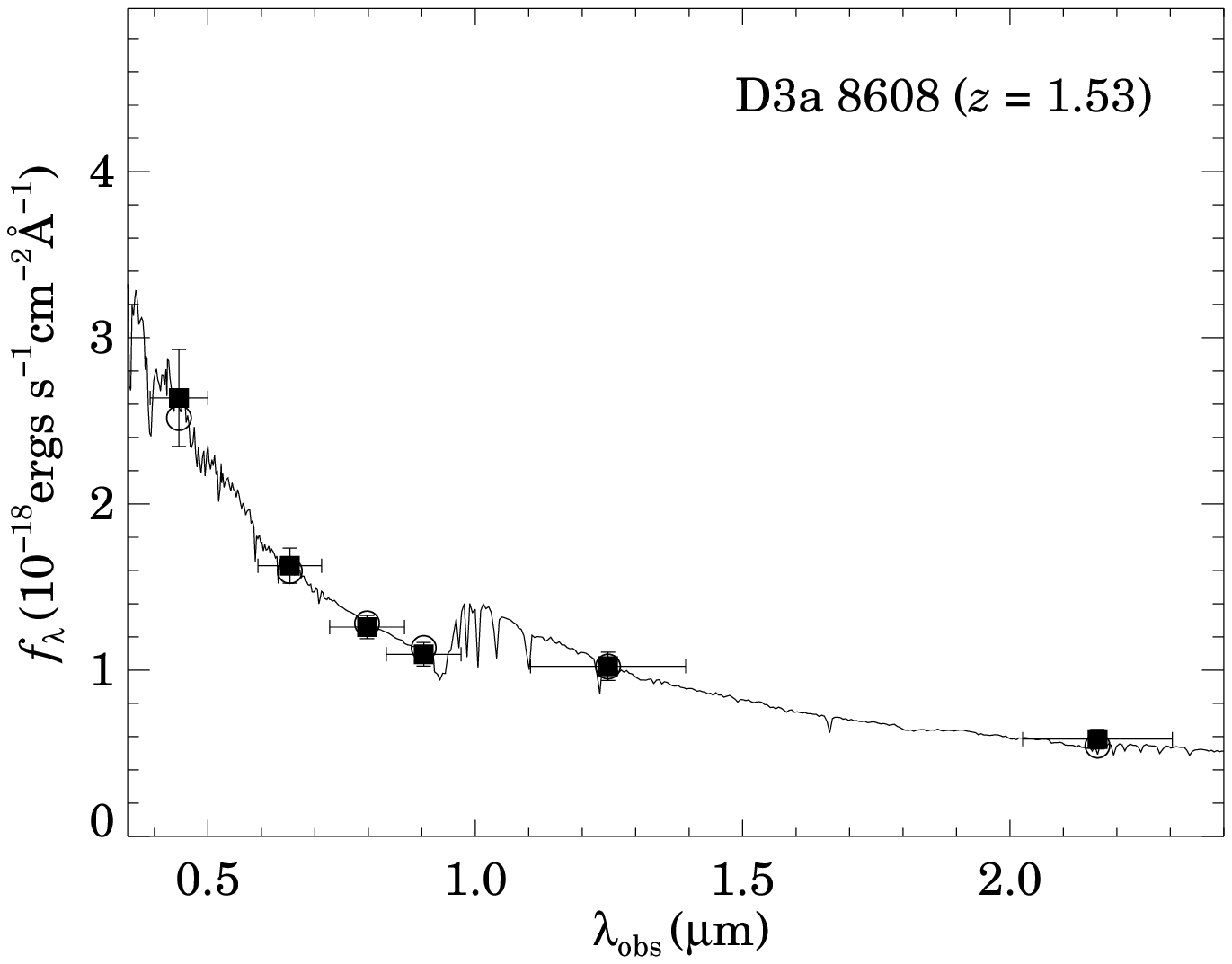}
    \includegraphics[width=0.32\linewidth]{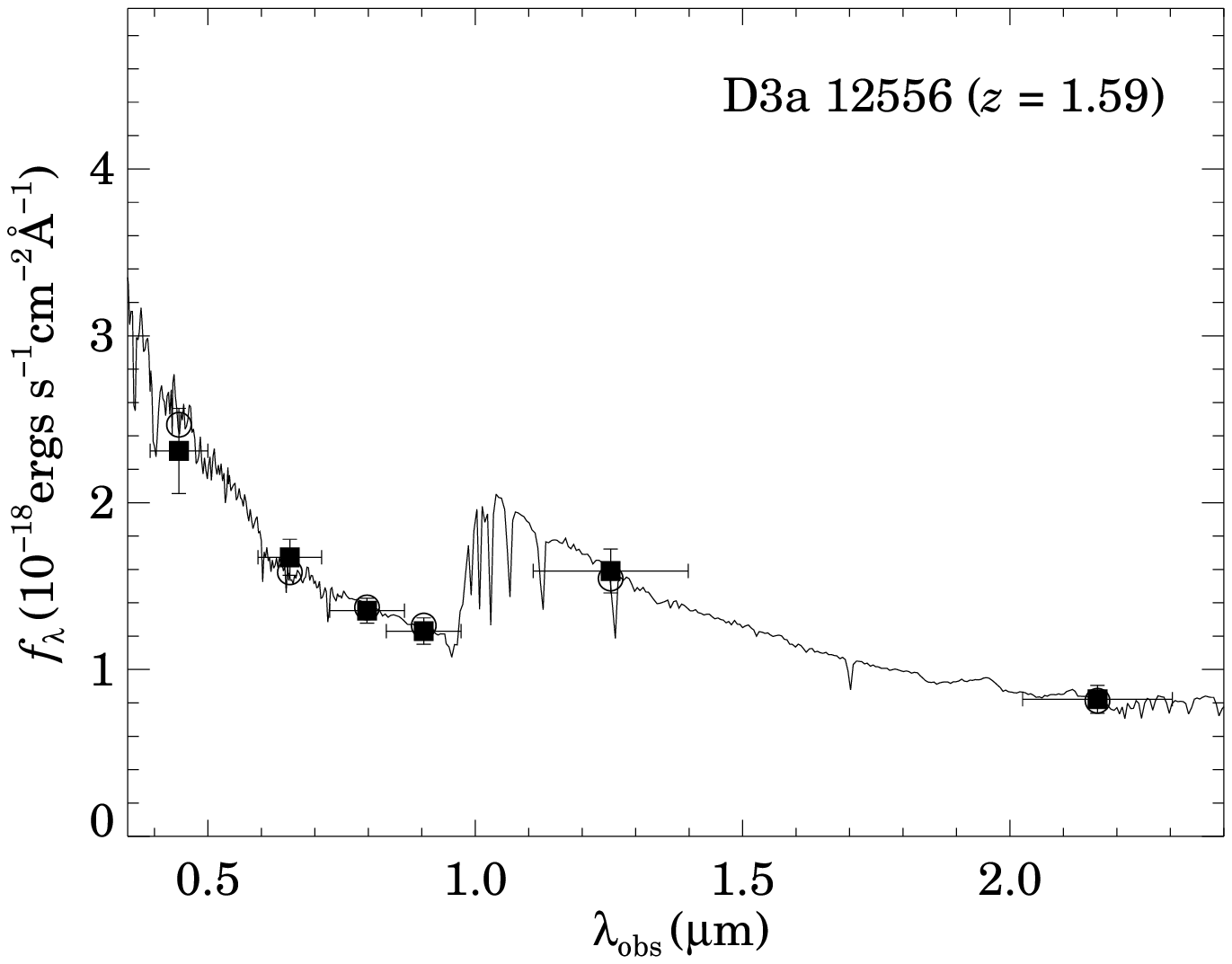}
    \includegraphics[width=0.32\linewidth]{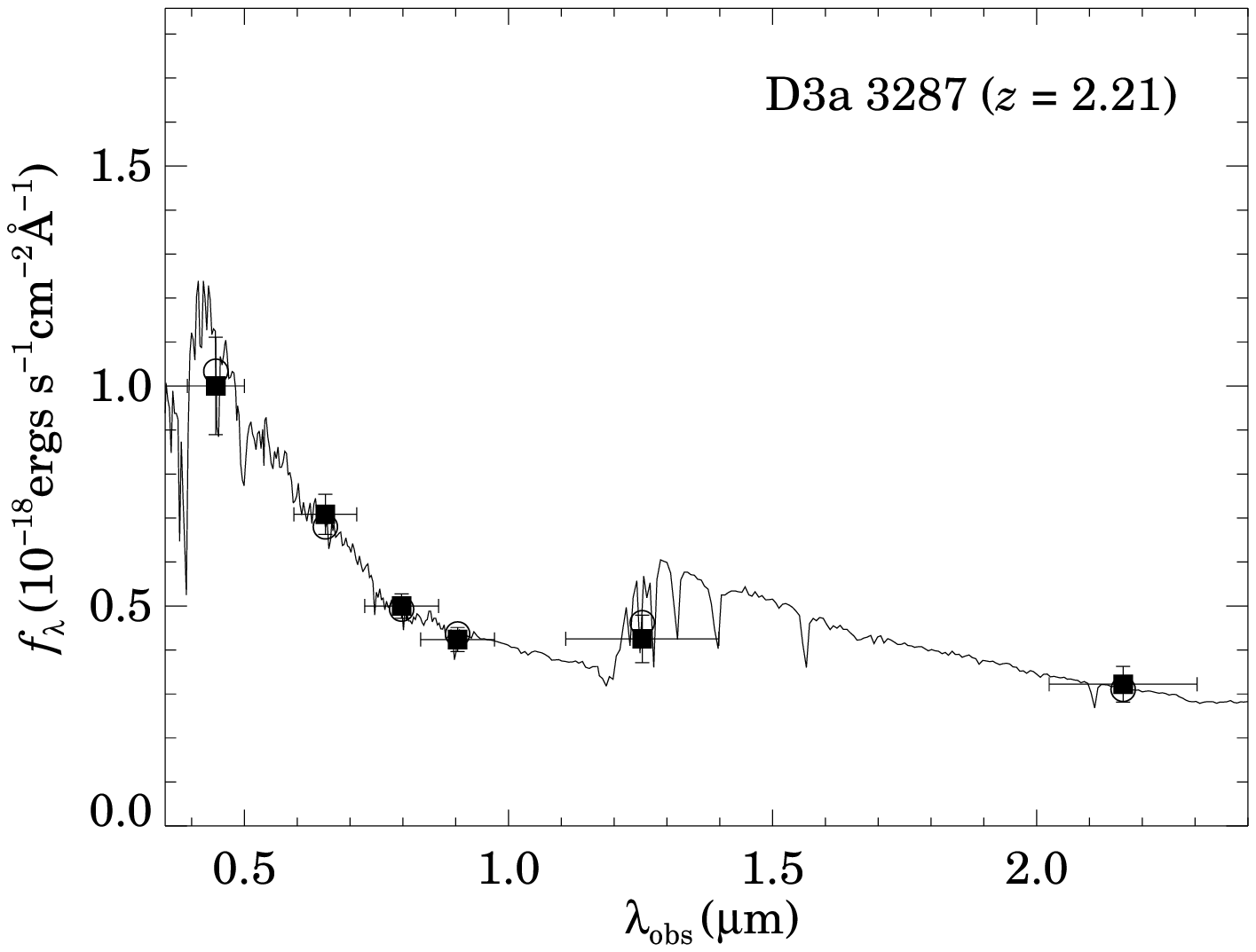}
    \includegraphics[width=0.32\linewidth]{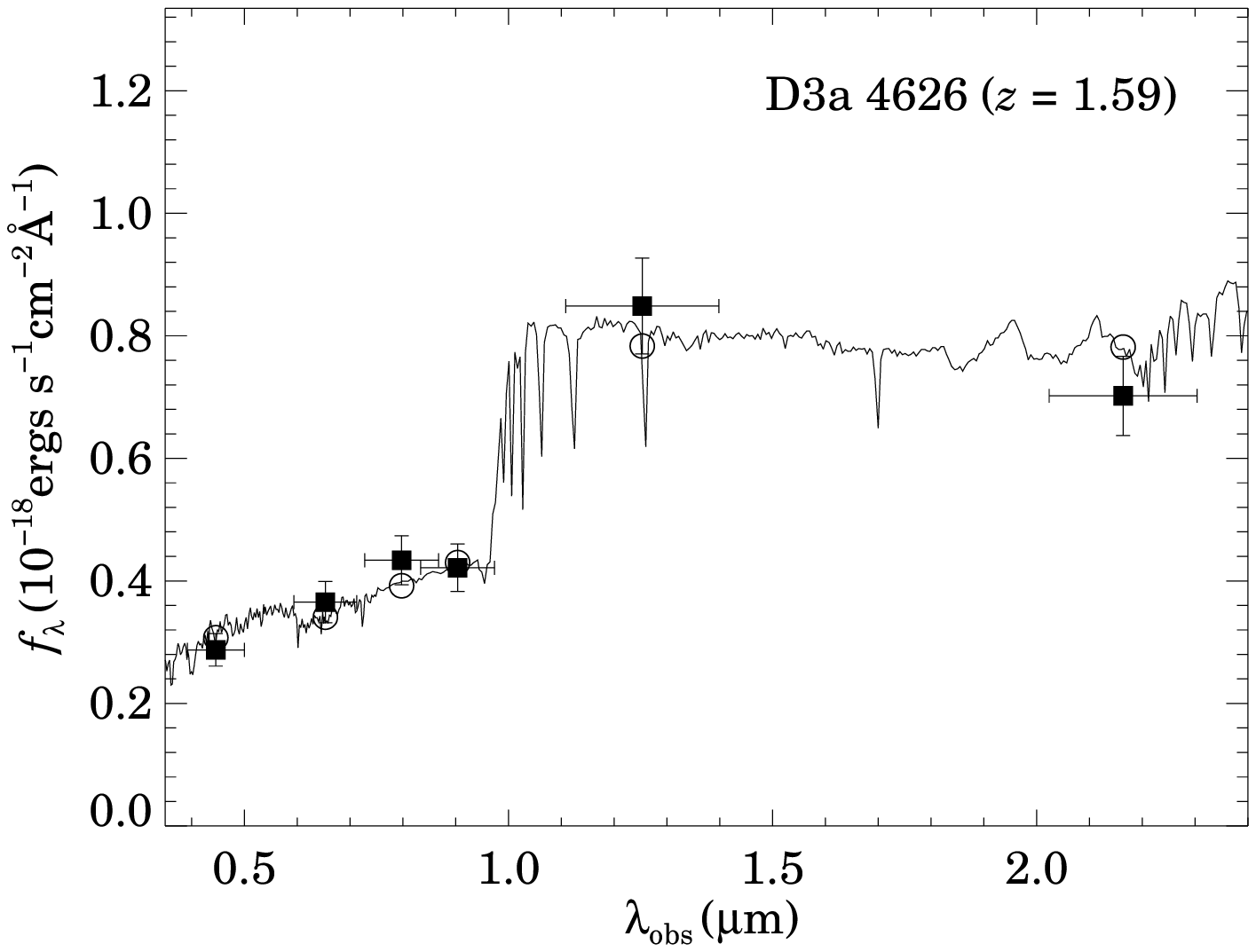}
    \includegraphics[width=0.32\linewidth]{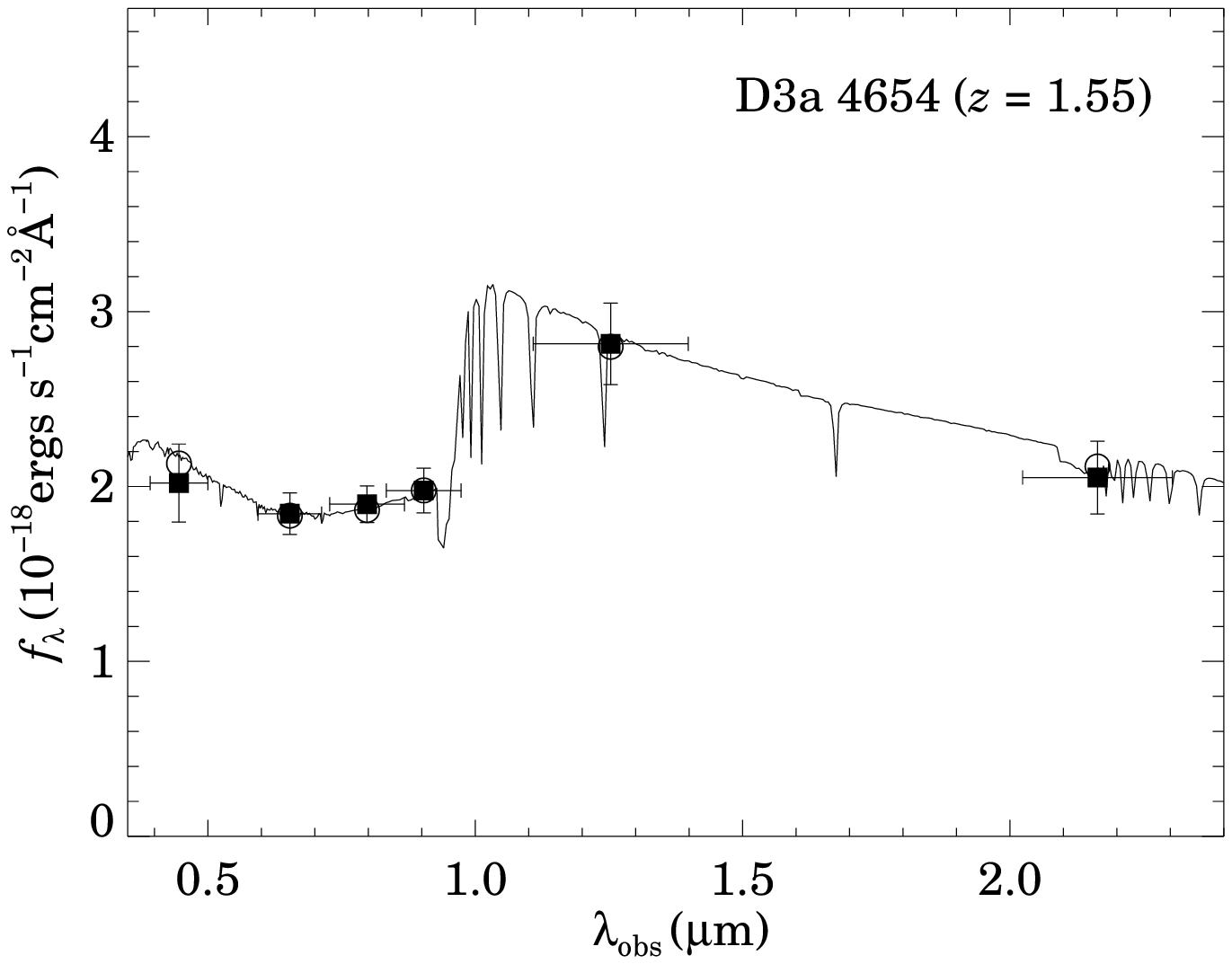}
    \includegraphics[width=0.32\linewidth]{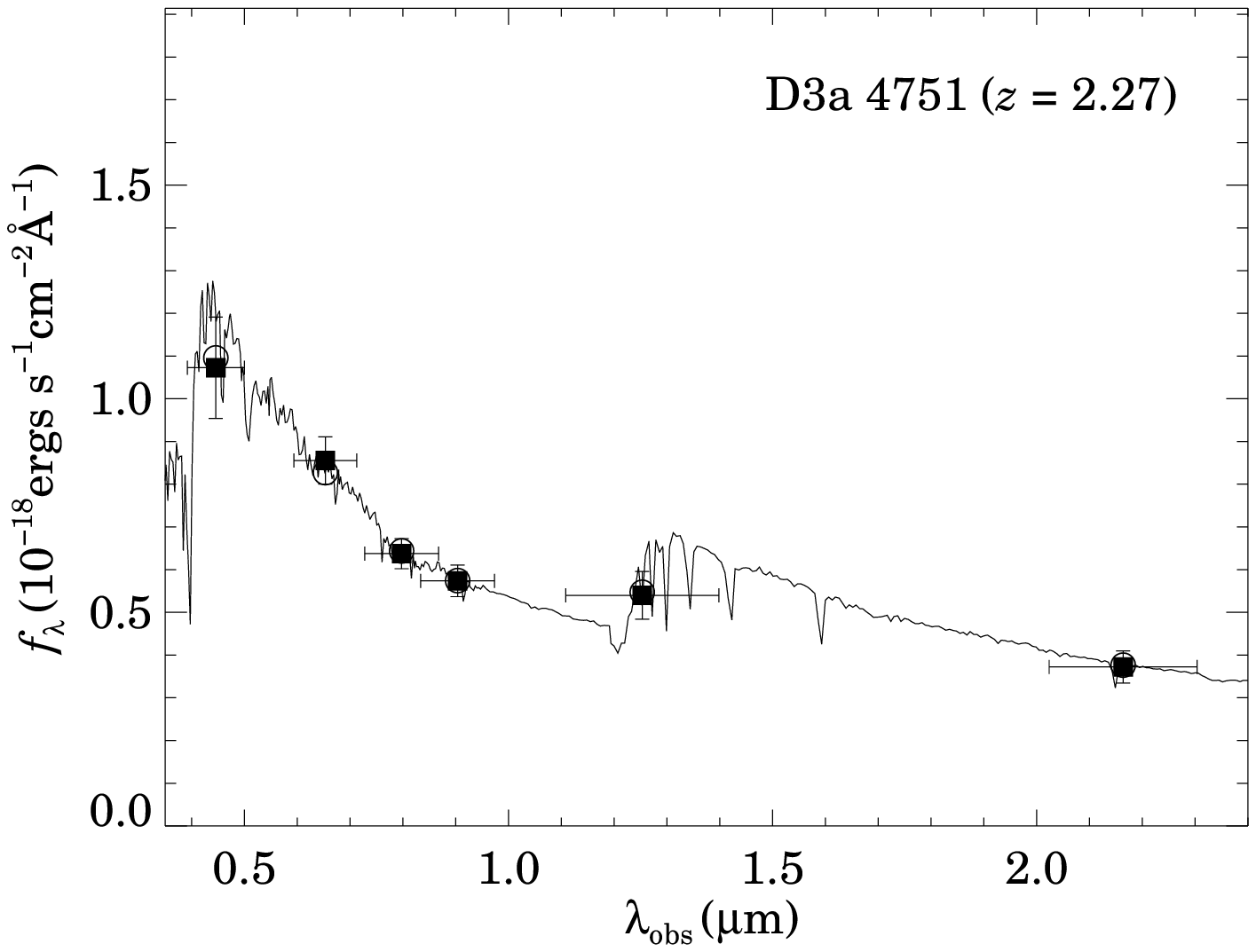}
    \includegraphics[width=0.32\linewidth]{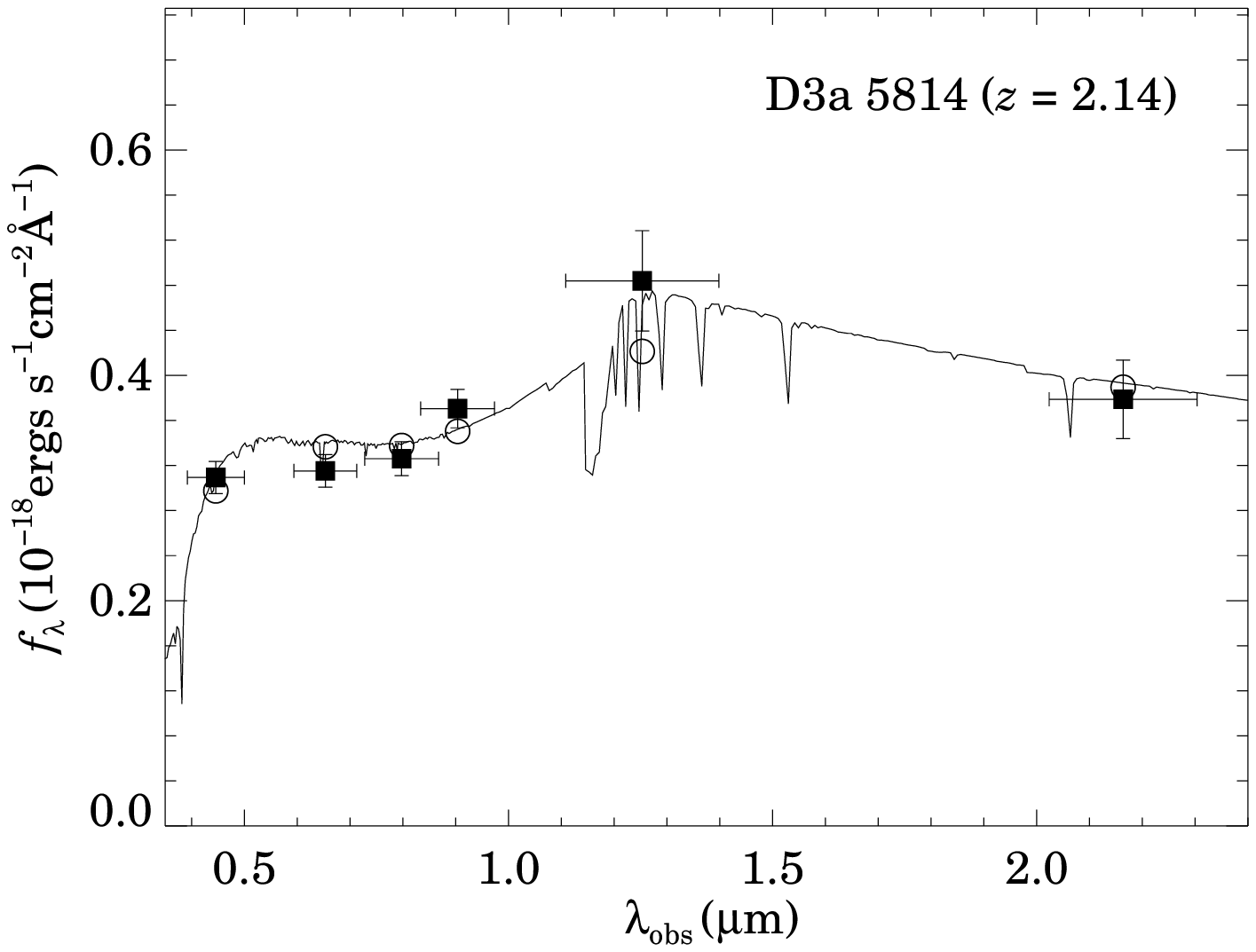}
    \includegraphics[width=0.32\linewidth]{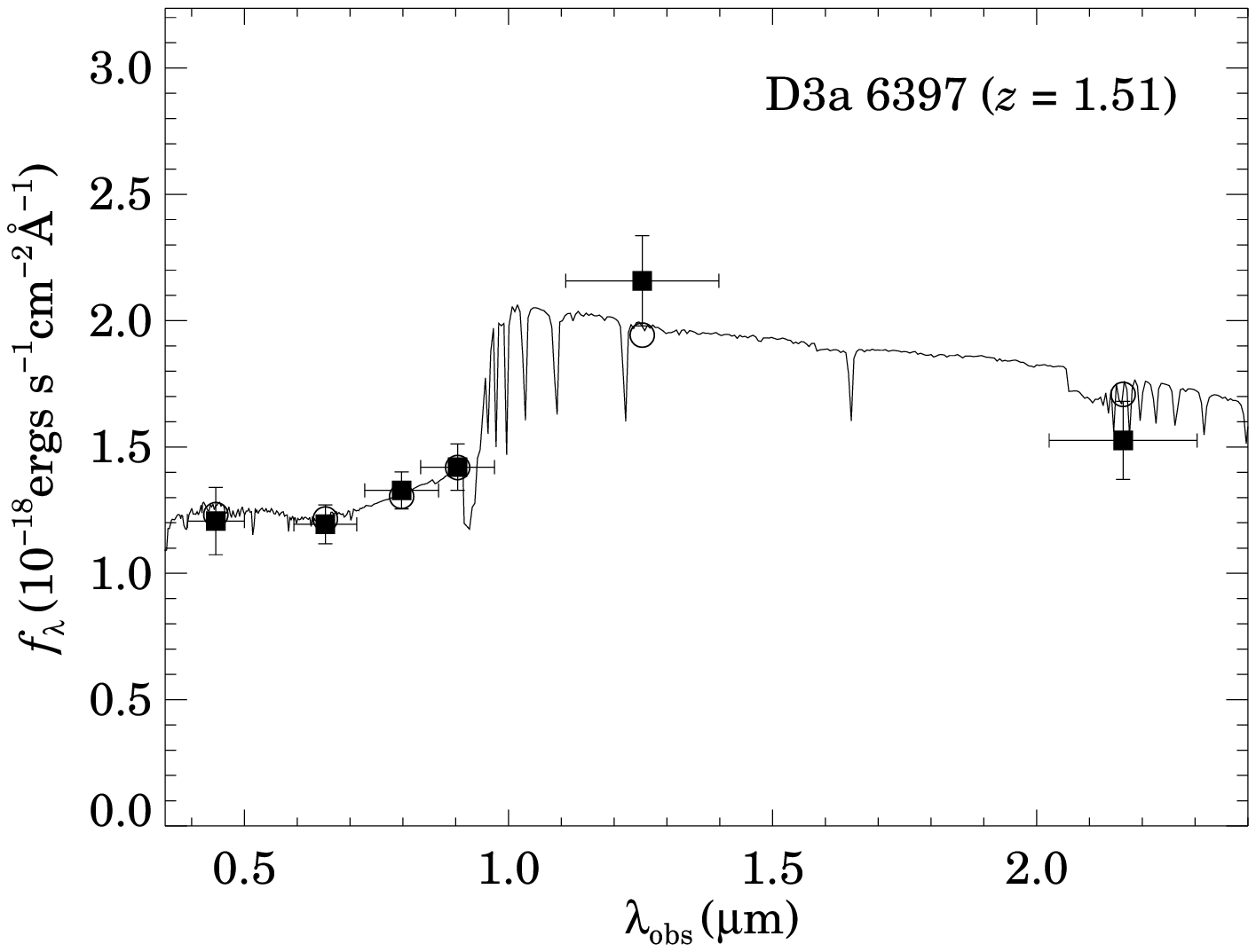}
    \includegraphics[width=0.32\linewidth]{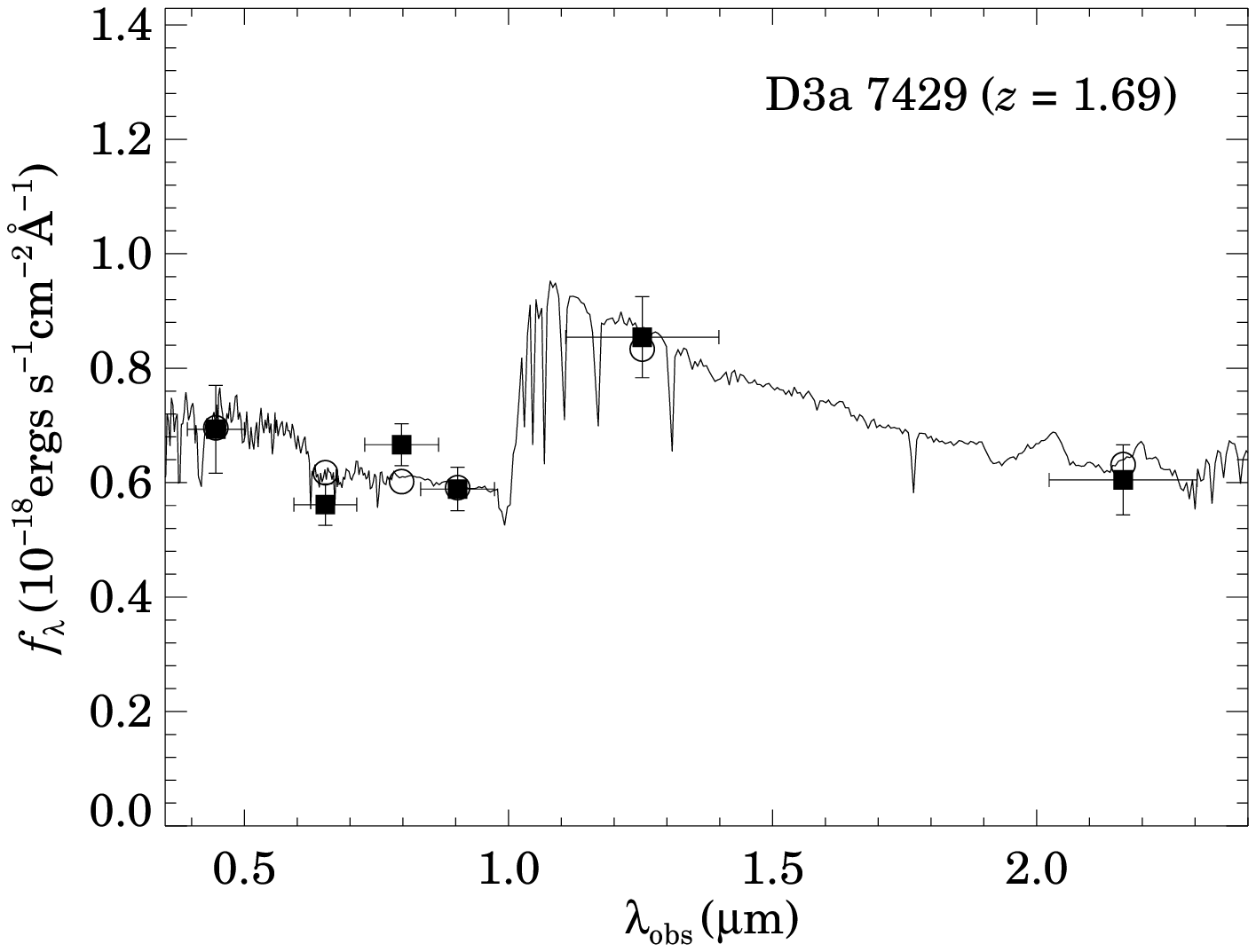}
    \includegraphics[width=0.32\linewidth]{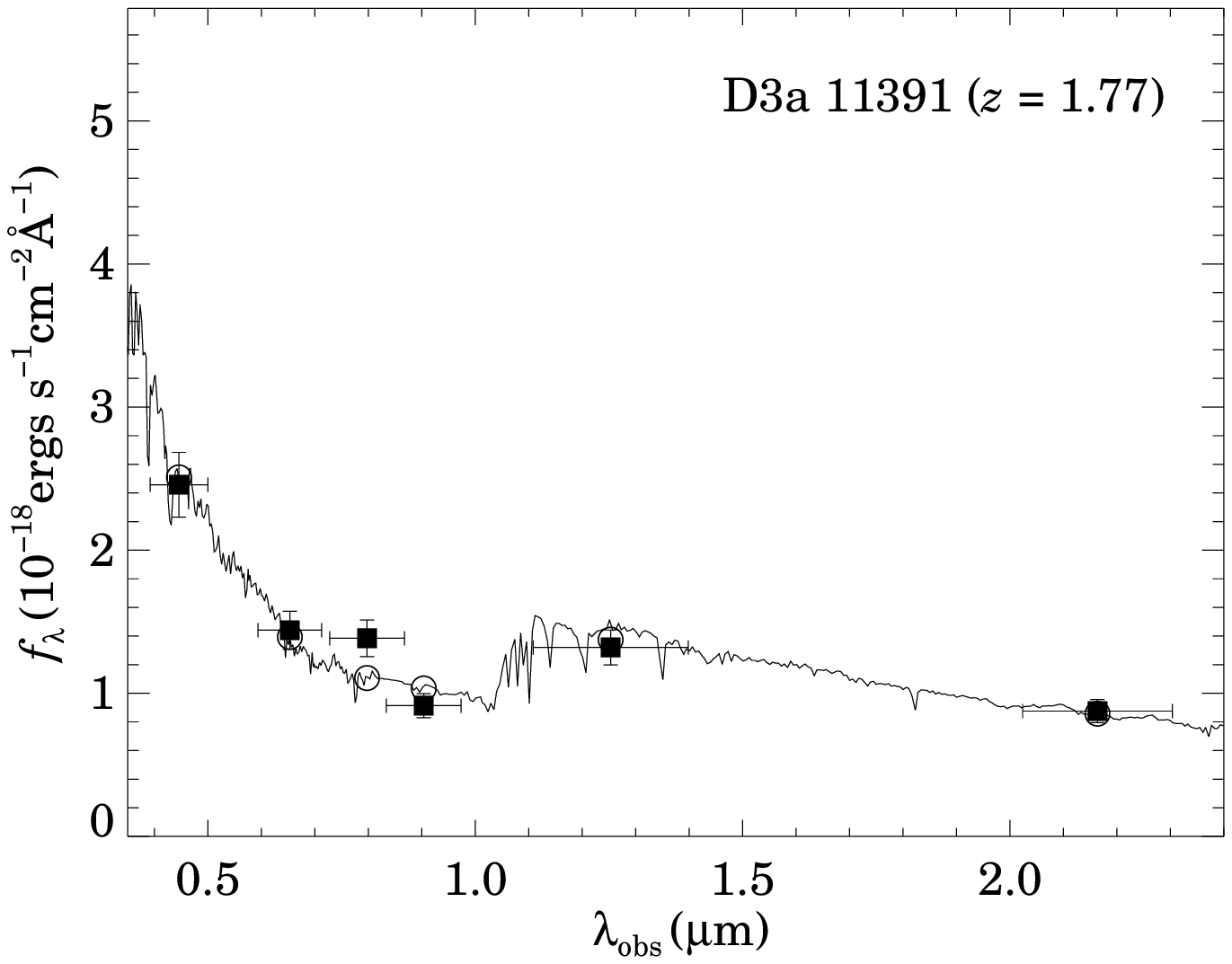}
    \figcaption{
      The SEDs of \hasp detected \sbzks: filled squares show
      the observed flux from broad-band photometry, while solid lines show the best-fit SED from \pegase{}
      models.  Open circles show model magnitudes derived by convolving
      best-fit SED with the filter response curves. 
      The agreement between the observed and fitted photometry is 
      typically very good and so often the open circles are hidden by 
      the filled circles. 
      Note that Dad 2426-b is excluded here because it is a serendipitous 
      object  not in our $K$-selected catalog. 
      \label{fig:bestsed}
    }
  \end{center}
\end{figure}

\begin{figure}
  \begin{center}
    \plotone{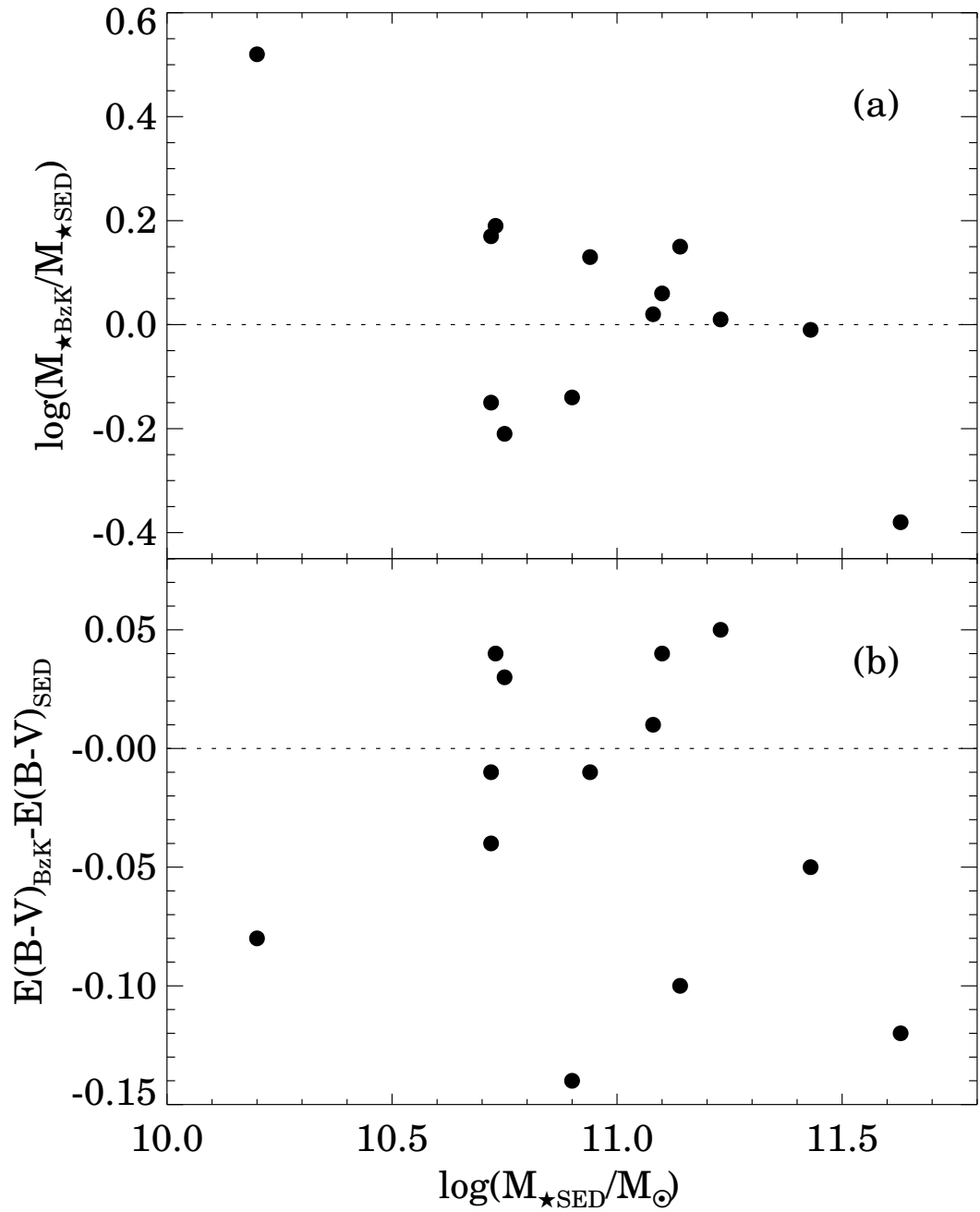}
    \figcaption{
      A comparisons of (a) stellar masses and 
      (b) reddenings as
      derived from \bzk-based formulae and from SED fitting, 
      as a function of stellar mass from the SED fitting. 
      \label{fig:compbzksed}
    }
  \end{center}
\end{figure}

The best-fit SEDs together with observed data points are
shown in Figure {\ref{fig:bestsed}} and the derived stellar masses and
reddenings are listed in Table
{\ref{table:propsed}}.  For a consistency check,  the  stellar
masses and reddenings are also calculated by using the
$BzK$-calibrations as given by Equation (4) and (6) in
\citet{daddi:2004bzk}:
\begin{equation}
  E(B-V) = 0.25 (B-z+0.1)_{\text{AB}}, \label{eq:bzkdust}
\end{equation}
\begin{eqnarray}
  \log(M_\star/10^{11}\,M_{\odot}) = && -0.4(K^{\text{tot}}_{\text{AB}}-21.38) 
  \nonumber \\
  && +  0.218[(z-K)_{\text{AB}}-2.29], \label{eq:bzkmass}
\end{eqnarray}
which are based on SED fits to the full $UBVRIzJHK$-band data
(cf. $BRIzJK$ in this study).  These $BzK$-based stellar masses and
reddening parameters are also reported in Table
{\ref{table:objsample}}.  
Figure \ref{fig:compbzksed}
shows the difference of stellar mass and reddening between the
{\bzk}-based values and those from SED fitting, for the \hasp detected
objects, as a function of stellar masses from SED fitting.  Average
offsets and dispersion between two estimators are
$\langle\log(M_{\star\text{SED}}/M_{\star{BzK}})\rangle=0.03$ and
$\sigma(\log(M_{\star\text{SED}}/M_{\star{BzK}}))=0.22$, respectively,
for the stellar masses, and $\langle\Delta E(B-V)\rangle=-0.03$ and
$\sigma (E(B-V))=0.07$, respectively, for the reddening.  Hence, they
agree reasonably well and without large systematic offsets.

\subsection{H$\alpha$ Star-formation Rates\label{section:hasfr}}

The \hasp luminosities are converted into SFRs 
following the relation in \citet{kennicutt:1998}:
\begin{equation} 
\text{SFR} (M_\odot \text{yr}^{-1}) = 7.9 \times 10^{-42} L_{\text{\ha}} 
(\text{ergs s}^{-1}), 
\end{equation}
which assumes solar abundance, a Salpeter IMF, 
and constant SFR within the last  100 Myr.  
Although the conversion factor may vary from $2.6 \times 10^{-42}$ to 
$8.7 \times 10^{-42}$, 
depending on metallicity, IMF, and star-formation history \citep{buat:2002}, 
we adopt Kennicutt's conversion as it is the most commonly used and it 
simplifies the comparison with results from the literature.  

Although \hasp is usually considered to be relatively unaffected by
dust extinction, dust extinction must be taken into account for
galaxies like \sbzks which undergo vigorous star-formation and are
likely to be heavily obscured by dust 
\citep[Table \ref{table:propsed};][]{daddi:2005b,pannella:2009}.  The \hbsp
emission line would enable us to estimate the dust extinction via the
Balmer decrement, but unfortunately it is not detected in our
spectra.  
The $3\sigma$ upper limits for the \hbsp flux of 
\hasp detected \sbzks is $(\text{\ha}/\text{\hb})_\text{limit}>2.86$, 
which cannot significantly constrain the reddening, 
except for D3a-4751 whose $\text{\ha}/\text{\hb}>5.0$
implies stellar $E(B-V)>0.2$, consistent with 
the stellar $E(B-V)=0.25$ from SED fitting. 
Correspondingly, the reddening values from SED
fitting are used for the extinction correction of the \hasp
luminosities.  One may expect that nebular emission lines from
\ion{H}{2} regions could suffer from larger dust extinction than the
stellar continuum emission \citep{calzetti:2001}. To cope with this
problem, \citet{cidfernandes:2005} compared the extinction of stellar
continuum with that of nebular emission by using star-forming galaxies
from the SDSS data set, and found a linear correlation between them.
This relation was combined with the Calzetti extinction curve by
\citet{savaglio:2005}, obtaining:
\begin{equation}
  A_V=3.173 + 1.841 A_V^\star - 6.418\log\left(\frac{\text{\ha}}{\text{\hb}}
\right)_{\text{th}},
  \label{eq:avgas}
\end{equation}
where $A_V$ and $A_V^\star$ are gas and stellar visual extinctions,
respectively, and (\ha/\hb)$_{\text{th}}$ is the line flux ratio
calculated from the atomic physics theory.  Here we assumed the case B
recombination with $T=10000$K and (\ha/\hb)$_{\text{th}}=2.86$
\citep{osterbrock:2005}.  Extinction corrected
\hasp SFRs are then derived by using $A_V$ from the above relation.
Visual extinctions of stars and gas, extinction at \ha, \hasp
luminosities, and SFRs with and without extinction corrections are 
listed in Table \ref{table:sfr}.

After extinction correction, Dad-2426 and D3a-6397 show
$\text{SFR}\gtrsim 1000 M_\odot\text{yr}^{-1}$.  Dad-2426 has been
detected at 1.2 mm at more than $3\sigma$ level, and its SFR from
FIR luminosity of $800$--$1900M_\odot\text{yr}^{-1}$ 
\citep{dannerbauer:2006} is consistent
with the \ha-based SFR.  In our sample
with \hasp detection, more than half of the \sbzks have $\text{SFR}>
100 M_\odot\text{yr}^{-1}$, after extinction correction.  
These \hasp SFRs $\gtrsim 100 M_\odot\text{yr}^{-1}$ agree 
with the SFRs derived by the combination of UV and MIR for the 
\sbzks with similar range of stellar mass in GOODS fields 
\citep[see \S\ref{section:compsfr};][]{daddi:2005b,daddi:2007sfr}.  
The comparison between SFRs from different indicators and different extinction 
corrections will be further discussed in \S\ref{section:compsfr}.

\subsection{Metallicities\label{section:metallicity}}

Accurate metallicities for the ionized gas can be obtained once the
electron temperature $T_{\rm e}$ is derived from the ratio of auroral
to nebular emission lines, such as
[\ion{O}{3}]$\lambda\lambda4959,5007$/[\ion{O}{3}]$\lambda4363$.
However, auroral lines are intrinsically faint, in particular in the
metal rich galaxies, since the electron temperature decreases due to
efficient cooling by metal lines. Hence, detection of such lines in
high-$z$ galaxies is not feasible with current  facilities 
except for strongly lensed galaxies \citep{yuan:2009}.
Alternatively, abundance indicators using strong emission lines are
widely used to determine the metallicity, being calibrated with the
$T_{\rm e}$ method and/or photoionization models.  One of the most commonly
used indicators is $R_{23}$ \citep{pagel:1979}, which is defined as
$\log R_{23} \equiv \log [(\text{[\ion{O}{2}]}\lambda3727 +
\text{[\ion{O}{3}]}\lambda4959 +
\text{[\ion{O}{3}]}\lambda5007)/\text{\hb}]$.  However, $R_{23}$
cannot be applied to our sample either, due to the lack of required emission
lines.  Here we use instead the N2 index \citep{storchibergmann:1994}
defined as $\text{N2} \equiv \log
([\text{\ion{N}{2}}]\lambda6583/\text{\ha})$.  Since \hasp and
[\ion{N}{2}]$\lambda6583$ are close to each other, the N2 index has
the advantage of being insensitive to dust extinction and flux
calibration, though the origin of nitrogen is rather complicated.

Metallicities derived from different calibrations are known to show
(almost systematic) discrepancies in the mass-metallicity relation of
the local star-forming galaxies \citep[][ hereafter \citetalias{tremonti:2004}]{tremonti:2004} of up to
$\simeq 0.5$ dex \citep{ellison:2006,kewley:2008}.  Thus, it is essential to
derive metallicities in a consistent way, as later we will compare the
metallicities of our \sbzk sample with those of galaxies at
different redshifts drawn from the literature.  Based on \citet{kewley:2002},
\citet[][hereafter KK04]{kobulnicky:2004} derived an analytical expression
for the conversion of N2 into \oh{} that is consistent  with
the $R_{23}$ calibration.  The relation is expressed as:

\begin{eqnarray}
  12  + \log(\text{O/H}) = && 7.04 + 5.28 X_{\text{N\tiny{II}}} + 6.28  X_{\text{N\tiny{II}}}^2 + 2.37 X_{\text{N\tiny{II}}}^3 \nonumber \\
  && - \log q (-2.44-2.01 X_{\text{N\tiny{II}}} \nonumber \\
  && - 0.325 X_{\text{N\tiny{II}}}^2 + 0.128 X_{\text{N\tiny{II}}}^3) \nonumber \\
  && + 10^{X_{\text{N\tiny{II}}}-0.2} \log q(-3.16+4.65 X_{\text{N\tiny{II}}}), \nonumber \\
  &&
  \label{eq:oh12n2}
\end{eqnarray}
where $X_{\text{N\tiny{II}}}\equiv\log
\text{EW([\ion{N}{2}]}\lambda6583)/\text{EW(\ha)}$ and $q$ is the
ionization parameter.  Since the EWs of [\ion{N}{2}] and \hasp were
derived by assuming a flat continuum, EWs in $X_{\text{N\tiny{II}}}$
can be replaced by the corresponding emission line fluxes.  According
to \citetalias{kobulnicky:2004}, the ionization parameter is derived
iteratively from the ionization-sensitive index $\log O_{32} = \log
[(\text{[\ion{O}{3}]}\lambda4959 +
\text{[\ion{O}{3}]}\lambda5007)/\text{[\ion{O}{2}]}\lambda3727]$.  However,
once more also $O_{32}$ cannot be derived for galaxies in our sample,
as some of the required lines are not available. We have then assumed a
constant ionization parameter, an assumption that is justified as follows
after comparing metallicities from the N2
index and those from the $R_{23}/O_{32}$ index.  To derive
metallicities from both indices, we used
emission line fluxes for 75,561   star-forming galaxies in the SDSS DR4 archive
compiled by MPA/JHU collaboration\footnote{http://www.mpa-garching.mpg.de/SDSS/}.
Then we calculated \oh for these SDSS star-forming galaxies using Equation
(\ref{eq:oh12n2}) and by the relation \citepalias{kobulnicky:2004}:
\begin{eqnarray}
\oh = && 9.11 -0.218x - 0.0587x^2 \nonumber \\
&& - 0.330x^3-0.199x^4 \nonumber \\
&& -y(0.00235 - 0.01105x - 0.051x^2 \nonumber \\
&& - 0.04085x^3 - 0.003585x^4),
\label{eq:oh12r23}
\end{eqnarray}
where $x=\log(R_{23})$ and $y=\log(O_{32})$.  

\begin{figure}[htbp]
  \begin{center}
    \plotone{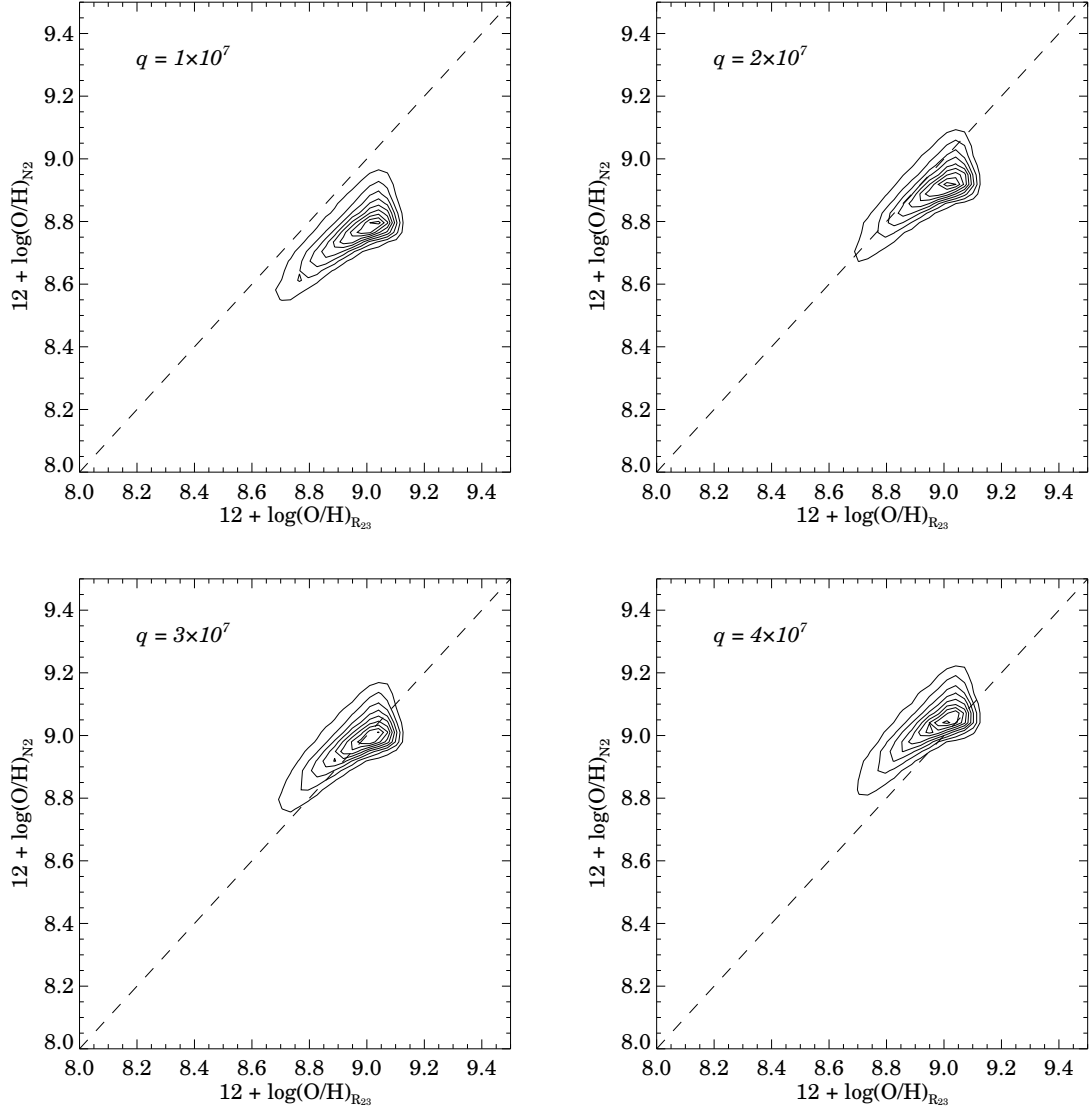}
    \figcaption{
      The correlation between the metallicity 
 derived from the $R_{23}/O_{32}$ parameter (Equation \ref{eq:oh12r23}) and from the N2
      indices (Equation \ref{eq:oh12n2}) for 75,561 star-forming galaxies in the
     SDSS DR4 release. The various panels refer to
different assumed values for the ionization parameter $q$ in  the N2 calibration,  as indicated.
 It appears that $q\simeq 3\times10^7$ is
      appropriate for metal-rich galaxies if a constant value is assumed.
      \label{fig:oh12n2r23sdss}
    }
  \end{center}
\end{figure}

Figure \ref{fig:oh12n2r23sdss} shows the
correlation between these two metallicity calibrators, with different
ionization parameters $q=1\times10^7$, $2\times10^7$, $3\times10^7$
and $4\times10^7$. This figure indicates that a value $q=3\times10^7$
makes the two metallicities to agree for galaxies near the peak of the
metallicity distribution, while the N2 calibration tends to
overestimate the metallicity for $\oh_{\rm R_{23}}\lesssim8.8$.
Given that the metallicities of
our sample are generally large, $\oh>8.8$ (see below), we have
adopted  $q=3\times10^{7}$ in Equation (\ref{eq:oh12n2}).

The resulting gas-phase oxygen abundances of our \sbzks{} are listed in
Table \ref{table:oh12bzk}.  Although D3a-8608 has [\ion{N}{2}]/\ha~$>0.7$, 
which indicates that the ionization may be  dominated by an AGN \citep{swinbank:2004}, 
we derived its metallicity 
with the same equation, just for a reference.  The average value of \oh{} 
in our sample is $8.97\pm0.21$ excluding objects with upper limits and 
AGN dominated features, i.e., D3a-3287, D3a-4626, D3a-8608, and D3a-11391.

\subsection{Dynamical Masses\label{section:mdyn}}
Thanks to the higher spectral resolution ($R\simeq1500$) of the
SINFONI instrument, \hasp and [\ion{N}{2}] emission lines are well
resolved, which enables us to measure the velocity width of individual
objects. In the following we refer to such velocity width as the {\it
velocity dispersion}, although we are aware that in many cases it is
likely due mostly to ordered rotation rather than true velocity
dispersion \citep[cf.][]{forsterschreiber:2009}. The measured velocity
dispersions,  after deconvolving in quadrature  the instrumental resolution ($R=1500$,
or $\sigma_{\text{instrument}}=85\kms$),  are listed in
Table \ref{table:lineprop}.  Then dynamical masses are derived using the equation
\citep[e.g., ][]{binney:1987,pettini:2001}:
\begin{equation}
  M_{\text{dyn}}=\frac{5r_{hl}\sigma^2}{G},
  \label{eq:mdyn}
\end{equation}
where for the half-light radius, $r_{hl}$ we have adopted 6
kpc, i.e., the average value for the \sbzks in the K20 survey measured by \citet{daddi:2004bzk1st}
on the $HST/ACS$ F850LP-band image. This assumes
that the \hasp emission comes
from \ion{H}{2} regions which are virialized within the potential well
of the host galaxy, and that the spatial
distribution of \ion{H}{2} regions represents that of
underlying light distribution.  Dynamical masses derived in this way are
listed in Table \ref{table:mdyn}.  
Note that the velocity dispersion of D3a-4751 is less than the width of the instrumental profile,
hence the velocity dispersion could have large uncertainty.  The
average of dynamical masses of 8 \sbzks with measured velocity
dispersion is $\langle M_{\text{dyn}}\rangle =2.3\times10^{11}M_{\odot}$ with
$1\sigma$ scatter of $1.5\times10^{11}M_{\odot}$.

There are several possible sources of error in these estimates of the
dynamical mass.  The half-light radius which is assumed as constant
must actually vary from one object to another. For example, in the
case of the UV-selected galaxies studied by \citet{erb:2006mass} the
half-light radius varies by more than a factor of 5 from, $\sim 2$ kpc
to $\sim 11$ kpc while the average is $\simeq 6$ kpc.  In a recent
integral field \hasp spectroscopic survey of $z\simeq 2$ galaxies
\citet{forsterschreiber:2009} find an average half-light radius of the
\hasp emitting region of 3.4 kpc. If we adopt this value, the derived
dynamical masses would be 0.6 times smaller than estimated above.  The
numerical factor in Equation (\ref{eq:mdyn}) could also be a dominant
source of uncertainty.  The adopted value of 5 is valid for a sphere
of uniform density, while the actual value of this factor depends on
various parameters, such as the mass distribution within the galaxy,
its velocity structure, and the relative contributions of rotation and 
pressure support for star-forming regions \citep{lanzoni:2003}.  
Considering a disk geometry, \citet{erb:2006mass} used 3.4 for the factor 
and obtained the average dynamical mass of UV-selected star-forming 
galaxies at $z\simeq2$ of $M_{\text{dyn}}\simeq 1\times10^{11}\,M_\odot$ 
with $1\sigma$ scatter of $8.5\times10^{10}\,M_\odot$, 
after accounting for differences in scaling factor. As mentioned above, 
several of these \sbzks are likely to be rotating disks \citep[e.g.,][]{genzel:2006}, 
in which case the reported dynamical masses are overestimated by $\sim50\%$.  
Therefore, the dynamical masses derived here could be taken as upper limits
unless half-light radius is larger than 6 kpc.

For most of the objects  these dynamical masses are $\sim2$--$3$ times larger
than the stellar masses, which in principle could be due
to the presence of large gas masses 
\citep[e.g.,][]{daddi:2008,daddi:2010}.  
However, spatially resolved integral field spectroscopy would be required 
for a more robust determination of the dynamical mass 
\citep[e.g.,][]{forsterschreiber:2009}.

\subsection{Composite Spectrum of \sbzks\label{section:composite}}
\begin{figure}
  \begin{center}
    \plotone{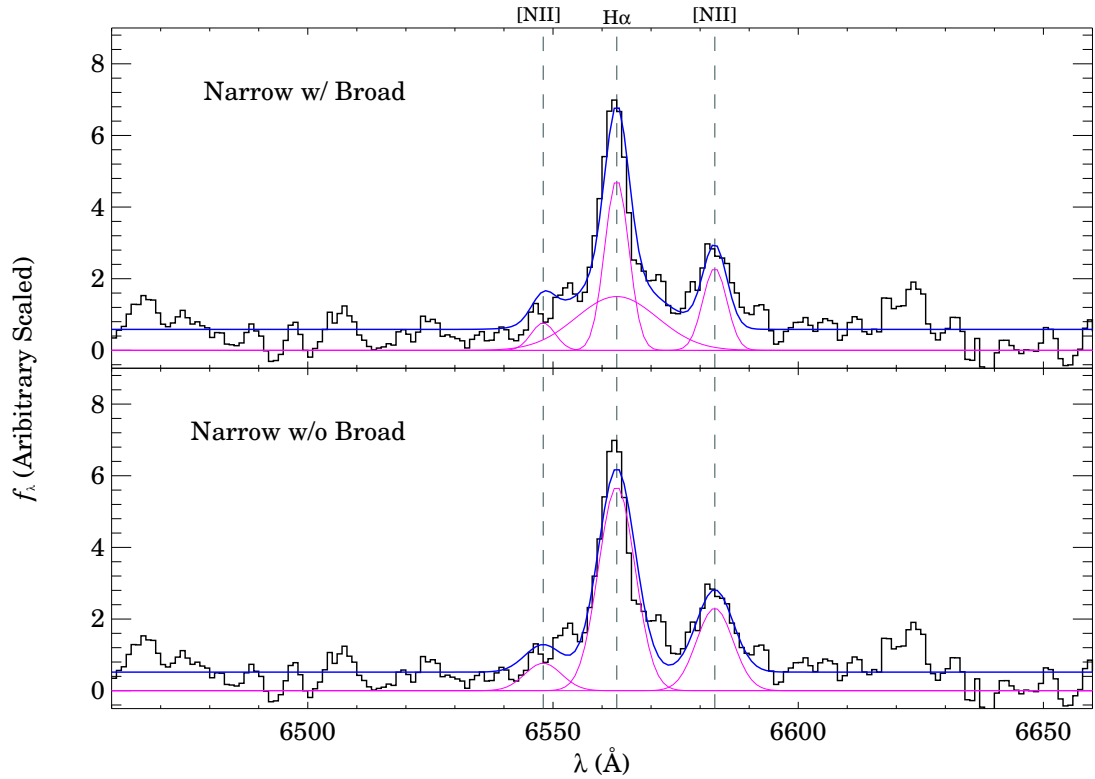}
    \figcaption{
      The stacked spectrum of the six SINFONI \hasp-detected objects (solid black lines).
      The positions of \hasp and the two [\ion{N}{2}] lines are marked at
      the top of the panel with dashed lines.  Best fit multi-Gaussian
      profiles are shown with solid blue lines for the  total spectrum, and with solid
      magenta lines for its individual components.  In the top panel, a
      broad \hasp component is included while only the narrow line
      components are used for the fit in the bottom panel.
      \label{fig:stackedspectrum}
    }
  \end{center}
\end{figure}

The composite spectrum shown in Figure \ref{fig:stackedspectrum} is
made by stacking the 6 SINFONI spectra of \sbzks with \hasp detection
and without AGN features, where \hasp and [\ion{N}{2}]$\lambda6583$ are
well resolved.  
Note that we have excluded D3a-11391 which shows 
broad-line \hasp component as well as the \sbzks with 
higher [\ion{N}{2}]/\hasp ratio as AGN candidates. 
Individual spectra were corrected to the rest-frame
wavelength and then stacked with weights inversely proportional to the square of
the $1\sigma$ rms versus wavelength.
Motivated by the finding of a broad-line \hasp component in the stacked 
spectra of $z\simeq2$ star-forming galaxies in the SINS survey 
\citep{shapiro:2009}, 
we also fit multiple Gaussian functions to the composite spectrum,
with and without a broad \hasp component, though due to the shorter 
exposure time for each object and smaller number of spectra 
used for the stacking, the S/N is lower than that of the SINS stacked spectrum.
Including broad a \hasp component makes the reduced-$\chi^2$ of the fit
about $25\%$ smaller.  The best fit Gaussian profiles are also shown
in Figure \ref{fig:stackedspectrum}. 

The spectral properties derived from this composite spectrum are
$\text{EW(\ha)}=102$~\AA, $\text{EW([\ion{N}{2}]}\lambda6583)=41$~\AA,
and $\text{FWHM}=350$~\kms{} or $\sigma=150$~\kms~ for the fitting
without a broad \hasp component, and $\text{EW(\ha; narrow)}=50$~\AA,
$\text{EW(\ha; broad)}=54$~\AA,
$\text{EW([\ion{N}{2}]}\lambda6583)=24$~\AA,
$\text{FWHM(narrow)}=170$~\kms{} or $\sigma(\text{narrow})=73$~\kms{}
and $\text{FWHM(broad)}=870$~\kms{} or
$\sigma(\text{broad})=370$~\kms{} for the fitting with a broad \hasp
component.  Velocity dispersions above, are corrected for the
instrumental resolution.  Gas-phase oxygen abundances derived from the
narrow line components of the composite spectrum can be calculated as
$\oh=9.12$ and $9.03$; the line widths lead to a dynamical mass of
$M_{\text{dyn}}=3.7\times10^{10}\,M_\odot$ and
$1.5\times10^{11}\,M_\odot$ for the fittings with and without a broad
component, respectively.  
If the broad component is real, the
dynamical mass is a factor $\sim 2$ lower than the average stellar
mass of the galaxies used for the stack ($7.6\times 10^{10}\,M_\odot$).  
However, stellar masses have been derived here
assuming a straight Salpeter IMF, with most of the mass being provided
by low-mass stars. Turning from a bottom-heavy IMF, to a bottom-light
one such as the IMF of Chabrier (2003), the average stellar mass drops
to $4.5\times 10^{10}\,M_\odot$, quite close to the estimated
dynamical mass, and we note that a bottom-light IMF appears to be more
appropriate to account for the observed mass to light ratio of local
galaxies \citep[e.g.,][]{renzini:2005}. In addition, we may have somewhat
overestimated stellar masses as our SED fits are based on stellar
population models that do not incorporate the TP-AGB phase of stellar
evolution. We conclude that the discrepancy between our dynamical and
stellar masses is primarily a result of the assumed IMF, and regard
the dynamical mass derived including the broad \hasp component as
quite plausible and close to the actual stellar mass as well.

\section{Discussion}

\subsection{\sbzks{} with and without H$\alpha$ Detection}
\begin{figure}
  \begin{center}
    \plotone{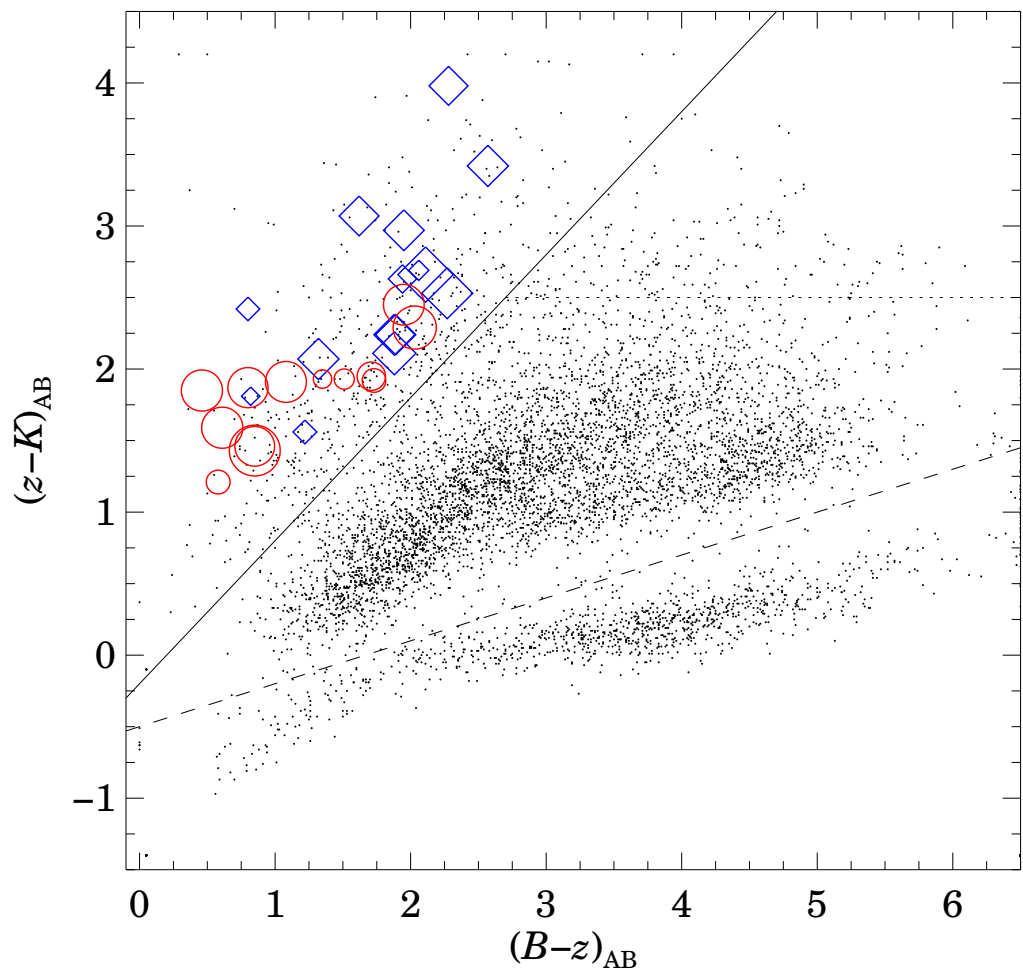}
    \figcaption{
      The $(B-z)$ vs. $(z-K)$ colors of the objects in the Deep3a field and the Daddi field. 
      Black dots show objects in the $K$-selected catalog, 
      while open circles (red) and open triangles (blue) 
      with sizes proportional to the $K$-band luminosity show \sbzks in the present study,
      with and without \hasp detection, respectively. 
      Solid, dotted, and dashed lines separate \sbzks, passive \bzk galaxies (\pbzks), 
      and stars (below the dashed line)  from other populations \citep[see][]{daddi:2004bzk1st,kong:2006}. 
      \label{fig:bzk_comp_detect}
    }
  \end{center}
\end{figure}

In most cases, non-detection of \hasp could well be due to a
combination of bad weather, poor seeing, mis-alignment during the
blind offset, and bad transmission or strong OH emission at the
wavelength of the lines.  However, it is interesting to see whether
there are systematic differences in colors and physical properties
between \sbzks with and without \hasp detection.  In Figure
\ref{fig:bzk_comp_detect} the $(B-z)$--$(z-K)$ color-color diagram for
the objects in our $K$-selected catalog is shown, using different
symbols for \hasp detected and non-detected objects.
\hasp detected \sbzks have on average $\sim 0.15$~mag bluer $BzK$
colors than those without \hasp detection, which indicates a systematically
lower dust extinction because the reddening vector runs
parallel to the diagonal line separating \sbzks from the other
objects \citep{daddi:2004bzk}.  \hasp detected \sbzks also appears to
be slightly brighter in $K$-band compared to those without \hasp
detections.  Therefore, \hasp detections seem to be biased in favor
of more massive, but less extincted objects.  

\begin{figure}
  \begin{center}
    \plotone{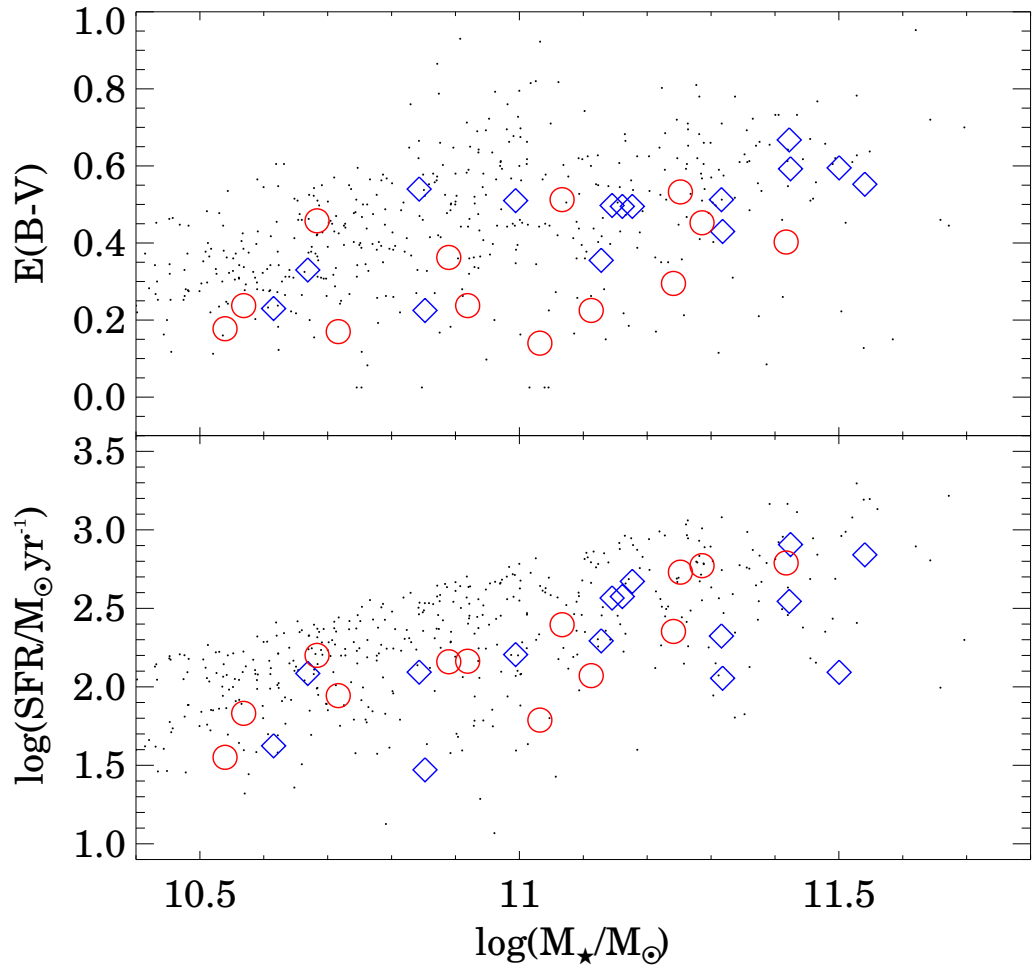}
    \figcaption{
      Top: the $BzK$-based $E(B-V)$ as a function of $BzK$-based stellar mass. 
      Symbols are as in Figure~\ref{fig:bzk_comp_detect}.
      Bottom: the same as in the top panel, but for the extinction corrected SFR derived 
      from the rest-frame 1500\AA{} luminosity, with extinction correction 
      based on the $E(B-V)$ values plotted on the top panel. 
      \label{fig:masssfrebv_comp_detect}
    }
  \end{center}
\end{figure}

The reddening $E(B-V)$ from Equation
(\ref{eq:bzkdust}) and the extinction corrected SFRs derived from the flux in the rest-frame UV
(observed-frame $B$-band)  according to the recipe in
\citet{daddi:2004bzk} are plotted in Figure
\ref{fig:masssfrebv_comp_detect} as a function of stellar masses
derived from Equation (\ref{eq:bzkmass}).  There are no significant differences in
stellar masses ($\sim 0.1$ dex) and SFRs ($20$\%, but uncertainties
are large) between \hasp detections and  non-detections, while the
average reddening is 0.1 mag higher for the \hasp
non-detections.  Therefore, the primary factor affecting the
detectability of \hasp emission lines from \sbzks 
appears to be the amount of dust extinction.  This is
also seen in \citet{hayashi:2009} where most of \sbzks with
non-detected emission lines have redder $BzK$ colors and large
$E(B-V)$ values, $\gtrsim 0.5$.

\subsection{Comparison with Other $z\simeq2$ \hasp Spectroscopic Survey\label{section:comphasurvey}}

\begin{figure}
  \begin{center}
    \includegraphics[]{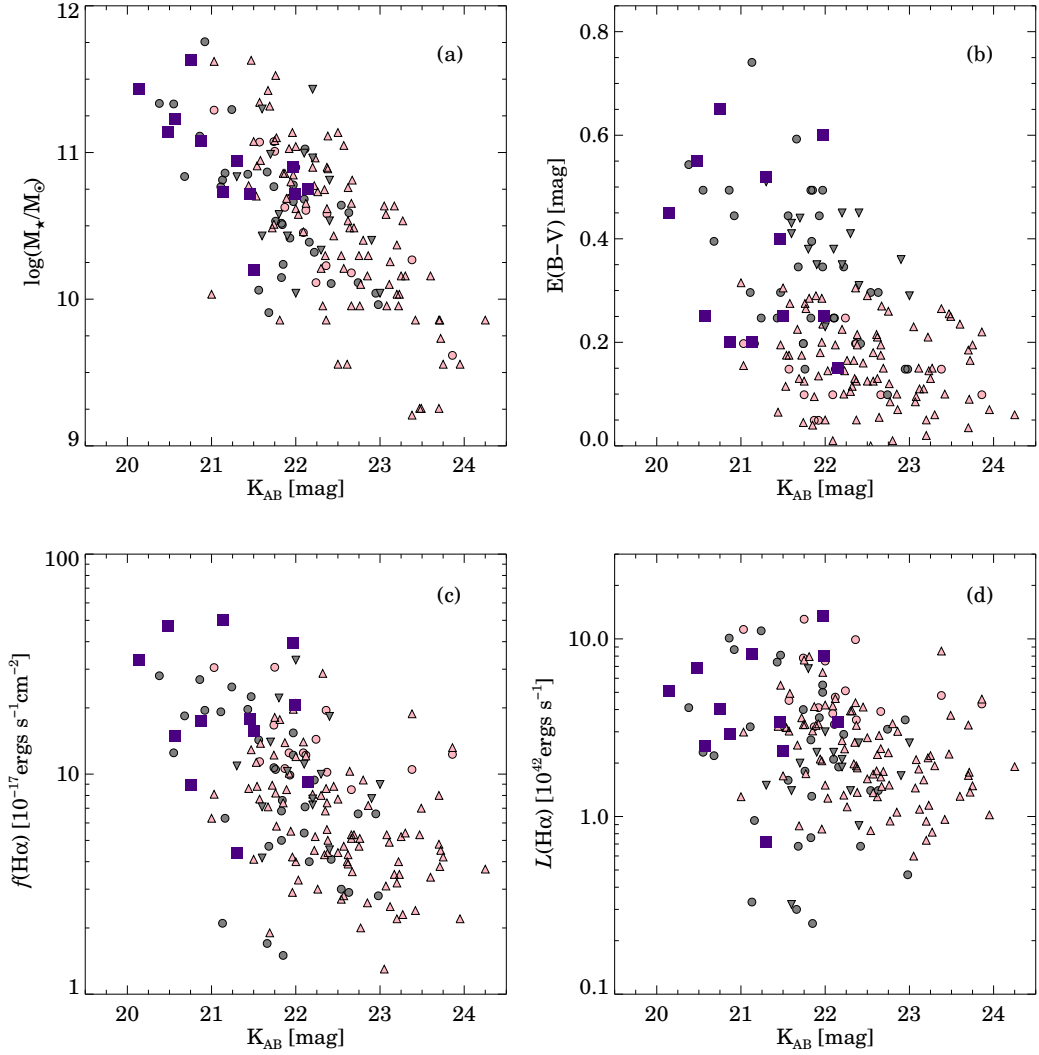} 
    \figcaption{
      Stellar mass, reddening and \hasp properties as a function of the $K$-band
      magnitude are plotted for the galaxies in the present study and
      those from other \hasp surveys. The various panels give: (a)
      stellar mass, (b) reddening, (c) \hasp flux, and (d) \hasp
      luminosity.  Our \sbzk sample is shown with filled indigo
      squares.  rest-frame UV-selected BX/BM galaxies from
      \citet{erb:2006mass,erb:2006sfr} and rest-frame optically
      selected \sbzks from \citet{hayashi:2009} are shown with pink
      triangles and gray upside-down triangles, respectively.  Circles
      represent $z\simeq2$ star-forming galaxies observed by SINS
      survey \citep{forsterschreiber:2009}.  For SINS galaxies, BX/BM
      galaxies and rest-frame optically selected galaxies are shown
      with pink and gray symbols, respectively.
      \label{fig:comphasurvey}
    }
  \end{center}
\end{figure}

Figure \ref{fig:comphasurvey} compares $K$-band magnitudes, stellar masses,
reddenings, \hasp fluxes and \hasp luminosities of our \bzk sample
with those from other \hasp spectroscopic surveys at $z\simeq2$.
The \hasp fluxes and luminosities are not corrected for extinction.
The galaxy samples include the rest-frame UV-selected BX/BM galaxies
\citep{erb:2006mass,erb:2006sfr},
rest-frame optically selected \sbzk galaxies \citep{hayashi:2009}, and the
$z\simeq2$ star-forming galaxies observed by the SINS survey \citep{forsterschreiber:2009}
which includes BX/BM galaxies as well as \bzk-selected galaxies.
Compared to the other \hasp detected star-forming galaxies, our sample contains
only $K$-bright, massive objects mainly because of the brighter $K$-band limiting magnitude
of the imaging survey from which they have been culled \citepalias{kong:2006}.
Apart from that, galaxies in our sample are distributed over  similar ranges of physical properties
(e.g., mass, SFR, extinction) compared to other samples based on rest-frame optical selections.
In contrast, BX/BM galaxies
tend to be
less obscured by dust as expected from their selection technique and
the fraction of objects with strong \hasp flux and luminosity
(e.g., $f_\lambda(\text{\hasp})\gtrsim10^{-16} \text{erg s}^{-1}\text{cm}^{-2}$
and $L(\text{\hasp})\gtrsim2\times10^{42}\text{erg s}^{-1}$)
appears to be smaller than among \sbzks.  Since dust extinction is not corrected in Figure
\ref{fig:comphasurvey} (c) and (d),
the difference between the rest-frame optically-selected population and rest-frame
UV-selected one
would become even larger once extinction correction is applied.

\subsection{Comparison of Star-formation Rates from UV, \ha, and Mid-infrared\label{section:compsfr}}

\begin{figure}
  \begin{center}
    \includegraphics[width=\linewidth]{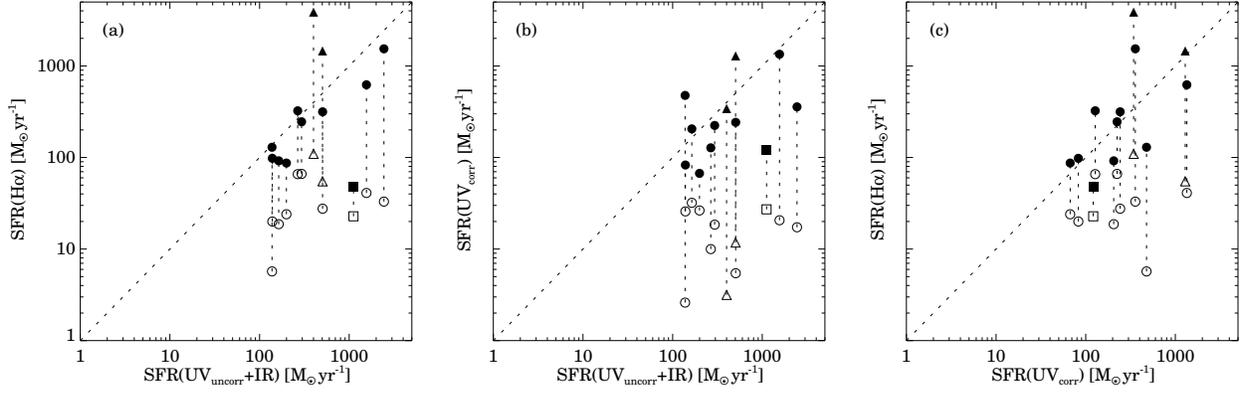}
    \figcaption{
      Comparisons of SFRs from different indicators:
      (a) the SFR from \hasp vs.\ the SFR from
      extinction uncorrected rest-frame UV luminosity and far-infrared
      luminosity derived from the $24\mu\text{m}$ flux:  (b) the SFR from
      rest-frame UV luminosity vs.\ SFR (UV+IR);  (c) the SFR(\ha)
      vs.\ extinction-corrected SFR(UV).  In all panels, open and filled
      symbols refer to SFRs before and after extinction correction,
      respectively, connected by vertical dotted lines.  
      An object with MIR-excess is shown with square symbol and 
      objects with SFR(\ha) much higher than other estimators 
      are shown as triangles.
      \label{fig:sfrcomp}
    }
  \end{center}
\end{figure}

In \S\ref{section:hasfr} we have derived \hasp star-formation rates
with extinction correction following, in which the dust extinction towards \ion{H}{2}
regions is larger than that resulting from the  SED fit to the stellar
continuum.  
To check the validity of this extinction correction, 
SFRs from \hasp are compared in Figure \ref{fig:sfrcomp} with  
those derived from the MIR and the UV 
using the same procedure as in \citet{daddi:2007sfr},  which is reproduced here below.
Following the conversion of \citet{kennicutt:1998}, the IR SFR, or SFR(IR), is derived from the total IR luminosity ($L_\text{IR}$) as 
\begin{equation}
  \text{SFR(IR)}\;[M_\odot\,\text{yr}^{-1}]=1.73\times 10^{-10}L_\text{IR}\;[L_\odot].
\end{equation}
The total IR luminosity is derived from the luminosity-dependent SED 
library of \citet{chary:2001} by using the rest-frame 8 \micron{} luminosity ($L_\text{8\micron}$),  
where the flux density at 24 \micron{} from \textit{Spitzer}/MIPS is used 
as a proxy of the rest-frame 8 \micron, since 24 \micron{} corresponds to $\simeq8$ \micron{} at $z\simeq2$. 
The conversion from $L_\text{8\micron}$ to $L_\text{IR}$ is then
\begin{equation}
  \log\left(\frac{L_\text{IR}}{L_\odot}\right)=1.50\log\left(\frac{\nu L_\text{8\micron}}{L_\odot}\right)-4.31,
\end{equation}
if $\log(\nu L_\text{8\micron})>9.75$,  and 
\begin{equation}
  \log\left(\frac{L_\text{IR}}{L_\odot}\right)=0.93\log\left(\frac{\nu L_\text{8\micron}}{L_\odot}\right)+1.23, 
\end{equation}
if $\log(\nu L_\text{8\micron})<9.75$ \citep{daddi:2007sfr}. 
The UV SFR, or SFR(UV), is derived from rest-frame 1500 \AA{} luminosity ($L_{1500}$) 
by using equation (5) of \citet{daddi:2004bzk}, i.e.,
\begin{equation}
  \text{SFR(UV)}\;[M_\odot\,\text{yr}^{-1}]=1.13\times 10^{-28}L_\text{1500}\;[\text{ergs}~\text{s}^{-1}]. 
\end{equation}
The observed-frame $B$-band is used to derive $L_{1500}$ after applying a $K$-correction based on the 
redshift and the UV-slope of each object.  
Finally, the reddening \ebv from Equation (\ref{eq:bzkdust}) is used for the correction of dust extinction 
to derive the extinction-corrected SFR. 
Figure \ref{fig:sfrcomp}(a) compares the total SFRs derived
from summing SFR(UV)  uncorrected for extinction and SFR(IR),
and SFRs derived from \hasp
luminosities with the extinction correction described in
\S\ref{section:hasfr}.  These two sets of SFRs agree well with each
other, with a few exceptions. 
One \sbzk, D3a-11391, having much lower
$\text{SFR(\ha)}=40M_\odot\text{yr}^{-1}$ compared to 
$\text{SFR(UV+IR)}=1100M_\odot\text{yr}^{-1}$ is a MIR-excess object
\citep{daddi:2007mirx}, defined as an object having SFR(UV+IR)
$\gtrsim 3$ times higher than the SFR(UV) corrected for extinction. 
The object has extinction corrected $\text{SFR(UV)}$ of $120M_\odot\text{yr}^{-1}$ 
which is also much smaller than SFR(UV+IR). 
\citet{daddi:2007mirx} suggested that MIR-excess objects
could be hosting an obscured AGN, being almost completely opaque in the
UV due to dust extinction, but emitting strongly in the MIR from
hot dust surrounding the nuclei.  

The outliers showing very large
excess in \hasp SFR, D3a-5814 and D3a-6397, are possibly due to
overestimates in \ebv from the SED fitting. The \ebv of these 2 \sbzks
from \bzk colors are smaller than those from SED fitting (Table
\ref{table:objsample} and \ref{table:propsed}).  
For large values of \ebv, even a relatively small difference in
\ebv makes a large difference in the resulting \hasp SFR.

In the original Calzetti's recipes for extinction correction it is
suggested that the emission lines from \ion{H}{2} regions suffer more
extinction than the stellar continuum by a constant factor of 2.3, or
$E(B-V)_\text{star} = 0.44E(B-V)_\text{gas}$ \citep{calzetti:2001}.
If we simply use $E(B-V)_\text{star}$ from SED fitting as
$E(B-V)_\text{gas}$, i.e., no additional obscuration toward \ion{H}{2}
regions, then the resulting \hasp SFRs are lower than those derived in
\S\ref{section:hasfr} by factors of $\sim2$ to 5.  On the other hand,
if $E(B-V)_\text{star}=0.44E(B-V)_\text{gas}$ is used for extinction
correction following \citet{calzetti:2001}, it generally overestimates
the \hasp SFR up to factor of $\sim2$.  
\hasp SFRs derived by \cite{hayashi:2009} by using $E(B-V)_\text{gas}$
according to the Calzetti's recipes are overestimated by a large
factor compared to the UV SFRs with extinction correction by
$E(B-V)_\text{star}$.  On the other hand, extinction corrected \hasp
SFRs and UV SFRs agree reasonably well for the UV-selected $z\simeq2$
galaxies \citep{erb:2006sfr} in which the amount of dust extinction is
not large on average.  This suggests that original Calzetti law could
produce over-correction at least for heavily obscured galaxies like
\sbzks.  Therefore, the two cases above,
$E(B-V)_\text{gas}=E(B-V)_\text{star}$ and
$E(B-V)_\text{gas}=E(B-V)_\text{star}/0.44$, bracket the \hasp SFR
derived in \S\ref{section:hasfr} by using the recipe of
\citet{cidfernandes:2005}, and additional extinction toward \ion{H}{2}
regions depending on the amount of extinction of stellar components
could be justified in correcting the \hasp fluxes.

In Figure \ref{fig:sfrcomp} UV SFR corrected for dust extinction and
\hasp SFR are plotted. After extinction correction for the \hasp
luminosities, both SFRs agree reasonably well except for D3a-5814
which is also an outlier in Figure \ref{fig:sfrcomp}(a).  The other
outlier in Figure \ref{fig:sfrcomp}(a), D3a-6397, is not an outlier in
Figure \ref{fig:sfrcomp}(c), possibly because of the smaller
difference between the $E(B-V)$ from SED fitting and that from \bzk
color, compared to that of D3a-5814.

\subsection{Broad \hasp Emission Lines and Super Massive Black Holes}

As shown in the above sections, D3a-11391 is the only galaxy
in our sample of \hasp detected \bzk{} sources that would be classified
as a MIR-excess galaxy following \citet{daddi:2007mirx}.
This galaxy shows a broad \hasp component in addition to the narrow component,
supporting the idea that it contains a powerful AGN.
The velocity width of D3a-11391, $\text{FWHM}\simeq2450\,\kms$, is 
in between that of stacked SINS galaxies at $z\simeq2$ \citep{shapiro:2009} with 
$M_\star\gtrsim10^{11}\,M_{\odot}$ ($\simeq2200\,\kms$) and that of
stacked SINS AGNs ($\simeq2900\,\kms$). 
Many of the most massive galaxies in \citet{shapiro:2009}
sample are also MIR excess galaxies.  
\citet{swinbank:2004} also detected a broad-line component in the stacked 
spectrum of submillimeter galaxies (SMGs), with $\text{FWHM}=1300$~\kms{} 
for the stacked spectrum of the whole SMG sample, and $\text{FWHM}=890$~\kms{} 
for that of the SMGs with no sign of an AGN component.
Moreover, some individual SMGs show a broad \hasp with $\text{FWHM}\simeq2000$--$4000$~\kms{} \citep{swinbank:2004,alexander:2008}. 
The velocity width of the broad-line component in our stacked spectrum, 
$\text{FWHM}=870$~\kms (cf. \S\ref{section:composite}) is close 
to the velocity width of the SMGs without obvious signs of AGN activity. 

Although strong supernova-driven winds can produce a
velocity width of $\text{FWHM}>2000\,\kms$, it seems difficult to
explain the MIR-excess as due to supernova remnants  because supernovae are directly related to the
star-formation activity which would also produce UV and \hasp emissions,
consistent with the MIR emission.  Thus, the
object could be hosting an AGN at the center and the MIR-excess can be
due to the surrounding dust torus which absorbs the UV emission
from the central nucleus. Although typical MIR-excess galaxies are
expected to contain very obscured, often Compton thick AGNs, the broad
\hasp emission can be explained as a leak from the broad-line region (BLR)
around the central nucleus, which would still be observable in the optical/NIR
where the optical depths is lower than in the UV. Alternatively, \hasp photons in the broad component
could have been scattered into 
the line-of-sight by either electrons or dust particles.  Unfortunately,  there are no X-ray 
data for this object,  and the \ha--X-ray diagnostics discussed 
in \citet{alexander:2008} cannot be applied.  
In the following discussion, we assume that the broad-line emission 
is not dominated by scattered light but is seen directly. 

By using the measured \hasp luminosity and the width of the broad line 
and assuming that it is coming from an AGN, we attempt here a rough estimate of 
the virial mass of the central super massive black hole (SMBH) 
by adopting the relation provided by \citet{greene:2005}: 
\begin{multline}
  M_{\text{BH}} = \left(2.0^{+0.4}_{-0.3}\right)\times10^6\\
  \times
  \left(\frac{L_{\text{\ha}}}{10^{44}\text{ergs\ s}^{-1}}\right)^{0.55\pm0.02}
  \left(\frac{\text{FWHM}_{\text{\ha}}}{10^3\kms}\right)^{2.06\pm0.06} M_\odot,
  \label{eq:mbh_greene}
\end{multline}
or the relation by \citet{kaspi:2005} as converted by \citet{greene:2005}
for the \hasp luminosity and width:
\begin{multline}
  M_{\text{BH}}=\left(1.3\pm0.3\right)\times10^6\\
  \times
  \left(\frac{L_{\text{\ha}}}{10^{44}\text{ergs\ s}^{-1}}\right)^{0.57\pm0.06}
  \left(\frac{\text{FWHM}_{\text{\ha}}}{10^3\kms}\right)^{2.06\pm0.06} M_\odot. 
  \label{eq:mbh_kaspi}
\end{multline}
            
For the broad component of D3a-11391,
$L_{\text{\ha}}=7.6\times10^{42}$ ergs\ s$^{-1}$ and
$\text{FWHM}_{\text{\ha}}=2450\kms$ lead to
$M_{\text{BH}}\simeq3\times10^7\,M_{\odot}$ by averaging
the values from these two equations.  If the geometry of
BLR is disk-like, the average correction factor for the
inclination would be $\simeq2.7$
\citep{mclure:2002,alexander:2008}, which gives
$M_{\text{BH}}\simeq9\times10^7\,M_{\odot}$.  The resulting
value is $\simeq 2$--$3$ times smaller than the average SMBH mass of
broad-line SMGs at $z\simeq2$ \citep{alexander:2008}, and a
factor $\gtrsim 10$ smaller than that of optically selected QSOs
in SDSS at $z=1.8$--$2.1$
\citep{mclure:2004,alexander:2008}.  On the other hand,
SINS galaxies with similar stellar mass have $\simeq 3$
times smaller black hole mass compared to what estimated here for D3a-11391
\citep{shapiro:2009}.  Dust extinction for the broad-line
\hasp emission is not considered here since \hbsp emission
line is out of the observed wavelength range.  If we
assume the same amount of extinction as that for the host
galaxy ($A_V=1$), it would increase the virial black hole
mass by about a factor of 2.

Rest-frame $2$--$10$ keV X-ray luminosity can be
inferred from rest-frame $8$ $\mu$m luminosity assuming
that the emission from AGN dominates at $8$ $\mu$m \citep{lutz:2004,alexander:2008}. 
Since 24 $\mu$m corresponds to rest-frame 8.7 $\mu$m at
$z=1.774$, MIPS $24\,\mu$m flux can be used as a proxy of
rest-frame $8\,\mu$m luminosity.  The flux of D3a-11391 is
$\sim 400\,\mu$Jy at $24\,\mu$m, corresponding to $\nu
L_{8\mu\text{m}}\simeq 10^{45}$ ergs\ s$^{-1}$, which is
translated into an absorption corrected 2--10 keV
luminosity of $L_{2\text{--}10\text{keV}}\simeq10^{44}$
ergs\ s$^{-1}$ 
by using the same method as in \citet{alexander:2005}. 
Alternatively, the \hasp luminosity can
be also used to derive $2$--$10$ keV luminosity
\citep{ward:1988}, and from Figure 5 of \citet{ward:1988} we derive
$L_{2\text{--}10\text{keV}}\simeq10^{44}$ ergs\ s$^{-1}$, in agreement with the estimate from
the $8\,\mu$m luminosity.  Given that a mass-accretion of
$1\,M_{\odot}$ yr$^{-1}$ corresponds to
$L_{2\text{--}10\text{keV}}\simeq 2.2 \times 10^{44}$~ergs\
s$^{-1}$, the mass-accretion rate of matter to the central
SMBH of D3a-11391 can be estimated as $\simeq
0.5\,M_{\odot}$~yr$^{-1}$.  While this estimate is obviously affected by large
uncertainties, we note that this SMBH accretion rate is about 1/2  of the value
estimated for 
broad-line SMGs \citep{alexander:2008} and one order of
magnitude smaller than the typical value of QSOs at $z=1.8$--$2.1$
\citep{mclure:2004,alexander:2008}.

Then we can proceed further, and crudely estimate the Eddington factor, $\eta$, by
comparing $M_{\text{BH}}$ and the accretion rate
\citep[Figure 2 of][]{alexander:2008}.  For D3a-11391 this gives 
$\eta\simeq0.3$--$0.9$, depending on the assumed
geometry,  somewhat higher than the typical Eddington ratio of
broad-line SMGs and lower than the average of SINS
galaxies. However, uncertainties in both black hole mass and mass-accretion
rate are large.

A broad-line component with $\text{FWHM}=870$~\kms{} 
was also  detected our stacked spectrum of the remaining \bzk{} galaxies
(Fig. \ref{fig:stackedspectrum}), although not with high significance. 
Its velocity width is
not as large as that of BLRs, but rather similar to that 
of narrow-line Seyfert galaxies.  Proceeding in the same way as for D3a-11391, 
we then estimate the black hole mass, X-ray luminosity, accretion rate 
and Eddington factor for the {\it average} galaxy represented by the stacked spectrum. 
We then obtain  $M_{\text{BH}}=3\times10^6\,M_\odot$
or $7\times10^6\,M_\odot$, respectively for spherical and
disk-like geometry, $L_{2\text{--}10\text{keV}}\simeq 5
\times 10^{43}$~ergs~s$^{-1}$, an accretion rate of
$0.2\,M_\odot$~yr$^{-1}$, and 
$\eta\gtrsim1$.  Here we have used the average redshift 
($z=1.96$), average \hasp luminosity 
($3.4\times10^{42}$~ergs~s$^{-1}$~cm$^{-2}$) and average
$24\mu\text{m}$ flux ($156\mu$Jy) of the galaxies in the stack.  
It is also assumed that the whole 
$24\mu\text{m}$ flux comes from
an obscured AGN, which would overestimate the
rest-frame $8\mu$m luminosity. 
By inferring the X-ray luminosity from 
the broad \hasp luminosity we get
$L_{2\text{--}10\text{keV}}\simeq3\times10^{43}$~erg~s$^{-1}$, quite consistent
with the value estimated from the $8\mu$m luminosity, and similar 
to the value estimated
by \citet{daddi:2007mirx} for the MIR-excess galaxies on the
basis of X-ray stacking.

\subsection{Mass-Metallicity Relation\label{section:mzrel}}

\begin{figure}
  \begin{center}
    \plotone{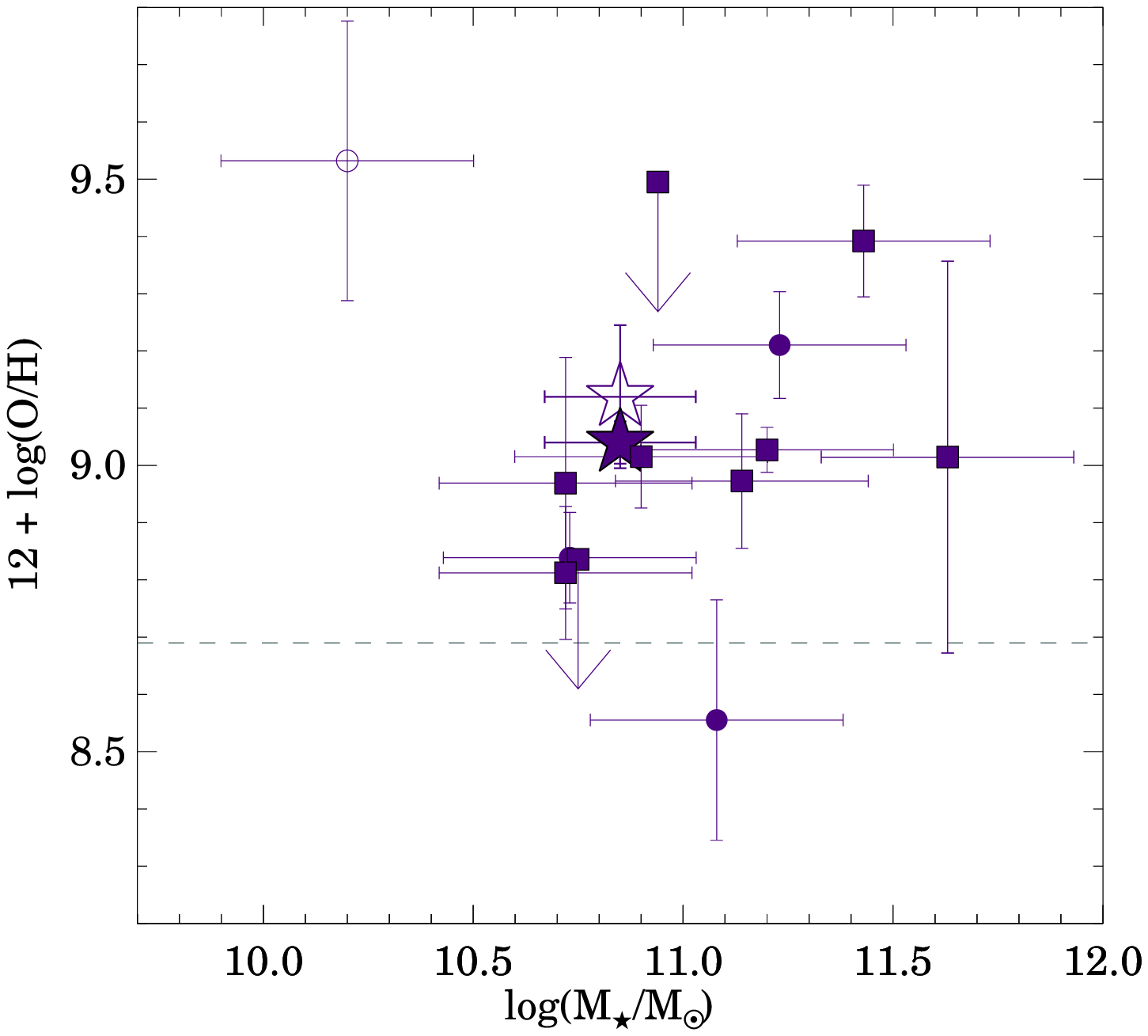}
    \figcaption{
      The \mz relation for \sbzks at $z\simeq2$.  
      Filled and open circles and squares 
      represent objects with [\ion{N}{2}]/\ha~$<0.7$ and [\ion{N}{2}]/\ha~$>0.7$,
      respectively.  Circles and squares represent metallicities from
      OHS/CISCO and SINFONI data, respectively.  A typical error bar for
      the stellar mass of 0.3 dex is indicated for each galaxy.  
      Stars represent the values derived from the stacked spectra 
      by means of profile fitting with and without a broad-line \hasp component 
      (open and filled star, respectively). 
      The horizontal dashed line shows the solar oxygen abundance,
      $12+\log(\text{O/H})_\odot=8.69$ {\citep{allendeprieto:2001}}.
      {\label{fig:mzrelbzk}}
    }
  \end{center}
\end{figure}

The gas-phase oxygen abundances of the galaxies in the present sample
as derived from Equation (\ref{eq:oh12n2}) are plotted as a function
of stellar mass in Figure{} \ref{fig:mzrelbzk}.  Error bars for
stellar mass are set to 0.3 dex, as the typical uncertainty of our
estimates.  AGN candidates, i.e., those showing large
[\ion{N}{2}]/\hasp ($>0.7$) ratios, are also plotted in Figure
\ref{fig:mzrelbzk} as open squares and circles.  The metallicity
derived from the stacked spectrum (with and without the broad line
component) is also shown along with the average stellar mass of the
stacked galaxies (star symbols).  The solar oxygen abundance is
indicated with the dashed line, hence almost all galaxies in the
present sample appear to have super-solar ISM abundances.  However,
the absolute values of the oxygen abundances should be taken with
caution having been obtained indirectly from the [\ion{N}{2}]/\hasp
ratio.  For example, by comparing various indicators of gas-phase
oxygen abundance, including the $T_{\text{e}}$ method,
\citet{shi:2005} using [\ion{O}{3}]$\lambda4363$ found that in the
luminosity-metallicity ($L$-$Z$) relation of the local blue compact
dwarfs, there is a systematic offset between the \lz relation derived
from the $T_{\text{e}}$ method and that from the $R_{23}$ method,
i.e., the method that we have used as a reference calibration by
scaling the ionization parameter in the N2-based calibration (see
\S\ref{section:metallicity}).  The $R_{23}$-based \lz relation has a
systematically higher zeropoint than the $T_{\rm e}$-based one, thus
the $R_{23}$ method may overestimate the true oxygen abundance.
Therefore, using the same calibrator is crucial to compare various
observations as shown below.

\begin{figure}
  \begin{center}
    \plotone{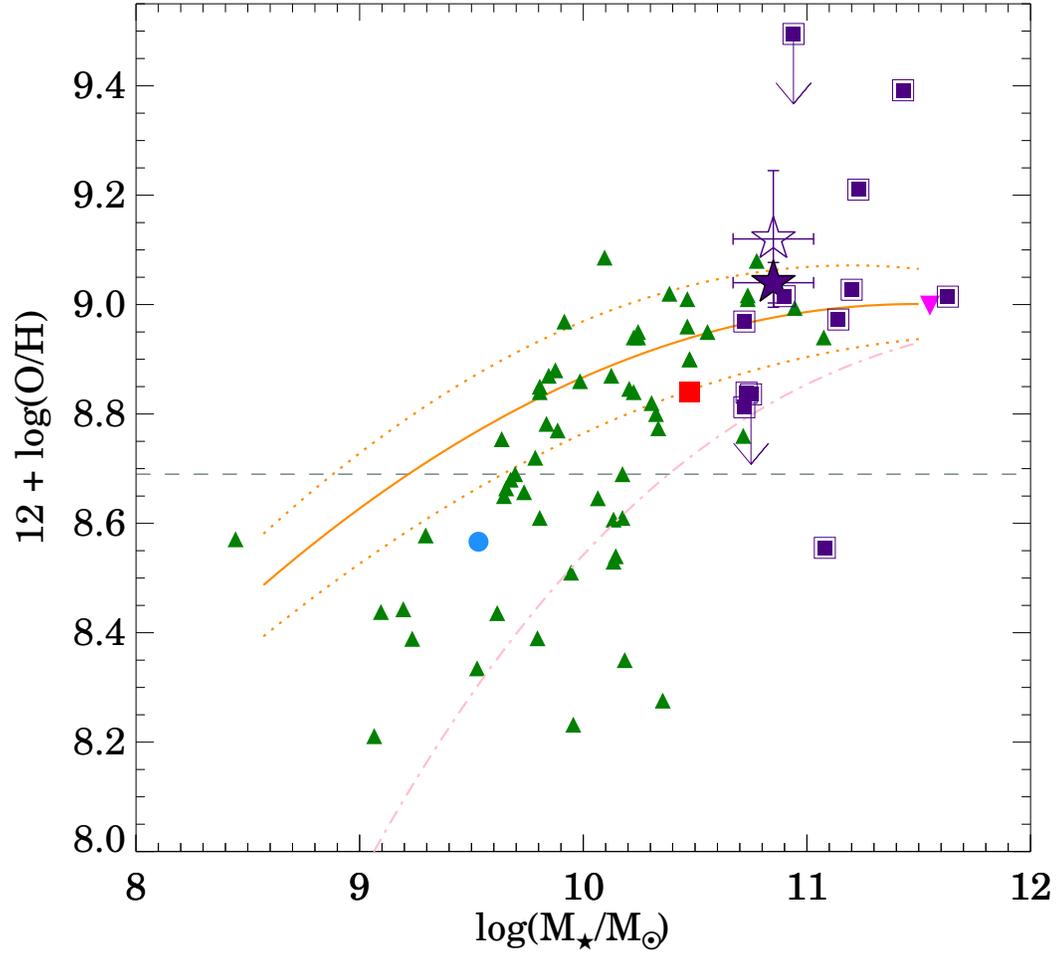}
    \figcaption{
    The  \mz relations for galaxies at $z=0.1$, $\simeq0.7$, and $\simeq2$ from various samples and sources.
      Filled indigo squares with border show the non-AGN \sbzks in our sample
      observed by OHS/CISCO and SINFONI, and filled star and the open stars refer to
      the stacked spectrum, as in Figure \ref{fig:mzrelbzk}.  
      Solid and dotted orange lines show the local \mz relation derived from SDSS data
      (see text), along with its $\pm1\sigma$ range.  Green filled
      triangles represent $z\simeq0.7$ galaxies in GDDS/CFRS sample
      \citep{savaglio:2005}.  A DRG \citep{vandokkum:2004} is plotted
      as a magenta upside-down triangle. The mean \mz relation for BX/BM galaxies is shown
      by the pink dot-dashed line.  Cyan circle and red square represent
      MS 1512-cB58 \citep{teplitz:2000,baker:2004} and the median value for the 
      SMGs \citep{smail:2004,swinbank:2004}, respectively.  In this plot error bars
      are omitted and stellar mass and metallicity are converted to the
      Salpeter IMF and to the \citetalias{kobulnicky:2004} metallicity
      calibration, respectively.  The horizontal dashed line indicates the
      solar oxygen abundance.
      \label{fig:mzrelall}
    }
  \end{center}
\end{figure}

\citetalias{tremonti:2004} derived the \mz relation in the local
universe ($z\sim 0.1$) by using 53,000 star-forming galaxies from the SDSS
DR2 release, and their \mz relation can be used as the reference relation for
$z\simeq0$.  However, their metallicity measurements are based on
model fitting to multiple emission lines \citep{charlot:2001}, and
therefore are not directly comparable to our results derived from
Equation (\ref{eq:oh12n2}).  To cope with this problem, we have used
SDSS DR4 data as described in \S\ref{section:metallicity}, and multiplied their
stellar masses by a factor 1.5 (to convert from Kroupa IMF to Salpeter
IMF).  Then the resulting \mz relation for SDSS galaxies with
$R_{23}/O_{32}$ calibrated abundances is fitted with the 2nd order
polynomial,
\begin{equation}
  \begin{split}
  12+\log(\text{O/H}) & = 1.0512 \\
                      & + 1.3836\log(M_\star/M_\odot) \\
                      &- 0.0602\left[\log(M_\star/M_\odot)\right]^2, 
  \end{split}
\end{equation}
which is finally used as the reference relation at $z\sim 0$ in Figure \ref{fig:mzrelall}.

The \mz relation for 57 star-forming galaxies from the GDDS/CFRS surveys
\citep{savaglio:2005} is then used as a reference at intermediate
redshift, $z\simeq0.7$. Since they used $R_{23}$ as in
\citetalias{kobulnicky:2004} to calibrate abundances and the
\citet{baldry:2003} IMF, we convert only their stellar masses to those
corresponding to the Salpeter IMF, by multiplying them by a
factor 1.8.

In addition to the \sbzks in this study, we also consider masses and
abundances of the following $z\sim 2$ objects: 6 distant red galaxies
(DRGs) at $2.4<z<3.2$ \citep{vandokkum:2004} where only one out of 6
presented in \citet{vandokkum:2004} is used here because 4 DRGs show
obvious AGN features (broad-line and high [\ion{N}{2}]/\hasp ratio),
and for another object neither N2 nor $R_{23}$ indicators could be
used for a metallicity measurement; 87 UV-selected (BX/BM)
star-forming galaxies at $\langle z\rangle =2.26$ \citep{erb:2006mz}
which are binned into 6 mass bins containing $\sim 15$ galaxies per
bin; the Lyman-break galaxy MS1512-cB58 at $z=2.73$
\citep{teplitz:2000,baker:2004}; and finally the average of 7 SMGs
with [\ion{N}{2}]/\hasp $<0.7$ at $\langle z \rangle =2.4$
\citep{smail:2004,swinbank:2004}.  The stellar masses are all
corrected to the Salpeter IMF, taking gravitational lensing into
account in the case of cB58.  All gas-phase oxygen abundances are
derived with the \citetalias{kobulnicky:2004} calibration by using the
[\ion{N}{2}]/\hasp emission line ratios as listed in the quoted
references so to ensure full homogeneity with the \sbzk abundances derived in this paper. 

Figure \ref{fig:mzrelall} compares masses and abundances of \sbzks
at $z\simeq2$ to those of the objects mentioned  above.  Most \sbzks
at $z\simeq2$ are already enriched to  abundances comparable to
those of SDSS and GDDS/CFRS galaxies in the same stellar mass range, i.e.,
$\log M_\star\gtrsim10.5$. The ISM abundances  of  \sbzks appears to be
consistent with those of $z\simeq2$ galaxies of comparable mass, although 
selected according to different  criteria (i.e., DRGs and
SMGs), while the \mz relation of UV-selected galaxies from \citet{erb:2006mz}
tend to show slightly lower ($\simeq 0.15$--$0.2$ dex) abundances at a given stellar mass.  
This offset from the \mz relation of UV-selected galaxies which is generated by 
the spectral stacking can be seen also when compared with the metallicities 
of the composite spectra of \sbzks at similar stellar mass. 
Since the metallicity uncertainties are rather 
large, this offset may not be very significant. Also
\citet{hayashi:2009} have reported a $\simeq0.2$ dex higher average metallicities 
of their \sbzks compared to UV-selected galaxies. Our metallicities 
are closer to those of \citet{hayashi:2009} than to those of \citet{erb:2006mz}, though uncertainties are large. 
\citet{hayashi:2009} also mentioned a possible bias towards higher 
metallicity objects, which would enable us to detect [\ion{N}{2}] emissions, 
hence to measure the metallicity. Indeed, their stacked spectrum of [\ion{N}{2}] 
undetected objects yields roughly the same 
metallicity as that of UV-selected galaxies. 

\citet{savaglio:2005} noted the differential redshift evolution of the
\mz relation from $z\sim 0.7$ to $z=0.1$,  with the most massive galaxies at $z\sim 0.7$  ($\log
M_\star \gtrsim 10.3$) being already
enriched to the ISM abundances of the local galaxies, while less
massive galaxies start leaving the local \mz relation already at redshift as low as
$z\sim 0.1$.  Our result extends this trend to higher redshifts,
finding that at the high mass end the chemical enrichment was
virtually complete by $z\sim 2$.  This indicates that massive galaxies
evolve faster than less massive galaxies, where star formation and
ensuing chemical enrichment are slow and last longer, i.e., yet another manifestation of 
{\it downsizing}.

\subsection{Relation between Stellar Masses and Specific Star-formation Rates}
The correlation between SFR and stellar mass has been extensively
investigated from the nearby universe to high redshift
\citep[e.g.,][]{brinchmann:2004,daddi:2007sfr,elbaz:2007,noeske:2007,dunne:2009,pannella:2009,santini:2009}.
In particular, the specific star-formation rate (SSFR), defined as the
SFR per unit stellar mass, is widely used to quantify the
contribution of current star-formation activity to the growth of the total
stellar mass, i.e., of how efficiently stars are formed.

\begin{figure}
  \begin{center}
    \plotone{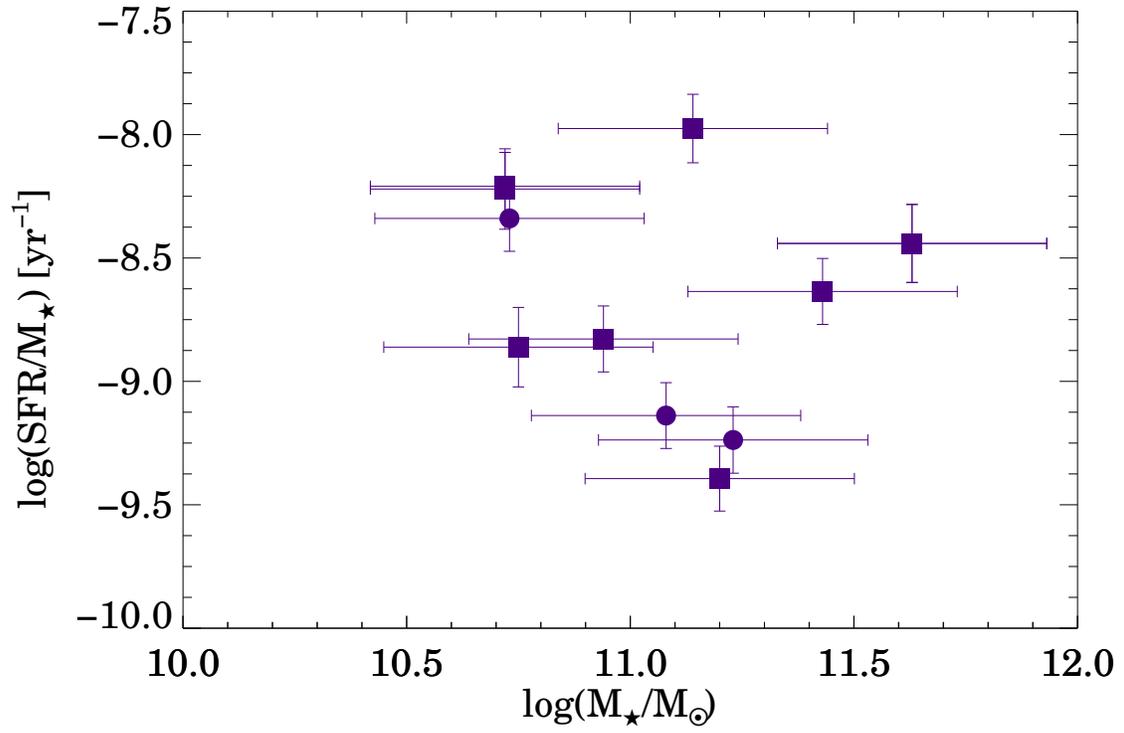}
    \figcaption{
      The SSFRs of \sbzks in this study are plotted as a function of 
      galaxy stellar mass.  Symbols are the same as in Figure {\ref{fig:mzrelbzk}}.
      A typical mass error of a factor of 2 (0.3 dex) is adopted
      for all galaxies. The objects possibly dominated by an AGN
      are not plotted in this figure.
      \label{fig:ssfr_bzk}
    }
  \end{center}
\end{figure}

\begin{figure}
  \begin{center}
    \plotone{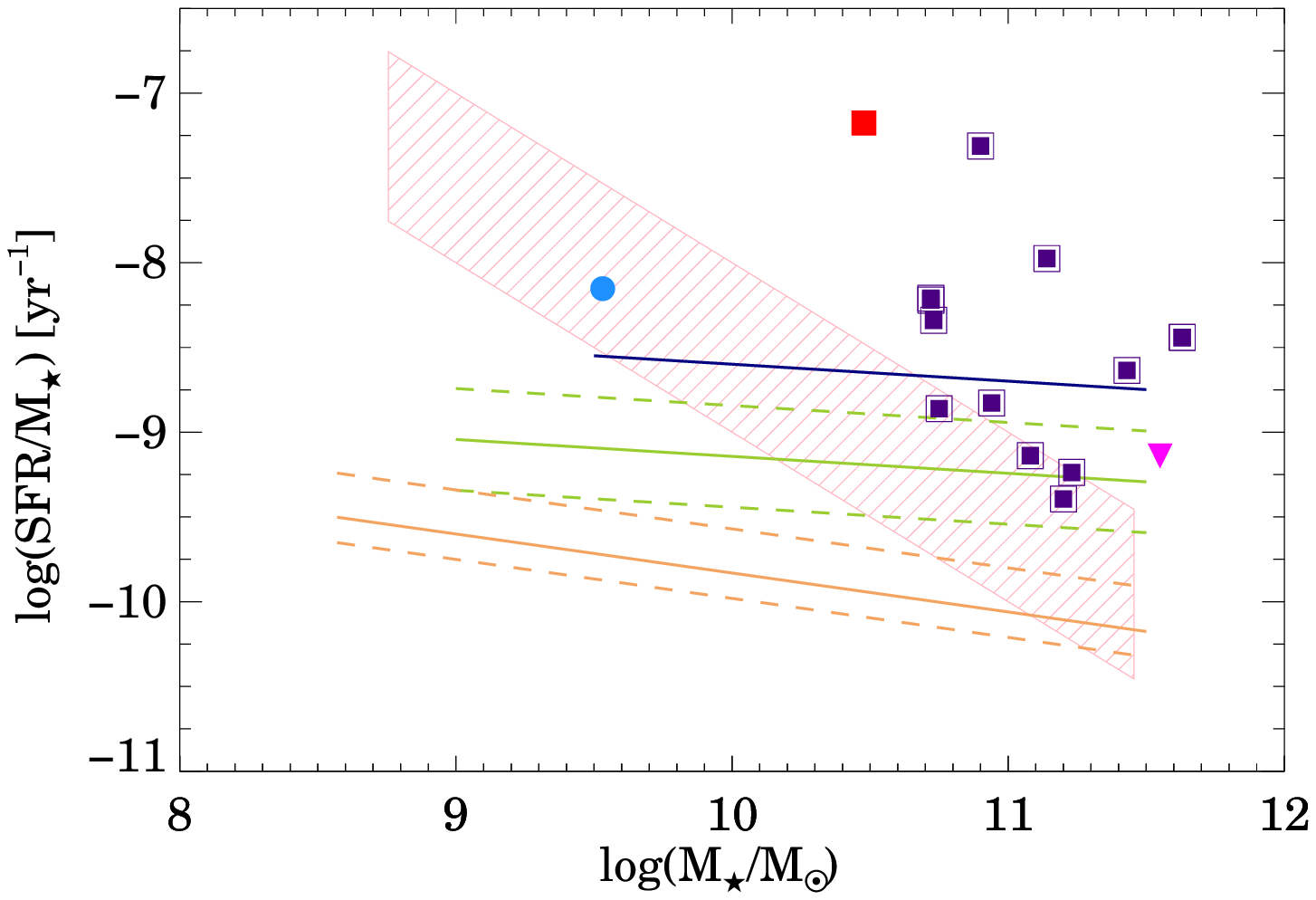}
    \figcaption{
      The SSFRs for different galaxy samples are plotted as a function of galaxy
      stellar mass.  
      Filled indigo squares with border show non-AGN 
      \sbzks in our sample. 
      Magenta square, red upside-down triangle, and cyan circle refer to a
      DRG \citep{vandokkum:2004}, to the median of SMGs
      \citep{smail:2004,swinbank:2004}, and to MS 1512-cB58
      \citep{teplitz:2000,baker:2004}, respectively.  The pink hatched
      band shows the region occupied by UV-selected BX/BM galaxies at $z\simeq2$
      from \citet{erb:2006sfr}.  
      Solid lines represent respectively the
      fits for the \mssfr  relations at $z\simeq2$ \citep[indigo;][]{daddi:2007sfr}, 
      and those at $z\simeq1$ and $z\simeq0$ \citep[green and orange, respectively;][]{elbaz:2007}. 
      Dotted and dashed lines associated with solid lines
      represent the $1\sigma$ scatter of the distributions.
      \label{fig:ssfr_all}
    }
  \end{center}
\end{figure}

In Figure \ref{fig:ssfr_bzk}, we plot the SSFR as a function of
stellar mass for \sbzk galaxies with \hasp detection. Though the
scatter and uncertainty in stellar mass are large, the average SSFR (in yr$^{-1}$) 
is around $\log \text{SSFR} \simeq -9$, and the SSFR$-M_\star$ does not show
a  detectable slope. 
A comparison of \mssfr relations from different
sources and at different redshifts is also shown in Figure
\ref{fig:ssfr_all}.  
Data from literature are taken from \citet{elbaz:2007} for 
star-forming galaxies at $z\simeq0$ in SDSS and $z\simeq1$ in GOODS, 
as well as $z\simeq2$ objects including \sbzks
\citep{daddi:2007sfr,pannella:2009}, UV-selected star-forming galaxies
\citep{erb:2006sfr}, DRGs \citep{vandokkum:2004}, and average of SMGs
\citep{smail:2004,swinbank:2004}.  
\citet{elbaz:2007} and
\citet{daddi:2007sfr} have estimated SFRs by adding extinction
uncorrected rest-frame UV and IR 
luminosities to trace both absorbed and unabsorbed star formation. 
SFRs for \sbzks in the study of
\citet{pannella:2009} have been derived from 1.4 GHz radio data (hence
independent of extinction) finding excellent agreement with
\citet{daddi:2007sfr}.  SFRs of DRGs and UV-selected
star-forming galaxies at $z\simeq2$ are derived from \hasp
luminosities with extinction correction described in
\citet{vandokkum:2004} and \citet{erb:2006sfr}, respectively. The SFR
of SMGs is derived from sub-mm emission.  Since these SFRs are taken
from the original papers, which besides different selection criteria have also used a variety of different
observables  (such as UV-luminosity, \hasp luminosity, FIR luminosity
and radio luminosity),  it is not a surprise to find
even large systematic differences.  

From low to high redshift, SSFRs at a given stellar mass
increase systematically as shown by e.g., \citet{elbaz:2007},
\citet{daddi:2007sfr} and \citet{pannella:2009}. The \sbzks presented
in this study distribute around the average SSFR-$M_\star$ relation for
\sbzks in these two latter studies.  
One \sbzk galaxy shows a SSFR as high as that typical of SMGs,
and can be regarded as a real outlier, 
though it is also possible that its \hasp SFR  may have been 
overestimated since the object shows an \hasp excess but appears normal 
in the SFR(UV)-SFR(IR) comparison (Figure \ref{fig:sfrcomp}).
Note that the \mssfr relations
for SDSS star-forming galaxies at $z\sim 0$, $z\simeq1$  and for \sbzks at $z\simeq2$ have almost flat slopes, ranging from
$\sim 0$ to $-0.23$, while the distribution of UV-selected galaxies in
\citet{erb:2006mz} shows a much steeper slope.  
A steeper slope, but systematically higher SSFRs for \sbzks in the \mssfr relation 
has been also reported by \citet{hayashi:2009}, having  used \hasp SFRs. 
However, their \hasp SFR are systematically higher than  SFRs from the UV, while no comparison with the total 
(e.g., UV+IR) SFR is available (see \S\ref{section:hasfr} and \S\ref{section:compsfr}). 
On average, \sbzks in
the present study, as well as in those of  \citet{daddi:2007sfr}, 
\citet{pannella:2009} and \citet{hayashi:2009},  have higher SSFRs at the massive end
compared to those for UV-selected galaxies. 

An almost flat \mssfr relation for $z\sim 2$ $BzK$-selected galaxies
which is almost indistinguishable from that of \citet{daddi:2007sfr} 
shown in Figure \ref{fig:ssfr_all} is indeed found by \citet{dunne:2009} and
\citet{pannella:2009}. The discrepancy with respect to other, steep
\mssfr relations is ascribed to selection effects
\citep{dunne:2009}, e.g., the UV selection being biased against
massive, highly obscured and intensively star-forming galaxies, and in
addition to a systematic underestimate of the dust extinction with
increasing mass \citep{pannella:2009}.  
Regarding the discrepancy of slopes between \hasp spectroscopic surveys, an
uncertain extinction correction for \hasp from SED fitting might 
be one of the major causes.  
Larger spectroscopic samples, exploring a wider range of stellar masses and SFRs, 
and detecting of H$\beta$ in addition to the \hasp (so
to estimate accurate dust extinctions from the Balmer decrement) would be 
crucial for an independent evaluation of the slope in \mssfr relation at $z\simeq2$.
Our results, though relative
to a very small sample, are in line with those of \citet{dunne:2009}
and \citet{pannella:2009}.

\subsection{An Interpretation of the Mass-metallicity and Mass-SSFR Relations with 
  Simple Evolutionary Population Synthesis Models}

In this section we try to interpret the \mz and \mssfr
relations with evolutionary population synthesis models. We use the
\pegase models that are also used for SED fitting to derive stellar
masses and dust extinction parameters.  These are one-zone models
which neglect interactions and merging of galaxies, that are naturally
expected in hierarchical structure formation scenario.  They also
neglect galactic winds driven by supernovae and/or AGN feedback, that
are known to exist and are responsible for the metal enrichment of the
intergalactic and intracluster media.  Although a galaxy rarely
evolves as is modeled by \pegase, its simple description may give us a
starting point towards understanding the \mz and \mssfr relations, and
indicate in which direction we should try to find more satisfactory
solutions. A detailed comparison with the full numerical simulation is
not the focus of this paper.

Two scenarios, the simple closed-box model and the infall model for chemical enrichment, are
considered here with a Salpeter IMF from $0.1M_\odot$ to $120
M_\odot$ and the B-series of chemical yields from
\citet{woosley:1995}. These models assume that a galaxy is initially a
purely gas cloud with zero metallicity, $Z=0$, and the SFR is assumed to
be proportional to the gas mass (i.e., a Schmidt law with $n=1$):
\begin{equation}
  \text{SFR}(t)=\frac{1}{\tausf} M_{\text{gas}}(t),
  \label{eq:pegasesfr}
\end{equation}
where $\tausf$ is the star-formation timescale.  Closed-box models assume no
gas infall and all gas resides in the system at the beginning.  This
is the classical {\it monolithic collapse} scenario. In the infall
models, initially all gas is assumed to be in an outer reservoir
and falls into the inner star-forming region at a rate of
\begin{equation}
  \xi_{\text{infall}}(t) = 
  \frac{1}{\tau_{\text{infall}}} \exp\left(-\frac{t}{\tau_{\text{infall}}}\right
), 
  \label{eq:pegaseinfall}
\end{equation}
where $\tau_{\text{infall}}$ is the infall time scale.  We use models
with $\tausf=0.1$ and $5$ Gyr and we assume
$\tausf=\tau_{\text{infall}}$ for infall models.  Short time scale
models with $\tausf=0.1$~Gyr can successfully reproduce the
color-magnitude (C-M) relations of elliptical galaxies in nearby
clusters of galaxies \citep{kodama:1997}, and long time scale models
with $\tausf=5$~Gyr can explain the photometric properties of nearby
late type, Sb--Sc, galaxies \citep{arimoto:1992,fioc:1997}.  
Following \citetalias{kobulnicky:2004},
the ISM oxygen abundance is derived from the gas-phase metal mass fraction 
using the relation:
\begin{equation}
  \oh=12+\log\left(\frac{Z}{29}\right),
\end{equation}
assuming the solar abundance pattern reported by \citet{anders:1989} and 
the solar oxygen abundance from \citet{allendeprieto:2001}. 

\begin{figure}
  \begin{center}
    \includegraphics[width=0.85\linewidth]{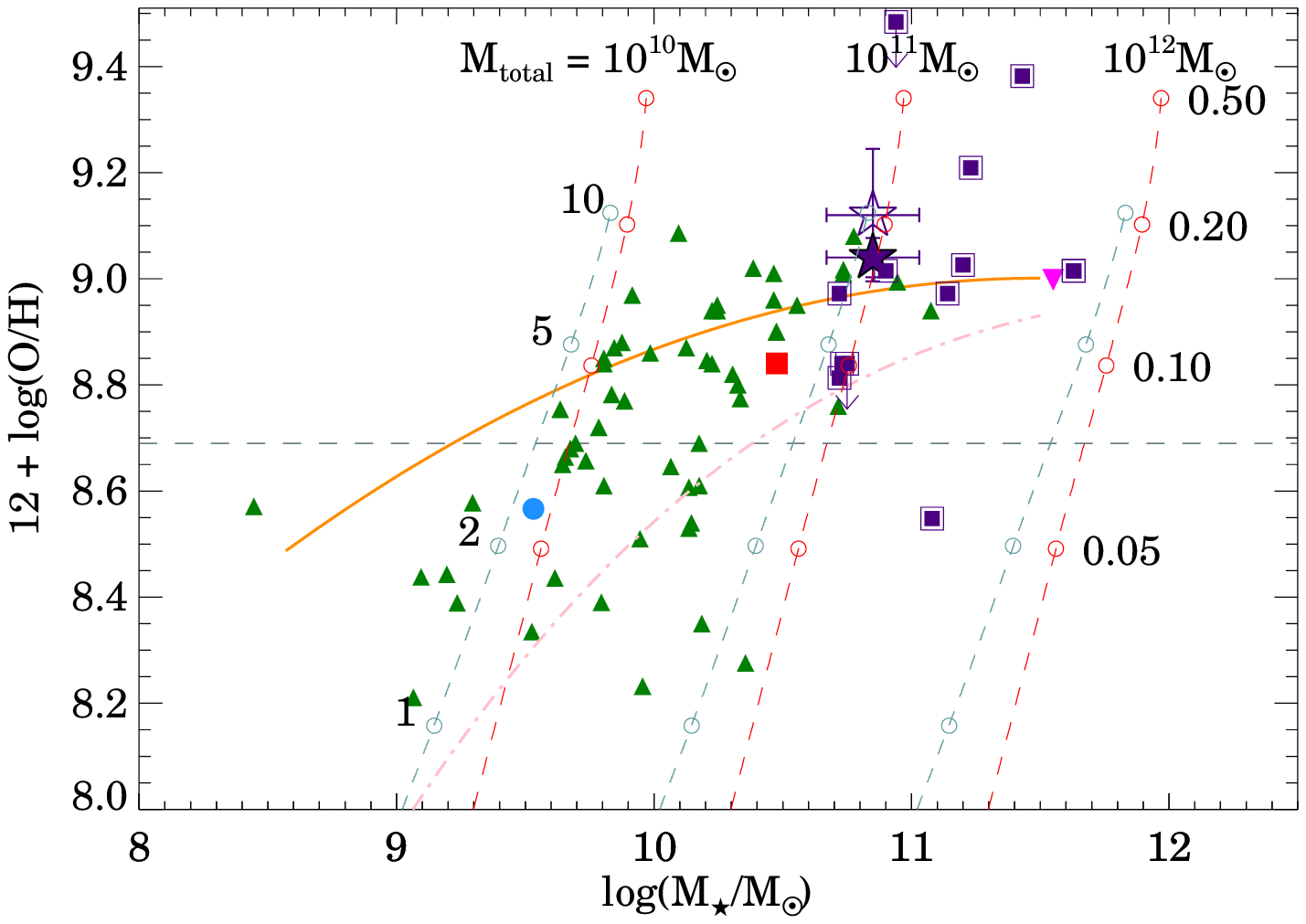}
    \includegraphics[width=0.85\linewidth]{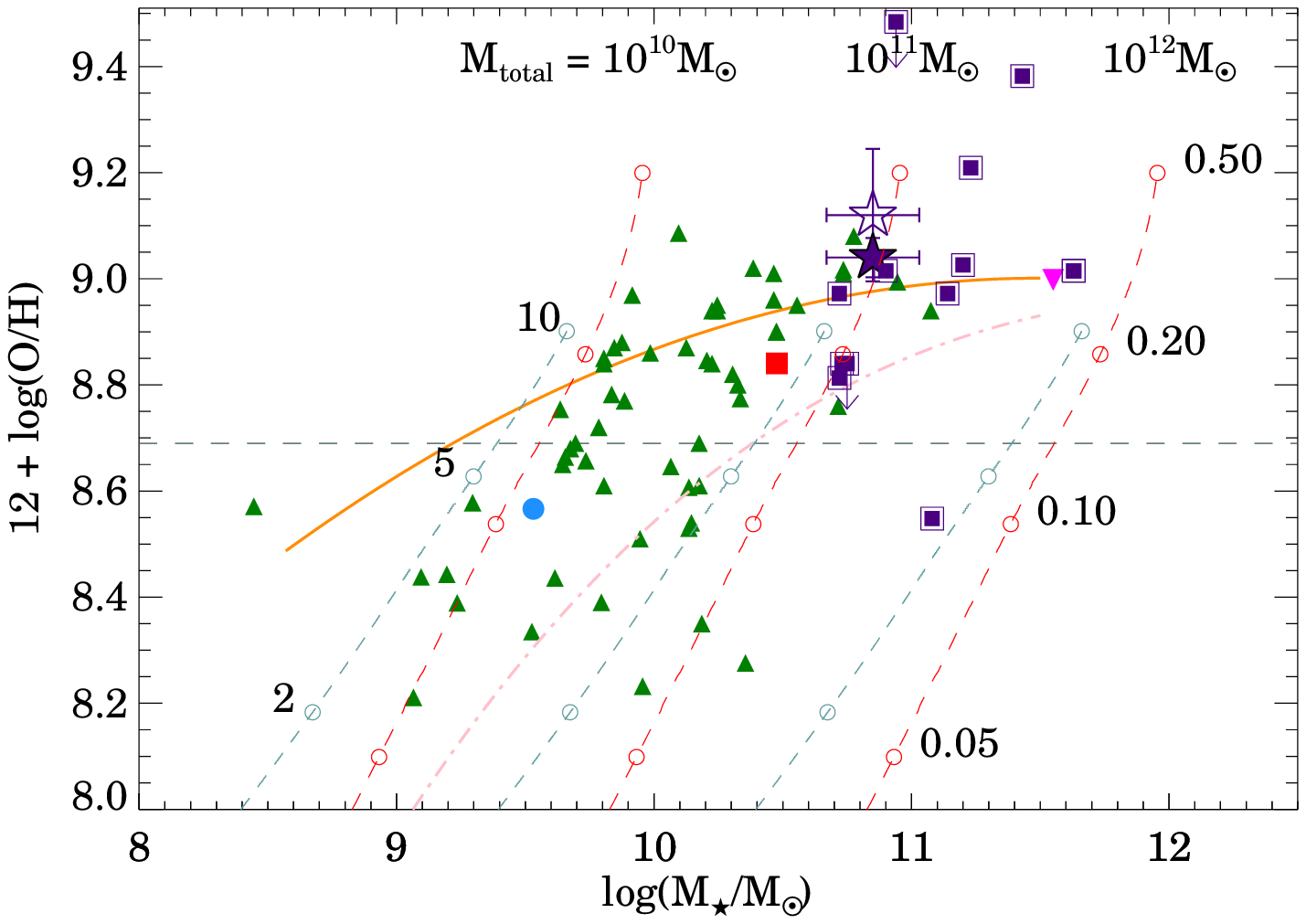}
    \figcaption{
      Model tracks from \pegase overplotted on the \mz
      relations presented in Figure {\ref{fig:mzrelall}}.  Top and
      bottom panels refer to closed-box and infall models,
      respectively.  Red and light-blue dashed lines correspond to
      $\tausf=0.1$ and 5 Gyr models, respectively.  Three parallel
      lines for each color represent the initial total gas mass as 
      indicated at the top of each panel.  The numbers near the small open circles 
     along the tracks indicate galaxy ages in Gyr.  The corresponding
      age for each open circle is the same for the lines with the
      same color.
      \label{fig:mzrelevo}
    }
  \end{center}
\end{figure}

\begin{figure}
  \begin{center}
    \includegraphics[width=0.85\linewidth]{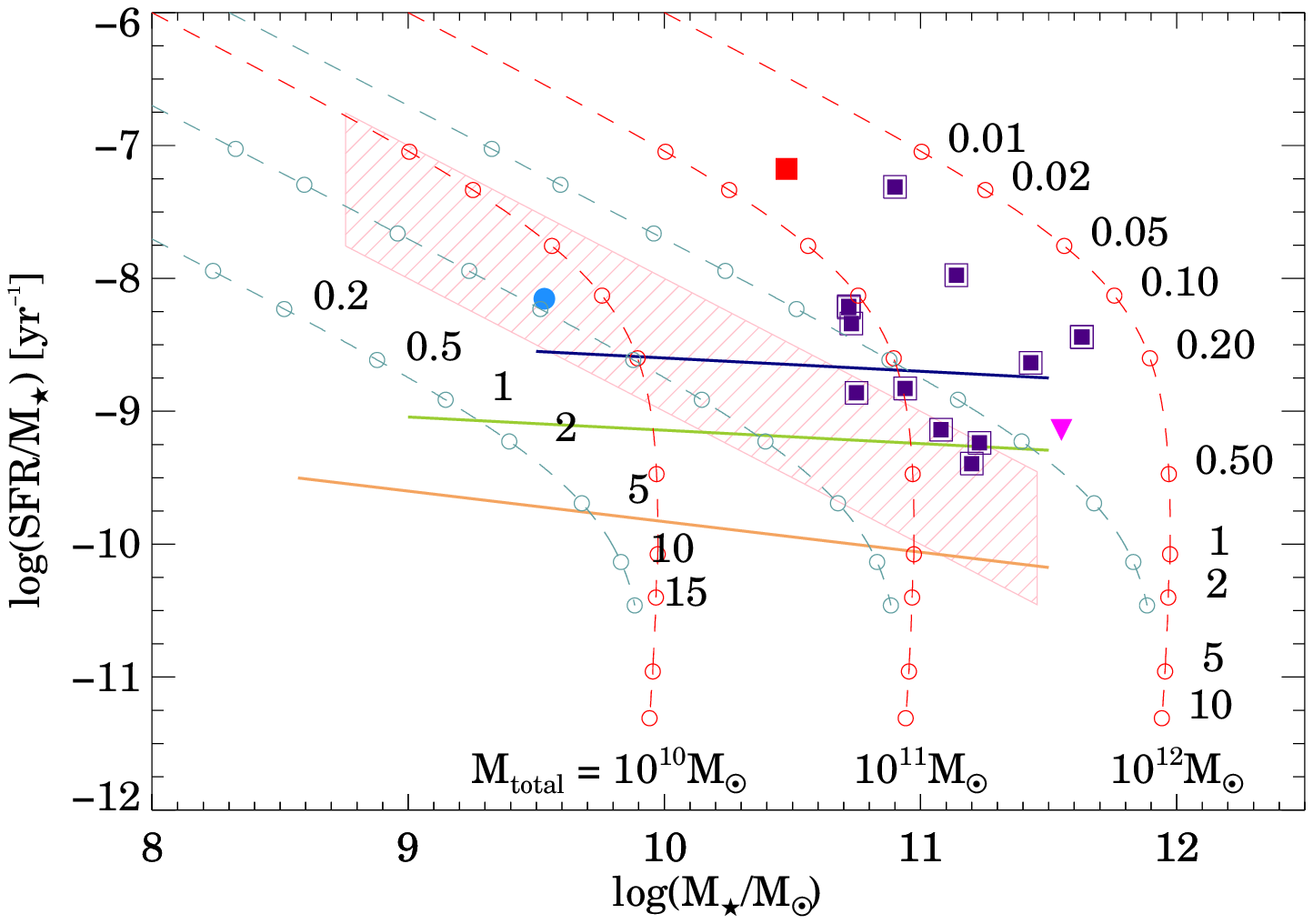}
    \includegraphics[width=0.85\linewidth]{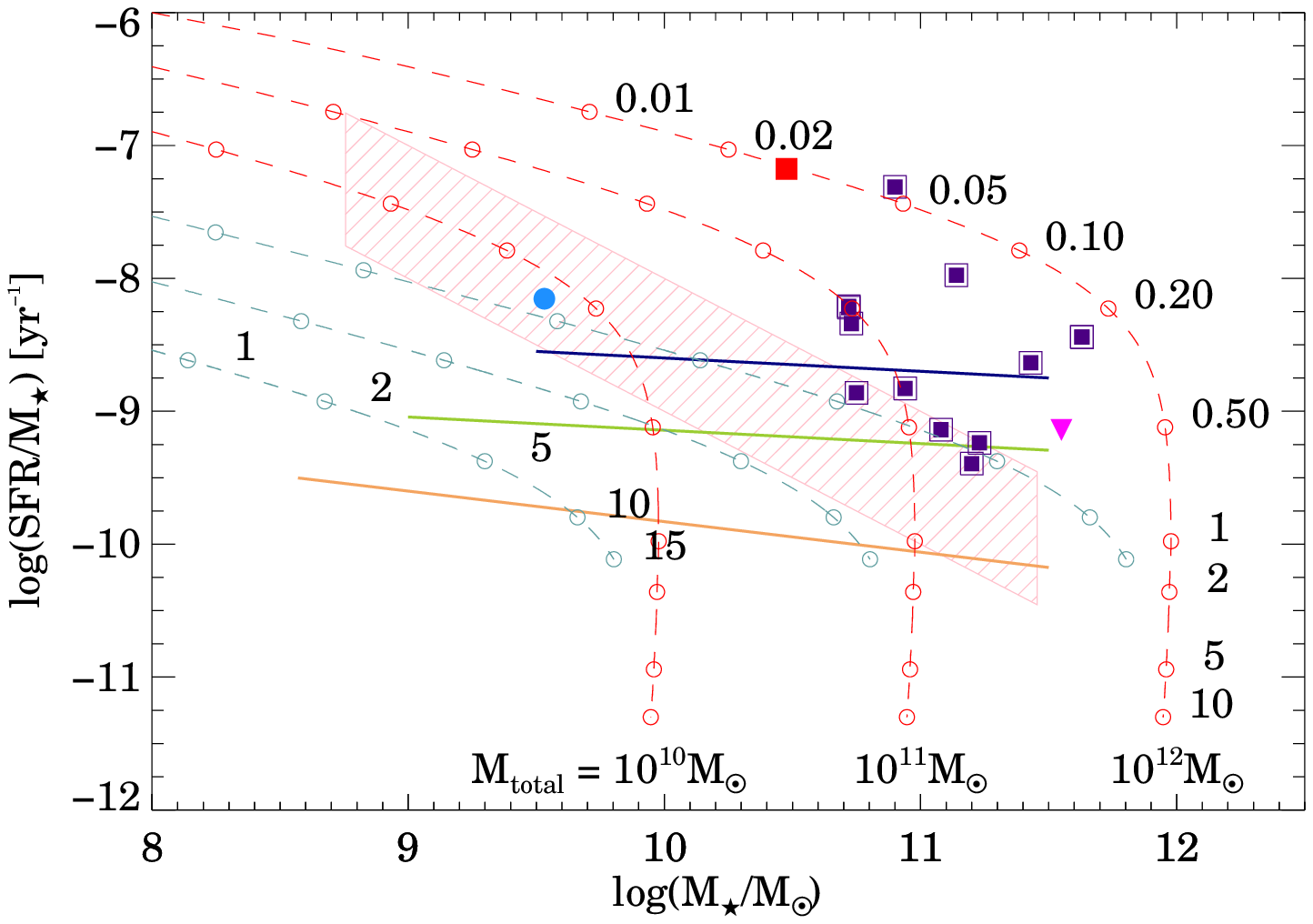}
    \figcaption{
      Model tracks from \pegase overplotted on Figure
      {\ref{fig:ssfr_all}}.  Top and bottom panels show tracks from
      closed-box and infall models, respectively.  Red and light-blue
      dashed lines correspond to $\tausf=0.1$ and 5 Gyr, respectively.
      Three parallel line sets of each color represent the initial total
      gas mass of as indicated.  Small open circles and texts
      along the tracks indicate galaxy ages in Gyr.  
      Note that the metallicity with given $\tausf$ and age is independent of $M_{\text{total}}$.
      \label{fig:ssfrevo}
    }
  \end{center}
\end{figure}

Figures {\ref{fig:mzrelevo}} and {\ref{fig:ssfrevo}} show that both
the closed-box and the infall models with $\tausf=0.1$ Gyr can roughly
reproduce the locations of \sbzks and other populations of $z\simeq2$
galaxies at the massive end of the \mz relation and of the
\mssfr relation, with $100\lesssim t \lesssim 500$ Myr ages
and $10^{11}\lesssim M_{\text{total}}/M_\odot \lesssim 10^{12}$ of gas
mass.  This range of ages is also roughly consistent with the onset
epoch of galactic winds needed by the models of \citet{kodama:1997} to
reproduce the C-M relations of elliptical galaxies in the Virgo and
Coma clusters.  On the other hand, models with longer time scale,
e.g., $\tausf=5$ Gyr, cannot fit both relations simultaneously. For
example, $\tausf=5$ Gyr infall models take $\gtrsim 5$ Gyr to reach
the observed metallicities of $z\simeq2$ galaxies, while the age of
the universe at $z\simeq2$ is only $\sim3.4$ Gyr for the adopted
cosmology. Only simple models can
reproduce the distribution of \sbzks and $z\simeq2$ galaxies in the
\mz relation with such long $\tausf$.  However, in the \mssfr 
relation, it turns out that simple models
with longer time scale require huge amount of initial gas,
$M_{\text{total}} > 10^{12}\,M_\odot$, which is larger than typical
stellar mass of the most massive early-type galaxies found in the local universe.
Moreover, the ages and stellar masses mentioned above imply an average 
$\text{SFR} > 10^3M_\odot {\rm yr}^{-1}$, much in excess of the observed 
SFRs of galaxies in such mass range.

On the other hand, Equations (\ref{eq:pegasesfr}) and (\ref{eq:pegaseinfall}) 
imply a secularly declining SFR,
such that all galaxies are assumed to have started with their maximum SFR
and to be caught at their
minimum SFR. This assumption is especially doubtful at $z\sim 2$, where
the SFR of many star-forming galaxies may actually increase with time,
rather than decrease \citep{renzini:2009a,maraston:2010}.

The high SFRs in the \bzk galaxies presented here require galaxies to host
a large amount of molecular gas to be used for star formation.
Recent millimeter and radio observations have been revealing 
that star-forming $BzK$ galaxies indeed harbor large amount of CO molecules,
hence also molecular hydrogen \citep{daddi:2008,daddi:2010,dannerbauer:2009}. 
These CO observations as well as multi-wavelength SED analysis 
\citep{daddi:2007sfr,daddi:2007mirx} suggest that the gas consumption 
or star-formation timescale could be up to 1~Gyr rather than 
$100$~Myr inferred in this study. In the framework of \pegase used here, 
$\tau\lesssim1$~Gyr is in fact an upper limit in order to reproduce 
the observed \mz and \mssfr relations at the same time, 
still being consistent with \citet{daddi:2007sfr,daddi:2008}. 

Again, note that closed-box or infall models are certainly
oversimplifications over the  real star-formation histories of galaxies.
Recent numerical simulations favor continuous (albeit fluctuating)
cold-stream accretion for the main driver of the galaxy growth, rather
than short-lived starbursts \citep[e.g.,][]{dekel:2009}. Thus, infall models
may capture this aspect, but the results must critically depend on the assumed
evolution with time of the infall rate.

SMGs, which are outlier in the \mssfr relation, require
extremely young ages, a few 10 Myr, while in the \mz relation they
occupy nearly the same position as other $z\simeq2$ star-forming
galaxies. This is possibly due to vigorous, merger-driven starbursts
rather than due to internal star-formation mode assumed
in the one-zone \pegase models. These violent star-formation processes could
push a galaxy almost vertically \citep{feulner:2005} in the
\mssfr diagram, with little change of its position in the \mz
diagram.

\section{Summary and Conclusions}
We have conducted NIR spectroscopic observations of 28 \sbzk galaxies 
at $\simeq2$ and detected \hasp emissions from 14 of them. By using the 
\hasp and [\ion{N}{2}] lines, we have derived \hasp SFRs and gas phase 
oxygen abundances.  Stellar masses and reddening parameters have been 
also derived by SED fitting to the multi-wavelength photometric data. 
Our results are summarized as follows:
\begin{itemize}
\item Stellar masses and reddening parameters derived from SED fitting 
agree well with those derived by $BzK$-based recipe introduced by \citet{daddi:2004bzk}.
\item A comparison of SFRs from different indicators (\hasp, extinction corrected UV, and 
UV+MIR) indicates that additional
extinction towards \ion{H}{2} regions over that derived from the SED
fitting of the stellar continuum is required for the \hasp SFR to recover the total SFR
inferred from UV+MIR. The required additional extinction is in
agreement with the recipe proposed by \citet{cidfernandes:2005}.
\item One object, D3a-11391, shows MIR-excess and a broad \hasp
component with $\text{FWHM}\simeq2500$\kms, as well as a narrow line
component, suggesting that it may host an AGN, which would be highly
obscured at UV and X-ray wavelengths. Although with large uncertainty,
we have estimated the mass of the SMBH and its intrinsic X-ray
luminosity following \citet{alexander:2008}, obtaining
$M_{\text{BH}}\simeq (3\text{--}9)\times10^7\,M_{\odot}$, depending on
the assumed geometry of the accretion disk, and
$L_{2-10\text{keV}}\simeq10^{44}$~ergs~s$^{-1}$.  From these quantities,
the mass accretion rate onto the central SMBH and Eddington factor are
calculated as $\dot{M}_{\rm BH}\simeq0.5\,M_{\odot}\,\text{yr}^{-1}$ and
$\eta\simeq0.3$--$0.9$, again depending on the geometry.  These properties are
in between those of normal star-forming galaxies and those of
broad-line SMGs, indicating a possible evolutionary connection between
normal UV/optical-selected galaxies, broad-line/MIR-excess \sbzks, and
star-forming broad-line SMGs that may result from an accelerated
growth of SMBH.
\item
Most of \sbzks presented here have already reached a metallicity similar
to that of local star-forming galaxies of the same mass. They tend to
have higher metallicity compared to UV-selected $z\sim 2$ galaxies,
even at the top mass end. This indicates that \sbzks are on average a
more evolved population compared to that of UV-selected galaxies at
the same redshift.
\item Specific SFRs of most \sbzks are consistent with a tight \mssfr
correlation \citep{daddi:2007sfr, pannella:2009}, with a couple of
outliers reaching very high SSFRs, similar to those of  SMGs
\item Within the framework of \pegase closed-box or infall models, the
\mz and \mssfr relations of $z\simeq2$ \sbzks can be reproduced
simultaneously assuming a very short star-formation or infall
timescale, $\tau\simeq100$~Myr, and large gas mass at the
beginning. However, the implied SFRs averaged over the life of the
galaxies appears to be excessively large, suggesting that secularly
declining SFRs may not be appropriate for star-forming galaxies at
$z\sim 2$.
\end{itemize}

\acknowledgments 
We thank the anonymous referee for providing the useful and constructive report 
Part of the materials presented in this paper is from the Ph.D.\ thesis of 
MO at the Astronomy Department of the University of Tokyo.  
We thank Kentaro Motohara for providing the reduction
pipeline used for the OHS/CISCO spectra and for giving advice about
data reductions.  Marcella Brusa is thanked for providing us upper
limits for the XMM-Newton/Chandra fluxes and Ranga-Ram Chary is
thanked for help with the MIPS $24\,\mu\text{m}$ data.  We thank the
staffs of the Subaru telescope and of the VLT for supporting the
observations.  This work was supported in part by a Grant-in-Aid for
Science Research (No. 19540245) by the Japanese Ministry of Education,
Culture, Sports, Science and Technology.  DMA thanks the Royal Society
and Leverhulme Trust for financial support.  AR express gratitude to
the INAF - Osservatorio Astronomico di Bologna for its hospitality and
support.


\clearpage

\clearpage                      

\begin{landscape} 
\begin{deluxetable}{lrrrrrrrrrrrrr}
  \tabletypesize{\scriptsize }
  \tablewidth{0pt}
  \tablecaption{Coordinates and Photometric Properties\label{table:objsample}}
  \tablehead{
    \colhead{ID} &
    \colhead{RA} &
    \colhead{Dec} &
    \colhead{$z_{\text{UV}}$} & 
    \colhead{$B$} & 
    \colhead{$R$} & 
    \colhead{$I$} & 
    \colhead{$z'$} &
    \colhead{$J$} & 
    \colhead{$Ks$} &
    \colhead{$B-z'$} &
    \colhead{$z'-Ks$} &
    \colhead{$\log M_{\star}$\tablenotemark{a}} &
    \colhead{$E(B-V)$\tablenotemark{a}}
    \\
    \colhead{} &
    \colhead{(J2000)} &
    \colhead{(J2000)} &
    \colhead{} &
    \colhead{(mag)} &
    \colhead{(mag)} &
    \colhead{(mag)} &
    \colhead{(mag)} &
    \colhead{(mag)} &
    \colhead{(mag)} &
    \colhead{(mag)} &
    \colhead{(mag)} &
    \colhead{($10^{11}M_\odot$)} &
    \colhead{(mag)}
  }
  \startdata
  \multicolumn{14}{c}{\textit{H$\alpha$ detection}} \\
  \tableline
  D3a 3287  & 11\ 24\ 37.7 & $-$21\ 49\ 37.0 & 2.20    & 24.35 & 23.89 & 23.84 & 23.74 & 23.03   & 22.15 & 0.61 & 1.59 & 10.54 & 0.18 \\ 
  D3a 4626  & 11\ 24\ 59.1 & $-$21\ 47\ 35.8 & \nodata & 25.70 & 24.61 & 23.99 & 23.75 & 22.28   & 21.30 & 1.95 & 2.45 & 11.07 & 0.51 \\ 
  D3a 4654  & 11\ 24\ 53.3 & $-$21\ 47\ 36.7 & \nodata & 23.58 & 22.85 & 22.39 & 22.07 & 20.98   & 20.14 & 1.51 & 1.93 & 11.42 & 0.40 \\ 
  D3a 4751  & 11\ 24\ 50.0 & $-$21\ 47\ 23.1 & 2.26    & 24.27 & 23.69 & 23.57 & 23.42 & 22.77   & 21.99 & 0.85 & 1.43 & 10.57 & 0.24 \\ 
  D3a 5814  & 11\ 24\ 34.4 & $-$21\ 45\ 49.4 & 2.14    & 25.62 & 24.77 & 24.30 & 23.89 & 22.89   & 21.97 & 1.73 & 1.92 & 10.76 & 0.46 \\ 
  D3a 6397  & 11\ 25\ 10.5 & $-$21\ 45\ 06.1 & 2.00    & 24.14 & 23.32 & 22.78 & 22.43 & 21.27   & 20.48 & 1.71 & 1.95 & 11.29 & 0.45 \\ 
  D3a 7429  & 11\ 25\ 12.9 & $-$21\ 43\ 30.1 & \nodata & 24.74 & 24.14 & 23.52 & 23.39 & 22.27   & 21.46 & 1.35 & 1.93 & 10.89 & 0.36 \\ 
  D3a 8608  & 11\ 25\ 06.3 & $-$21\ 41\ 52.7 & 1.53    & 23.29 & 22.99 & 22.83 & 22.71 & 22.09   & 21.50 & 0.58 & 1.21 & 10.72 & 0.17 \\ 
  D3a 11391 & 11\ 24\ 54.7 & $-$21\ 34\ 13.3 & \nodata & 23.37 & 23.12 & 22.73 & 22.91 & 21.80   & 21.06 & 0.46 & 1.85 & 11.16 & 0.14 \\ 
  D3a 12556 & 11\ 25\ 11.6 & $-$21\ 35\ 48.6 & 1.58    & 23.44 & 22.96 & 22.76 & 22.59 & 21.60   & 21.13 & 0.85 & 1.46 & 10.92 & 0.24 \\ 
  Dad 1901  & 14\ 49\ 41.4 &     8\ 59\ 50.5 & 1.60    & 23.56 & 23.02 & 22.90 & 22.48 & \nodata & 20.57 & 1.08 & 1.91 & 11.24 & 0.30 \\ 
  Dad 2426  & 14\ 49\ 29.0 &     8\ 51\ 52.9 & 2.36    & 25.07 & 23.92 & 23.56 & 23.04 & \nodata & 20.75 & 2.03 & 2.29 & 11.25 & 0.53 \\ 
  Dad 2426-b\tablenotemark{b} &              
                           &               &\nodata&\nodata&\nodata&\nodata&\nodata&\nodata&\nodata&\nodata&\nodata&\nodata&\nodata \\ 
  Dad 3551  & 14\ 48\ 59.6 &     8\ 52\ 06.6 & 1.60    & 23.54 & 22.98 & 22.93 & 22.74 & \nodata & 20.87 & 0.80 & 1.87 & 11.10 & 0.21 \\ 
  \tableline
  \multicolumn{14}{c}{\textit{H$\alpha$ non-detection}} \\
  \tableline
  D3a 6004  & 11\ 25\ 03.8 & $-$21\ 45\ 32.7 & 2.39    & 25.88 & 24.69 & 24.31 & 23.93 & 22.79   & 20.96 & 1.95 & 2.97 & 11.32 & 0.51 \\ 
  D3a 6048  & 11\ 25\ 23.1 & $-$21\ 45\ 32.2 & 1.41    & 24.33 & 23.88 & 23.28 & 23.01 & 21.99   & 20.94 & 1.32 & 2.07 & 11.13 & 0.36 \\ 
  D3a 7182  & 11\ 25\ 12.4 & $-$21\ 43\ 48.1 & \nodata& 26.74 &\nodata&\nodata & 24.68 & 23.41   & 21.99 & 2.06 & 2.69 & 10.84 & 0.54 \\ 
  D3a 8249  & 11\ 25\ 15.9 & $-$21\ 42\ 23.3 & 2.06    & 25.25 & 24.38 & 23.47 & 22.98 & 21.54   & 20.45 & 2.27 & 2.53 & 11.42 & 0.59 \\ 
  D3a 12153\tablenotemark{c} 
            & 11\ 24\ 40.7 & $-$21\ 36\ 19.2 & 2.34    & 25.04 & 24.58 & 24.29 & 24.24 & \nodata & 21.82 & 0.80 & 2.42 & 10.85 & 0.23 \\ 
  D3a 13557 & 11\ 25\ 04.9 & $-$21\ 37\ 09.1 & 2.40    & 26.15 & 25.30 & 24.53 & 24.21 & 22.75   & 21.58 & 1.94 & 2.63 & 10.99 & 0.51 \\ 
  D3a 13600 & 11\ 25\ 05.9 & $-$21\ 37\ 13.1 & 2.27    & 24.71 & 24.28 & 24.02 & 23.89 & 22.78   & 22.08 & 0.82 & 1.81 & 10.62 & 0.23 \\ 
  D3a 14009\tablenotemark{c} 
            & 11\ 24\ 34.4 & $-$21\ 38\ 34.1 & 2.43    & 24.59 & 23.65 & 23.56 & 23.37 & \nodata & 21.81 & 1.22 & 1.56 & 10.67 & 0.33 \\ 
  Dad 759   & 14\ 50\ 11.4 &     9\ 07\ 51.3 & \nodata & 26.93 & 25.21 & 25.16 & 24.36 & \nodata & 20.94 & 2.57 & 3.42 & 11.42 & 0.67 \\ 
  Dad 1250  & 14\ 49\ 59.2 &     8\ 56\ 18.0 & \nodata & 25.70 & 24.88 & 24.54 & 24.08 & \nodata & 21.01 & 1.62 & 3.07 & 11.32 & 0.43 \\ 
  Dad 2079  & 14\ 49\ 37.3 &     8\ 58\ 16.6 & 0.73\tablenotemark{d} 
                                                       & 25.07 & 24.12 & 23.45 & 23.19 & \nodata & 20.95 & 1.88 & 2.24 & 11.16 & 0.50 \\ 
  Dad 2742  & 14\ 49\ 20.5 &     8\ 50\ 52.4 & 2.12    & 24.83 & 24.02 & 23.67 & 22.95 & \nodata & 20.84 & 1.88 & 2.11 & 11.18 & 0.50 \\ 
  Dad 3882  & 14\ 48\ 49.0 &     8\ 53\ 15.1 & 2.25    & 25.12 & 24.01 & 23.76 & 23.23 & \nodata & 20.99 & 1.89 & 2.24 & 11.15 & 0.50 \\ 
  Dad 3977  & 14\ 48\ 45.8 &     8\ 57\ 45.8 & \nodata & 25.00 & 23.85 & 23.50 & 22.89 & \nodata & 20.23 & 2.11 & 2.66 & 11.54 & 0.55 \\ 
  Dad 4008  & 14\ 48\ 44.9 &     8\ 53\ 46.9 & 2.22    & 27.31 & 25.97 & 25.84 & 25.03 & \nodata & 21.05 & 2.28 & 3.98 & 11.50 & 0.59  
  \enddata
  \tablenotetext{a}{Derived from the \bzk-calibration of \citet{daddi:2004bzk}. 
    See Equations (\ref{eq:bzkdust}) and (\ref{eq:bzkmass}) in \S\ref{section:massdust}.}
  \tablenotetext{b}{This object is not included in our $K$-selected catalog, 
    but it was serendipitously found in the FoV of SINFONI.}
  \tablenotetext{c}{This object is not included in our current $K$-selected catalog. 
    Quoted values are from an older version of the catalog.}%
  \tablenotetext{d}{This object was considered at $z\simeq2$ based on its photometric redshift, 
    but it was later spectroscopically confirmed to be at $z=0.73$.}
  \tablecomments{All magnitudes are shown in the AB system and are measured within a $2''$ aperture \citepalias{kong:2006}.}
\end{deluxetable}
\clearpage
\end{landscape}

\begin{deluxetable}{llccr}
  \tablewidth{0pt}
  \tabletypesize{\scriptsize}
  \tablecaption{Observational Information\label{table:obslog}}
  \tablehead{
    \colhead{ID} & 
    \colhead{Date} & 
    \colhead{Instrument} & 
    \colhead{Grism} & 
    \colhead{Exposure} 
    \\
    \colhead{} & 
    \colhead{} &
    \colhead{} & 
    \colhead{} & 
    \colhead{(seconds)}
  }
  \startdata
  \multicolumn{5}{c}{\textit{H$\alpha$ detection}} \\
  \tableline
  D3a 3287  & 2005 April 15 & SINFONI                  & $H+K$ &  3600 \\
  D3a 4624  & 2005 April 16 & SINFONI\tablenotemark{a} & $H+K$ &  3600 \\
  D3a 4654  & 2005 April 16 & SINFONI                  & $H+K$ &  3600 \\
  D3a 4751  & 2005 April 16 & SINFONI                  & $H+K$ &  3600 \\
  D3a 5814  & 2005 April 14 & SINFONI                  & $H+K$ &  5400 \\
  D3a 6397  & 2005 April 15 & SINFONI                  & $H+K$ &  3600 \\
  D3a 7429  & 2005 April 15 & SINFONI                  & $H+K$ &  3600 \\
  D3a 8608  & 2005 April 25 & OHS                      & $JH$  &  8400 \\
  D3a 11391 & 2005 April 16 & SINFONI\tablenotemark{a} & $H+K$ &  1800 \\
  D3a 12556 & 2005 April 25 & OHS                      & $JH$  &  3600\tablenotemark{b} \\
            & 2005 April 26 & OHS                      & $JH$  &  3600 \\
  Dad 1901  & 2004 May    1 & OHS                      & $JH$  &  8000 \\
  Dad 2426  & 2004 May    6 & CISCO                    & $wK$  &  6000 \\
            & 2005 April 14 & SINFONI                  & $H+K$ &  7200 \\
  Dad 3551  & 2005 April 24 & OHS                      & $JH$  & 14400 \\
            & 2005 April 26 & OHS                      & $JH$  &  7200\tablenotemark{c} \\
  \tableline
  \multicolumn{5}{c}{\textit{H$\alpha$ non-detection}} \\
  \tableline
  D3a 6004                   & 2005 April 16 & SINFONI & $H+K$                  &  3600 \\
  D3a 6048                   & 2005 April 24 & OHS     & $JH$                   & 10800 \\
  D3a 7182                   & 2005 April 16 & SINFONI & $H+K$\tablenotemark{a} &  3600 \\
  D3a 8249                   & 2005 April 16 & SINFONI & $H+K$                  &  3600 \\
  D3a 12153                  & 2005 April 14 & SINFONI & $H+K$                  &  3600 \\ 
  D3a 13557\tablenotemark{d} & 2004 May    6 & CISCO   & $wK$                   &  7000 \\
                             & 2005 April 30 & CISCO   & $wK$                   &  7000 \\
  D3a 13600\tablenotemark{d} & 2004 May    6 & CISCO   & $wK$                   &  7000 \\
  D3a 14009                  & 2005 April 30 & CISCO   & $wK$                   &  5000 \\ 
  Dad  759                   & 2005 April 16 & SINFONI & $H+K$\tablenotemark{a} &  1800 \\
  Dad 1250                   & 2005 April 16 & SINFONI & $H+K$\tablenotemark{a} &  1800 \\
  Dad 2079                   & 2004 May    6 & CISCO   & $wK$                   &  3000 \\
  Dad 2742                   & 2005 April 15 & SINFONI & $H+K$                  &  7200 \\
  Dad 3882                   & 2005 April 16 & SINFONI & $H+K$                  &  1800 \\
  Dad 3977                   & 2005 April 25 & OHS     & $JH$                   &  2400 \\
  Dad 4008                   & 2005 April 25 & OHS     & $JH$                   &  3600
  \enddata
  \tablenotetext{a}{AO-module was used.}
  \tablenotetext{b}{This exposure was not used 
    because it does not improve S/N ratio due to worse seeing.}
  \tablenotetext{c}{This exposure was not used because it  
    does not improve S/N ratio due to reduced signal by cloud passage.}
  \tablenotetext{d}{D3a 13557 and D3a 13600 were supposed to be in the same slit, 
    but the objects were not correctly placed.}
  \tablenotetext{e}{The target acquisition was failed and the object was not correctly placed in the slit.}
\end{deluxetable}


\clearpage
\begin{landscape} 
\begin{deluxetable}{lcccccccc}
  \tabletypesize{\scriptsize} 
  \tablewidth{0pt}
  \tablecaption{Emission Line Properties of \ha{} and [\ion{N}{2}]$\lambda6583$ \label{table:lineprop}}
  \tablehead{
    \colhead{ID} & 
    \colhead{$z_{\text{\ha}}$} & 
    \colhead{$f(\text{\ha})$} & 
    \colhead{EW$_{\text{rest}}$(\text{\ha})\tablenotemark{a,b}} & 
    \colhead{$f$([N$\;${\tiny{II}}])} & 
    \colhead{EW$_{\text{rest}}$([N$\;${\tiny{II}}])\tablenotemark{a}} & 
    \colhead{[N$\;${\tiny{II}}]/\ha} &
    \colhead{$\sigma$\tablenotemark{c}}
    \\
    \colhead{} &
    \colhead{} &
    \colhead{($10^{-17}$ergs\ s$^{-1}$cm$^{-2}$)} &
    \colhead{(\AA)} &
    \colhead{($10^{-17}$ergs\ s$^{-1}$cm$^{-2}$)} &
    \colhead{(\AA)} &
    \colhead{} &
    \colhead{km s$^{-1}$}
  }
  \startdata
  \multicolumn{8}{l}{\hspace{-1.5ex}{\textsc{Ohs}}} \\[0.5ex]
  Dad 1901            & $1.600$ & $15 \pm 1$  & $ 38 \pm  8$ & $7.9 \pm 0.7$ & $20 \pm  4$ & $0.53 \pm 0.06$ & \nodata \\
  Dad 3551            & $1.603$ & $17 \pm 1$  & $ 55 \pm 12$ & $1.6 \pm 0.9$ & $ 5 \pm  3$ & $0.09 \pm 0.05$ & \nodata \\
  \textsl{D3a 8608}   & $1.528$ & $16 \pm 2$  & $ 86 \pm 20$ & $11 \pm 1$    & $60 \pm 13$ & $0.70 \pm 0.11$ & \nodata \\
  D3a 12556           & $1.588$ & $50 \pm 3$  & $184 \pm 38$ & $11 \pm 3$    & $40 \pm 14$ & $0.22 \pm 0.06$ & \nodata \\[1.5ex]
  \multicolumn{8}{l}{\hspace{-1.5ex}{\textsc{Cisco}}} \\[0.5ex]
  Dad 2426            & $2.397$ & $9.0\pm1.8$ & $ 22 \pm  6$ & $3.4 \pm 2.7$ & $ 9 \pm  7$ & $0.38 \pm 0.31$ & \nodata \\[1.5ex]
  \multicolumn{8}{l}{\hspace{-1.5ex}{\textsc{Sinfoni}}} \\[0.5ex]
  Dad 2426-b          & $1.772$ & $31 \pm 1$  & $>194$       & $12 \pm 1$    & $>76$       & $0.39 \pm 0.04$ & 245 \\
  D3a 3287            & $2.205$ & $9.2\pm2.0$ & $ 89 \pm 26$ & $<2.0$        & $<19$       & $<0.22$         & 118 \\
  D3a 4626            & $1.586$ & $4.4\pm0.3$ & $ 22 \pm  6$ & $<3.0$        & $<15$       & $<0.68$         & 212 \\
  D3a 4654            & $1.551$ & $33 \pm 2$  & $ 52 \pm 11$ & $21\pm1$      & $33 \pm  7$ & $0.63 \pm 0.05$ & 170 \\
  D3a 4751            & $2.266$ & $21 \pm 2$  & $165 \pm 37$ & $4.1\pm1.6$   & $33 \pm 14$ & $0.20 \pm 0.08$ &  74:\tablenotemark{d}\\
  D3a 5814            & $2.141$ & $39 \pm 3$  & $316 \pm 68$ & $15\pm3$      & $120\pm 34$ & $0.38 \pm 0.08$ & 169 \\
  D3a 6397            & $1.513$ & $47 \pm 5$  & $ 99 \pm 23$ & $16\pm5$      & $34 \pm 13$ & $0.34 \pm 0.11$ & 265 \\
  D3a 7429            & $1.694$ & $17 \pm 4$  & $ 96 \pm 29$ & $6.0\pm3.5$   & $32 \pm 20$ & $0.34 \pm 0.21$ & 125 \\
  \textsl{D3a 11391} narrow&$1.774$  & $13 \pm 1$ & $46  \pm 10$ & \nodata & \nodata   & \nodata &  187 \\
  \phantom{\textsl{D3a 11391}} broad& \nodata & $35 \pm 2$ & $123 \pm 26$ & \nodata & \nodata   & \nodata & 1040
  \enddata
  \tablenotetext{a}{Equivalent widths are calculated by assuming the continuum flux from the best-fit SED. 
    For Dad-2426-b, $K_{\text{AB}}=21.5$ \citepalias{kong:2006} is used for upper limit of the continuum}
  \tablenotetext{b}{\hasp equivalent widths are corrected for underlying stellar \hasp absorption calculated from the best-fit SED. 
    For Dad-2426-b, the average of the \hasp absorption line equivalent widths of the other objects, 4.4~\AA{}, is assumed.}
  \tablenotetext{c}{The velocity dispersions are corrected for the instrumental resolution, 
    $\sigma_{\text{instrument}}=85$~km s$^{-1}$.}
  \tablenotetext{d}{This is smaller than instrumental resolution, hence the uncertainty is large.}
  \tablecomments{
    [N$\;${\tiny{II}}] stands for  [N$\;${\tiny{II}}]$\lambda6583$.  
    The object IDs with slanted fonts  are considered to be 
    AGN-dominated \sbzks judged from a large [N$\;${\tiny{II}}]/\hasp ratio (D3a-8608)  
    or a broad-line component (D3a-11391).}
\end{deluxetable}
\clearpage
\end{landscape}

\begin{deluxetable}{lcc}
  \tabletypesize{\scriptsize}
  \tablewidth{0pt}
  \tablecaption{SED Fitting Results  \label{table:propsed}}
  \tablehead{
    \colhead{ID} & 
    \colhead{$E(B-V)_{\text{SED}}$\tablenotemark{a,b}} & 
    \colhead{$\log(M_{\star\text{SED}}/M_\sun)$\tablenotemark{b}} \\
    \colhead{} & 
    \colhead{(mag)} & 
    \colhead{}
  }
  \startdata
  \multicolumn{3}{l}{\hspace{-1.5ex}{\textsc{Ohs}}} \\[0.5ex]
  Dad 1901  & 0.25 & 11.23 \\
  Dad 3551  & 0.20 & 11.08 \\
  D3a 8608  & 0.25 & 10.20 \\
  D3a 12556 & 0.20 & 10.73 \\
  \multicolumn{3}{l}{\hspace{-1.5ex}{\textsc{Cisco}}} \\[0.5ex]
  Dad 2426  & 0.65 & 11.63 \\
  \multicolumn{3}{l}{\hspace{-1.5ex}{\textsc{Sinfoni}}} \\[0.5ex]
  Dad 2426-b& \nodata & \nodata \\
  D3a 3287  & 0.15    & 10.75   \\
  D3a 4626  & 0.52    & 10.94   \\
  D3a 4654  & 0.45    & 11.43   \\
  D3a 4751  & 0.25    & 10.72   \\
  D3a 5814  & 0.60    & 10.90   \\
  D3a 6397  & 0.55    & 11.14   \\
  D3a 7429  & 0.40    & 10.72   \\
  D3a 11391 & 0.10    & 11.10
  \enddata
  \tablenotetext{a}{$E(B-V)$ from the stellar continuum.}
  \tablenotetext{b}{Derived by SED fitting.}
\end{deluxetable}

\begin{deluxetable}{lcccccc}
  \tabletypesize{\scriptsize}
  \tablewidth{0pt}
  \tablecaption{Extinctions and \ha{} Star-formation Rates  \label{table:sfr}}
  \tablehead{
    \colhead{ID} & 
    \colhead{$A_V^\star$} &
    \colhead{$A_V^{\text{gas}}$} & 
    \colhead{$A_{\text{\ha}}$} &
    \colhead{$L_{\text{\ha}}$\tablenotemark{a}} & 
    \colhead{SFR$_{\text{raw}}$\tablenotemark{a}} &
    \colhead{SFR$_{\text{corr}}$\tablenotemark{b}} \\
    \colhead{} &
    \colhead{(mag)} & 
    \colhead{(mag)} &
    \colhead{(mag)} & 
    \colhead{($10^{42}$ ergs s$^{-1}$)} &
    \colhead{($M_\sun$ yr$^{-1}$)} & 
    \colhead{($M_\sun$ yr$^{-1}$)}
  }
  \startdata
  \multicolumn{5}{l}{\hspace{-1.5ex}{\textsc{Ohs}}} \\[0.5ex]
  Dad 1901  & 1.0 & 2.1 & 1.7 & $2.2 \pm 0.2$ & $20 \pm 1$ & $  98 \pm  7$ \\
  Dad 3551  & 0.6 & 1.4 & 1.1 & $2.7 \pm 0.2$ & $24 \pm 1$ & $  87 \pm  5$ \\
  D3a 8608  & 1.0 & 2.1 & 1.7 & $2.2 \pm 0.2$ & $19 \pm 2$ & $  92 \pm 12$ \\
  D3a 12556 & 0.8 & 1.7 & 1.4 & $8.1 \pm 0.6$ & $66 \pm 4$ & $ 246 \pm 15$ \\
  \multicolumn{5}{l}{\hspace{-1.5ex}{\textsc{Cisco}}} \\[0.5ex]
  Dad 2426  & 2.6 & 5.1 & 4.2 & $3.3 \pm 0.8$ & $33 \pm 7$ & $1540 \pm 320$ \\
  \multicolumn{5}{l}{\hspace{-1.5ex}{\textsc{Sinfoni}}} \\[0.5ex]
  Dad 2426-b&\nodata&\nodata&\nodata&$6.1\pm0.1$&$53\pm 2$ &   \nodata \\
  D3a 3287  & 0.6 & 1.4 & 1.1 & $3.2 \pm 0.7$ & $28 \pm 6$ & $  77 \pm  17$ \\
  D3a 4626  & 2.1 & 4.1 & 3.4 & $0.6 \pm 0.3$ & $5.7\pm0.4$& $ 129 \pm   8$ \\
  D3a 4654  & 1.8 & 3.6 & 3.0 & $4.6 \pm 0.3$ & $41 \pm 3$ & $ 622 \pm  39$ \\
  D3a 4751  & 1.0 & 2.1 & 1.7 & $7.9 \pm 0.6$ & $66 \pm 6$ & $ 323 \pm  31$ \\
  D3a 5814  & 2.4 & 4.7 & 3.9 & $13  \pm 1.0$ & $110 \pm 8$& $3880 \pm 290$ \\
  D3a 6397  & 2.2 & 4.3 & 3.6 & $6.5 \pm 0.7$ & $55 \pm 6$ & $1460 \pm 160$ \\
  D3a 7429  & 1.6 & 3.2 & 2.6 & $3.3 \pm 0.7$ & $28 \pm 6$ & $ 315 \pm  71$ \\
  D3a 11391\tablenotemark{c} & 0.4 & 1.0 & 0.8 & $2.5 \pm 0.3$ & $23\pm2$ & $48 \pm 4$
  \enddata
  \tablenotetext{a}{Values not corrected for extinction.}
  \tablenotetext{b}{Extinction corrected values.}
  \tablenotetext{c}{Only the contribution from the narrow-line component is listed and 
    the narrow-line is assumed to come from star-forming regions.}
\end{deluxetable}

\begin{deluxetable}{lc}
  \tabletypesize{\scriptsize}
  \tablewidth{0pt}
  \tablecaption{Gas-phase oxygen abundances  \label{table:oh12bzk}}
  \tablehead{
    \colhead{ID} & 
    \colhead{$12+\log(\text{O/H})$}
  }
  \startdata
  \multicolumn{2}{l}{\hspace{-1.5ex}{\textsc{Ohs}}} \\[0.5ex]
  Dad 1901  & $ 9.21 \pm 0.09$ \\
  Dad 3551  & $ 8.55 \pm 0.21$ \\
  D3a 8608  & $(9.53 \pm 0.24)$ \\
  D3a 12556 & $ 8.84 \pm 0.08$ \\
  \multicolumn{2}{l}{\hspace{-1.5ex}{\textsc{Cisco}}} \\[0.5ex]
  Dad 2426  & $ 9.02 \pm 0.34$ \\
  \multicolumn{2}{l}{\hspace{-1.5ex}{\textsc{Sinfoni}}} \\[0.5ex]
  Dad 2426-b& $ 9.03 \pm 0.04$ \\
  D3a 3287  & $<8.84$ \\
  D3a 4626  & $<9.48$ \\
  D3a 4654  & $ 9.38 \pm 0.10$ \\
  D3a 4751  & $ 8.81 \pm 0.12$ \\
  D3a 5814  & $ 9.02 \pm 0.09$ \\
  D3a 6397  & $ 8.97 \pm 0.12$ \\
  D3a 7429  & $ 8.97 \pm 0.22$ \\
  D3a 11391 & \nodata 
  \enddata
  \tablecomments{D3a 8608 which has [N$\;${\tiny{II}}]/\ha~$\ge 0.7$ is 
    considered to host an AGN. Thus the derived metallicity is listed within parentheses.}
\end{deluxetable}

\tabletypesize{\scriptsize}
\begin{deluxetable}{lc}
  \tablewidth{0pt}
  \tablecaption{Dynamical masses for SINFONI objects  \label{table:mdyn}}
  \tablehead{
    \colhead{ID} & 
    \colhead{$\log(M_{\text{dyn}}/M_\sun)$}
  }
  \startdata
  Dad 2426-b& 11.6 \\
  D3a 3287  & 11.1 \\
  D3a 4626  & 11.5 \\
  D3a 4654  & 11.3 \\
  D3a 4751  & 10.6 \\
  D3a 5814  & 11.4 \\
  D3a 6397  & 11.7 \\
  D3a 7429  & 11.0 \\
  D3a 11391 & 11.4\tablenotemark{a}
  \enddata
  \tablenotetext{a}{Calculated from the velocity dispersion of the narrow-line component.}
\end{deluxetable}

\end{document}